\documentclass[nofootinbib,
    reprint,
    aps,
    pra,
    superscriptaddress,
    longbibliography
]{revtex4-2}

\usepackage{xcolor, booktabs, tabularx, array}
\usepackage{graphicx}
\usepackage{amsmath,amssymb}
\usepackage{bm}
\usepackage{physics}
\usepackage[scaled=0.92]{helvet}
\usepackage{dsfont}
\usepackage{tikz}
\usepackage{pgfplots}
\usepackage{pgfplotstable}
\usepackage{hyperref}
\usepackage{cleveref}
\usepgfplotslibrary{groupplots}
\usetikzlibrary{calc}

\pgfplotsset{compat=1.18}
\usetikzlibrary{arrows.meta,backgrounds,calc,decorations.pathmorphing,fit,positioning}

\definecolor{exactblack}{HTML}{111111}
\definecolor{mone}{HTML}{E31A1C}
\definecolor{mtwo}{HTML}{0072FF}
\definecolor{mthree}{HTML}{FF9800}
\definecolor{mfour}{HTML}{8A2BE2}

\newcommand{\OH}{\ensuremath{\mathbf{O}_H}}
\newcommand{\OS}{\ensuremath{\mathbf{O}_S}}
\newcommand{\h}{\mathbf{H}}
\newcommand{\normsp}[2]{\ensuremath{\left\lVert #1\right\rVert_{#2}}}
\newcommand{\genH}{\ensuremath{\bm{\mathcal{L}}}}

\newtheorem{lemma}{Lemma}
\newtheorem{remark}{Remark}

\begin{document}

\title{Generalized \texorpdfstring{$m$}{m}-body mean-field projections for classical simulation of \texorpdfstring{$d$}{d}-dimensional Heisenberg picture dynamics}

\author{Federico Tom{\'a}s B. P{\'e}rez}
\email{ftbperez.quantum@gmail.com}
\affiliation{Institut für Theoretische Physik, Albert-Einstein-Allee 11, Universität Ulm, D-89081 Ulm, Germany}
\affiliation{Center for Integrated Quantum Science and Technology (IQST), 89081 Ulm, Germany}
\affiliation{Information Sciences, Los Alamos National Laboratory, Los Alamos, NM 87545, USA}
\affiliation{Instituto de Física La Plata (IFLP), CONICET--Universidad Nacional de La Plata, Departamento de Física, Facultad de Ciencias Exactas, C.C. 67, 1900 La Plata, Argentina}
\author{Gabriela W{\'o}jtowicz}
\email{gabriela.wojtowicz@uni-ulm.de}
\affiliation{Institut für Theoretische Physik, Albert-Einstein-Allee 11, Universität Ulm, D-89081 Ulm, Germany}
\affiliation{Center for Integrated Quantum Science and Technology (IQST), 89081 Ulm, Germany}

\author{Matthias Kost}
\email{matthias.kost@uni-ulm.de}
\affiliation{Institut für Theoretische Physik, Albert-Einstein-Allee 11, Universität Ulm, D-89081 Ulm, Germany}
\affiliation{Center for Integrated Quantum Science and Technology (IQST), 89081 Ulm, Germany}

\author{Martin B. Plenio}
\email{martin.plenio@uni-ulm.de}
\affiliation{Institut für Theoretische Physik, Albert-Einstein-Allee 11, Universität Ulm, D-89081 Ulm, Germany}
\affiliation{Center for Integrated Quantum Science and Technology (IQST), 89081 Ulm, Germany}

\date{\today}

\begin{abstract}
Classical simulation of quantum many-body dynamics is challenged by the rapid growth of operator complexity, particularly in higher dimensions where tensor-network methods become increasingly demanding.
We develop state-adapted Pauli-weight propagation based on generalized $m$-body mean-field projections, which compress many-body observables into optimal low-body representations in a covariance geometry induced by the initial state.
Combined with adaptive Krylov-like restricted Heisenberg dynamics, this provides an alternative to conventional Hilbert--Schmidt Pauli-weight truncation.
We derive necessary and conditional sufficient bounds on low-body approximation accuracy using standard operator complexity measures and characterize possible dynamical failure mechanisms of the state-adapted method.
Numerical benchmarks demonstrate substantial state-dependent accuracy gains, while identifying regimes of reduced accuracy or violations of Pauli spectral bounds.
For a three-dimensional system of $80$ interacting spins, our method achieves strong body-order convergence, close agreement with independently converging tensor-network simulations, and substantially smaller sparse Pauli representations.
These results establish state-adapted Pauli propagation as a promising approach to classical simulation of higher-dimensional coherent many-body dynamics.
\end{abstract}
\maketitle

\section{Introduction}
\label{sec:introduction}

Programmable quantum simulators now access non-equilibrium many-body dynamics at system sizes and evolution times that increasingly challenge direct classical calculation~\cite{King_2025, Andersen_2025, Manovitz_2025, Vetter_2026}.
Classical simulation nevertheless remains essential for benchmarking experiments, establishing convergence, and determining where quantum simulations exceed known numerical approaches.
As exact calculations become unavailable, these tasks increasingly rely on controlled approximation hierarchies and comparisons between independent classical methods~\cite{King_2025,Andersen_2025,Manetsch_2025,Vetter_2026}.

The underlying difficulty is the curse of dimensionality, arising from the exponential growth of Hilbert space and the dynamical generation of correlations under interacting evolution~\cite{eisert2010,haegeman_time-dependent_2011}.
Classical methods circumvent this growth only when the dynamics admit a suitably compressed representation.
Tensor-network (TN) methods compress many-body states and operators through structured low-rank representations, including matrix-product states and operators, respectively, ~\cite{Oestlund1995, Rommer1997, vidal2003, schollwoeck2011,orus_tensor_2019}.
Their representation cost is governed by bond dimensions, which are constrained by bipartite entanglement for states and operator-space entanglement for observables.
More general TN geometries can improve representational efficiency but typically increase the computational cost of contraction~\cite{vidal_class_2008,orus_practical_2014,vlaad_simulation_2021}.
Furthermore, efficient truncation beyond one-dimensional TN representations is generally nontrivial~\cite{Haegeman_2016,Yang_2020}.
During interacting time evolution, entanglement growth can require increasing bond dimensions to maintain accuracy~\cite{haegeman_time-dependent_2011,Haegeman_2016}.
Working in the Heisenberg picture can mitigate this difficulty in favorable regimes, but does not eliminate it for generic many-body dynamics~\cite{hartmann2009,clark_exact_2010,pizorn_real_time_2014,hemery_matrix_2017}.

A complementary simulation method relies on sparse representations in operator space using a selected operator basis.
For spin systems, Pauli propagation (PP) represents a Heisenberg
observable as a linear combination of Pauli strings, retaining only a select dynamically relevant subset during evolution.
Its computational cost is therefore governed by the complexity
of the evolving representation in the Pauli basis and by the
truncation criterion employed.
Existing strategies include coefficient thresholds, top-$K$
selection, Pauli-path expansions, Pauli-weight truncation,
and criteria based on locality or interaction-graph
geometry~\cite{shao2026_topK,Angrisani_2025,Angrisani_2026,
GGarcia_Pauli_path_2025,clifford_pert_theo,Loizeau_2025,
pauli-qite,Kemper2026,D_Anna_2026}, with recent advances in state-informed criteria~\cite{xSPD2026}.
The experimental relevance of such low-body truncations is
illustrated by their application to modeling pulsed nuclear-spin
dynamics in atomically thin diamond layers, involving an
effectively two-dimensional interacting spin ensemble~\cite{Vetter_2026}.

Standard PP schemes lead to rigorous efficient-simulation guarantees for Clifford or weakly non-Clifford dynamics, noisy circuits, and selected circuit ensembles~\cite{clifford_pert_theo,schuster2024,Angrisani_2026,GGarcia_Pauli_path_2025}.

This is not guaranteed for generic coherent Hamiltonian dynamics where no mechanism controlling the Pauli weight exists. 
There, repeatedly commuting with the Hamiltonian causes higher-weight components to proliferate, and a component that is negligible at one time may not stay dynamically unimportant at later times. 
Coefficient- and weight-based truncations may therefore remain useful numerical approximations, but their truncation criteria do not, by themselves, provide general finite-time accuracy guarantees for coherent continuous-time simulation of Hamiltonian dynamics~\cite{Loizeau_2025,pauli-qite}.
Independently of accuracy, approximate Heisenberg dynamics need not preserve the spectrum of the original observable.
Exact unitary evolution preserves the spectrum of a Pauli observable, maintaining its operator norm and confining expectation values to $[-1,1]$ for every physical state.
However, preservation or approximate control of the Hilbert--Schmidt norm does not guarantee these spectral constraints, and truncated observables may exhibit expectation values outside the exact spectral interval~\cite{Angrisani_2026,xSPD2026,Li_2026}.

Fundamentally, most truncation criteria are state agnostic: they assess operator components through coefficient magnitude, support, body order, path amplitude, geometric distance, or a state-independent norm.
These notions need not coincide with relevance to an expectation value in a particular initial state, an idea recently explored in state-informed PP schemes~\cite{xSPD2026}.
This motivates the question addressed in this work and in Ref.~\cite{Perez_short_paper_2026}: given an initial state \(\sigma_0\), how should dynamically-generated many-body operator content be compressed into a retained low-body space when the objective is to reproduce expectation values in \(\sigma_0\), rather than reconstruct the full Heisenberg operator?
A high-body component can be large in a state-independent norm while being weakly resolved by \(\sigma_0\), whereas another can possess nontrivial low-body representatives once contracted against the correlations of the state.
Discarding all information above a body-order threshold therefore generally will not retain the information most relevant to the observable being calculated.

In this work, we develop state-adapted Pauli-weight propagation to address this problem, which we call covariance dynamics for short.
Building on the generalized mean-field and restricted maximum-entropy constructions introduced in Refs.~\cite{perez_quantum_2024, Perez_2026, Perez_short_paper_2026}, we define low-body representations through a covariance geometry induced by a tractable reference state.
The retained linear space remains the familiar space $\mathcal A_m$ of observables of body order at most $m$, but higher-body operators are represented by their optimal low-body representatives in the state-adapted geometry rather than simply discarded.
Related low-order closures arise throughout many-body physics, including mean-field, BBGKY, and cumulant-based constructions~\cite{Weiss-curie,Plankensteiner_2022}; the present formulation instead derives the closure from an explicit geometric projection, for which representation errors can be characterized independently of the subsequent dynamical approximation.

We combine this projection with sparse Pauli propagation and an adaptive Krylov-like representation of restricted Heisenberg dynamics.
For full-rank reference states, the covariance form defines a genuine operator geometry on the retained space.
For rank-deficient states it instead defines a seminorm, introducing covariance-null directions and a quotient-space structure whose interaction with Hamiltonian evolution leads to a nontrivial dynamical-compatibility problem.
This distinction is central both mathematically and numerically, since directions that are initially invisible to the covariance geometry can become dynamically coupled to retained bright directions.

The present article extends the accompanying Letter~\cite{Perez_short_paper_2026} by analyzing the dynamical compatibility of rank-deficient covariance geometry, deriving necessary and conditional sufficient bounds on low-body approximations, and detailing the adaptive numerical implementation.
We focus on product reference states, for which the covariance projection admits an explicit construction entirely in terms of local expectation values.
Our numerical benchmarks consider non-integrable transverse-field Ising models, varying the purity and orientation of the initial state to identify favorable and unfavorable regimes for state-adapted Pauli-weight propagation.
In three dimensions, we further demonstrate strong body-order convergence and agreement with independently converging tensor-network simulations for a system of $80$ interacting spins.

The manuscript is organized as follows.
In Sec.~\ref{sec:heisenberg-picture}, we recap on the Heisenberg-picture formulation of state dynamics. 
In Sec.~\ref{sec:state-adapted-projection}, we explain the covariance geometry and related state-adapted projection to fixed operator subspace.
In Sec.~\ref{sec:projected-dynamics}, we introduce the restricted Heisenberg dynamics and discuss dynamical compatibility for rank-deficient reference states.
In Sec.~\ref{sec:complexity-errors}, we develop complexity and error bounds. 
In Sec.~\ref{sec:numerical-implementation}, we describe the numerical implementation. 
Finally, in Sec.~\ref{sec:results}, we present the numerical results and discuss the comparison between our scheme and complementary approaches. 

\section{Heisenberg-picture simulation}
\label{sec:heisenberg-picture}

We work in the Heisenberg picture, where the initial state is kept fixed and only the observables of interest are propagated.
For an initial state $\rho_0$ and a Schr\"odinger-picture observable $\OS$,
\begin{equation}
\Tr[\OS\rho(t)] = \Tr[\rho_0\OH(t)],
\label{eq:heisenberg-expectation}
\end{equation}
where
\begin{equation}
\dot{\mathbf O}_H(t) = \genH\!\left(\OH(t)\right),
\qquad \OH(0)=\OS .
\label{eq:heisenberg-eom}
\end{equation}
For the unitary coherent dynamics considered here, 
\begin{equation}
\genH(\OH) = i[\h,\OH],
\label{eq:unitary}
\end{equation}
with $\h$ the many-body Hamiltonian.

For an initially local observable evolving under a local Hamiltonian, Lieb--Robinson (LR) locality constrains the growth of the spatial region over which $\OH(t)$ can develop appreciable support~\cite{lieb1972}.
Up to exponentially suppressed tails, the dynamics may therefore be restricted to an effective causal region $B(t)$ whose volume grows polynomially with time.
This spatial restriction, however, does not imply a corresponding restriction in operator correlation or body order.
Within $B(t)$, repeated commutation with the Hamiltonian can generate an increasing number of higher-body Pauli components, and the complete operator space available on the active region still grows as $\mathcal O(4^{|B(t)|})$.
The relevant compression problem is, therefore, not only to restrict the spatial support of the Heisenberg observable, but to control the growth of its operator content within that support.

\section{State-adapted low-body projection}
\label{sec:state-adapted-projection}

This section introduces the static operator geometry and the associated state-adapted low-body projection used throughout this work.
The discussion is deliberately independent of the subsequent time evolution.
Our purpose here is to make precise three separate objects: the low-body space that is retained, the state-dependent notion of statistical distinguishability, and the representative selected when a higher-body operator is compressed into the retained space.

\subsection{Pauli propagation and low-body hierarchy}
\label{subsec:pauli-low-body}

For a system consisting of $N$ spin-\(1/2\) sites, the complete set of Pauli strings reads $\mathcal P_N=\{I,X,Y,Z\}^{\otimes N}$, where the set forms an orthogonal basis of operator space in the Hilbert--Schmidt (HS) geometry.

The basis can be categorized into a hierarchy of $m$ operator spaces based on Pauli weight,
\begin{equation}
    \mathcal A_0
    \subset
    \mathcal A_1
    \subset
    \cdots
    \subset
    \mathcal A_N,\label{eq:hierarchy_Am}
\end{equation}
where \(\mathcal A_0=\operatorname{span}\{I\}\) is the lowest subspace, ${\cal A}_N$ is the entire space of all operators and any up-to-$m$-body space is defined as
\begin{equation}
    \mathcal A_m
    =
    \operatorname{span}
    \left\{
        P\in\mathcal P_N
        \,\middle|\,
        \operatorname{wt}(P)\leq m
    \right\},
    \label{eq:am-definition}
\end{equation}
where \(\operatorname{wt}(P)\), the Pauli weight is given by a number of nonidentity local factors in the Pauli string.
The space dimension depends on the system size and the maximal body order
\begin{equation}
    D_m(N) \equiv \dim(\mathcal A_m) =
    \sum_{k=0}^{m}\binom Nk 3^k.
    \label{eq:dim-am}
\end{equation}
For fixed \(m\) and large $N$, $D_m(N)\leq\mathcal O(N^m)$ implies polynomial growth in terms of \(N\).
This growth pattern of storing and representing ${\cal A}_m$ does not imply efficient classical simulation because the body order required to achieve a prescribed accuracy and the sparsity of the evolving representation are separate dynamical questions.

In the PP simulation, any Heisenberg-picture observable is expressed in terms of Pauli basis
\begin{equation}
    \OH(t) = 
    \sum_{P\in\mathcal P_N} c_P(t)P,
    \label{eq:pauli-expansion}
\end{equation}
where 
\begin{equation}
c_P(t)=2^{-N}\Tr[P\OH(t)]
\end{equation}
is a dynamical coefficient corresponding to the Pauli operator $P\in\mathcal P_N$. 

The construction developed in this article and in \cite{Perez_short_paper_2026} keeps \(\mathcal A_m\) as the retained operator space.
What changes relative to ordinary HS-based Pauli-weight truncation is how an operator outside of \(\mathcal A_m\) is represented inside it.
Rather than discarding every Pauli observable whose original weight exceeds \(m\) we use a systematic projection to low-body representative that is optimal in a geometry induced by a physically sensible reference state.

\subsection{State-adapted covariance geometry}
\label{subsec:covariance-geometry}

On the real vector space of Hermitian operators, we use the (symmetrized and uncentered) covariance (covar) bilinear form
\begin{equation}
    (\mathbf A,\mathbf B)_{\mathrm{covar}(\sigma_0)} =
    \Tr\!\left[
        \frac{\sigma_0}{2}\{\mathbf A,\mathbf B\}
    \right],
    \label{eq:covar-sp}
\end{equation}
which for Hermitian operators $\mathbf A$ and $\mathbf B$ is $(\mathbf A,\mathbf B)_{\mathrm{covar}(\sigma_0)}\in\mathbb R$ and $\sigma_0$ is the reference state for the bilinear form. 
In the limit $\sigma_0 \propto I$, the HS scalar product is recovered.

The expectation values are recovered when the reference state is the same as the initial state. 
Motivated by this we keep the reference state to be the initial state unless specified explicitly. 
The relation is directly given through
\begin{equation}
    \Tr(\sigma_0\mathbf O) = 
    (I,\mathbf O)_{\mathrm{covar}(\sigma_0)}.
    \label{eq:expectation-as-covar-inner-product}
\end{equation}

The associated covariance quadratic form  for Hermitian $\mathbf A$ is
\begin{equation}
    \normsp{\mathbf A}{\mathrm{covar}(\sigma_0)}^2 =
    (\mathbf A,\mathbf A)_{\mathrm{covar}(\sigma_0)} =
    \Tr(\sigma_0\mathbf A^2),
    \label{eq:covar-norm}
\end{equation}
and the following hierarchy holds
\begin{equation}
    \normsp{\mathbf A}{\mathrm{covar}(\sigma_0)}
    \leq
    \normsp{\mathbf A}{\mathrm{op}}
    \leq
    \normsp{\mathbf A}{\mathrm{Fro}}.
    \label{eq:covar-norm-bound}
\end{equation}
Therefore, the covariance geometry can strongly suppress directions that are globally large as operators but weakly resolved by the reference state.

For the density matrix $\sigma_0 = \sum_i p_i \ket{i}\bra{i}$ where $p_i$ are eigenvalues and $\ket{i}$ eigenstates of the density matrix, the quadratic form
$\|\mathbf N\|_{\mathrm{covar}(\sigma_0)}^2 = \sum_i p_i\|\mathbf N\ket{i}\|^2$.
For full-rank state \(\sigma_0\), the quadratic form is uniquely zero, that is, $\normsp{\mathbf A}{\mathrm{covar}(\sigma_0)}=0$, if and only if $\mathbf A= \mathbf 0$ and $\normsp{\mathbf A}{\mathrm{covar}(\sigma_0)}>0$, otherwise. 
In this case \((\cdot,\cdot)_{\mathrm{covar}(\sigma_0)}\) is a genuine scalar product.
If instead \(\sigma_0\) is rank-deficient, there exist $\mathbf N\neq\mathbf 0$ for which all eigenstates of $\sigma_0$ lie in the kernel of $\mathbf N$ and the covariance product is not positive definite. 
Its null space is non-trivial and given by
\begin{equation}
\mathcal N^{\sigma_0} = \left\{
\mathbf N=\mathbf N^\dagger
\,\middle|\,
\normsp{\mathbf N}{\mathrm{covar}(\sigma_0)}=0
\right\},
\label{eq:covariance-null-space}
\end{equation}
and the covariant form induces only a seminorm. 

For a pure state, \(\sigma_0=\ket{\psi}\!\bra{\psi}\), implies $\mathbf N\in\mathcal N^{\sigma_0}$ such that $\mathbf N\ket{\psi}=0$. 

The covariance form becomes a genuine scalar product after quotienting out the null space,
\begin{equation}
    \mathcal A/\mathcal N^{\sigma_0},
    \label{eq:global-covariance-quotient}
\end{equation}
or, equivalently, removing all null-space directions from the operator space. 

In similarity to the operator spaces in Eq.~\eqref{eq:hierarchy_Am}, we define hierarchy of up-to-$m$-body operators in the null space
\begin{equation}
    \mathcal N_{0}^{\sigma_0}
    \subset
    \mathcal N_{1}^{\sigma_0}
    \subset
    \cdots
    \subset
    \mathcal N_{N}^{\sigma_0},
\end{equation}
where $\mathcal N_{N}^{\sigma_0}=\mathcal N^{\sigma_0}$ is the full null space and ${\cal N}_m^{\sigma_0}=\mathcal A_m\cap\mathcal N_{\sigma_0}$. 

Consequently, for rank-deficient states, the covariance seminorm induces an equivalence relation on operators
\begin{equation}
    \mathbf A\simeq_{\mathrm{covar}(\sigma_0)}\mathbf B
    \quad\Longleftrightarrow\quad
    \mathbf A-\mathbf B\in\mathcal N^{\sigma_0},
    \label{eq:covariance-equivalence}
\end{equation}
or equivalently $\normsp{\mathbf A-\mathbf B}{\mathrm{covar}(\sigma_0)}=0$. 
The equivalence relation in Eq.~\eqref{eq:covariance-equivalence} implies that two distinct operators are statistically indistinguishable in the reference-state geometry whenever their difference lies in the null space of the covariance form. 

The equivalence relation in Eq.~\eqref{eq:covariance-equivalence} is relevant from the practical point of view, since the null space may not be preserved by the simulation. 
This is further discussed in Sec.~\ref{sec:dynamical-compatibility}. 

The exact equivalence is nontrivial only for rank-deficient states. 
For full-rank \(\sigma_0\), the covariance-null space is trivial, so distinct observables cannot be exactly identified in the covariance quotient, however, directions may become near-null as a rank-deficient limit is approached.

\subsection{Covariance geometry and the 
\texorpdfstring{\(m\)}{m}-body projection}
\label{subsec:projection-forms}

For full-rank $\sigma_0$, the covariance form is a scalar product and the state-adapted projection onto $\mathcal A_m$ is the unique minimizer of the variational problem
\begin{equation}
    \pi_m^{\sigma_0}\mathbf{O}
    =
    \underset{\mathbf O_m\in\mathcal A_m}{\operatorname{argmin}}
    \;\normsp{\mathbf O-\mathbf O_m}{\mathrm{covar}(\sigma_0)}.
    \label{eq:pim-variational-definition}
\end{equation}
For rank-deficient $\sigma_0$, the same variational problem determines only an equivalence class of minimizing representatives, or equivalently a unique element of
\begin{equation}
\mathcal A_m/\mathcal N_m^{\sigma_0}.
\end{equation}
Any two minimizing representatives may differ by an element of $N_m^{\sigma_0}$, therefore the projected equivalence class is unique even when its operator representative is not, that is, we can always choose the $N_m^{\sigma_0}$ elements freely and independent of the full $\mathbf O$.

\subsubsection{General $m$-body projection}
\label{sec:general_pim_projection}

For an arbitrary reference state $\sigma_0$, the projection onto $\mathcal A_m$ is defined by the variational problem in Eq.~\eqref{eq:pim-variational-definition}.
For a basis $\{\mathbf Q_a\}_{a=1}^{D_m(N)}$ of $\mathcal A_m$, any minimizing representative
\[
\mathbf O^\star=\sum_{a=1}^{D_m(N)}c_a^\star\mathbf Q_a
\]
satisfies the covariance-orthogonality conditions
\[
\left(
\mathbf Q_a,\mathbf O-\mathbf O^\star
\right)_{\mathrm{covar}(\sigma_0)}
=0,
\qquad
a=1,\ldots,D_m(N).
\]
Defining the Gram matrix and the projection coefficients as
\[
G_{ab} = (\mathbf Q_a,\mathbf Q_b)_{\mathrm{covar}(\sigma_0)},
\qquad
b_a = (\mathbf Q_a,\mathbf O)_{\mathrm{covar}(\sigma_0)},
\]
with $\mathbf b=(b_a)_{a=1}^{D_m(N)}$ and $\mathbf c^\star=(c_a^\star)_{a=1}^{D_m(N)}$, the projection is obtained from the Gram system
\begin{equation}
G\,\mathbf c^\star=\mathbf b.
\label{eq:projm_general}
\end{equation}
Although the basis operators $\mathbf Q_a$ belong to $\mathcal A_m$, the original observable $\mathbf O$ is potentially many-body, so $\mathbf b$ generally contains covariance overlaps with operator contributions of body order greater than $m$.
These higher-body contributions are precisely what distinguishes the covariance projection from ordinary HS Pauli-weight truncation.
For full-rank $\sigma_0$, the Gram matrix is positive definite and the solution is unique.
For singular Gram matrices, the minimizing representative is defined only up to an element of $\mathcal N_m^{\sigma_0}$.

A direct construction requires evaluating up to $\mathcal O(D_m(N)^2)$ Gram-matrix elements and is therefore generally expensive as the retained body order or system size increases.
This cost can be reduced when additional structure of the reference state, operator basis, or Gram matrix is available.
For the product reference states considered in this work, the factorization of the covariance contractions avoids the global Gram construction entirely and yields the closed expression derived below.

\subsection{State-adapted projection for product states}
\label{subsec:product-state-projection}

In this paper, we restrict the reference states to be product states~\cite{perez_quantum_2024,Perez_short_paper_2026},
\begin{equation}
\label{eq:product_state}
\sigma_0=\bigotimes_{i=1}^{N}\sigma_i ,
\end{equation}
where $\sigma_i$ is a single-site density matrix.
For product reference states, the covariance projection can be evaluated entirely in terms of local expectation values and factorized contractions, without evaluating the general Gram construction in Eq.~\eqref{eq:projm_general}.

Importantly, operators with different or even disjoint spatial supports are not in general orthogonal with respect to the covariance scalar product in Eq.~\eqref{eq:covar-sp}.
For two product operators $\mathbf A_S$ and $\mathbf B_T$ with disjoint supports $S\cap T=\varnothing$, the product structure of $\sigma_0$ gives
\[
(\mathbf A_S,\mathbf B_T)_{\mathrm{covar}(\sigma_0)} = \Tr(\sigma_0\mathbf A_S)\Tr(\sigma_0\mathbf B_T),
\]
which is generally nonzero.
Therefore, the ordinary uncentered product basis of ${\cal A}_m$ does not directly provide an orthogonal decomposition of operator space with respect to the covariance geometry.

The orthogonal structure becomes explicit after centering the local observables, that is, separating on every site the expectation value of the local observable from its centered contribution.
For a local operator $\mathbf p_i$, we define
\begin{equation}
\mu_i=\Tr(\sigma_i\mathbf p_i),
\qquad
\mathbf r_i=\mathbf p_i-\mu_iI_i,
\label{eq:centered-local-operators}
\end{equation}
such that $\Tr(\sigma_i\mathbf r_i)=0$.
For each support $S\subseteq\Lambda$, let $\mathcal C_S$ denote the space spanned by product operators with centered factors on the sites in $S$ and identities elsewhere.
The full operator space can then be decomposed as
\[
\mathcal A=\bigoplus_{S\subseteq\Lambda}\mathcal C_S.
\]
The same at-most-$m$-body operator space introduced through Pauli weight ${\rm wt(P)}$ in Eq.~\eqref{eq:am-definition} can then be written as
\[
\mathcal A_m=\bigoplus_{|S|\leq m}\mathcal C_S.
\]
Thus, $\mathcal A_m$ is the same operator space introduced in Eq.~\eqref{eq:am-definition}; the reference state changes the orthogonal decomposition of the observable space and the geometry used to define the projection, but not the dimension of the retained space.

For product $\sigma_0$, centered-support sectors with different supports are covariance orthogonal.
Indeed, if $\mathbf A_S\in\mathcal C_S$ and $\mathbf B_T\in\mathcal C_T$ with $S\neq T$, there exists at least one site $j\in S\triangle T$ where one operator contains a centered factor while the other contains the identity. 
The corresponding local contraction is then $\Tr(\sigma_j\mathbf r_j)=0$, and therefore
\begin{equation}
(\mathbf A_S,\mathbf B_T)_{\mathrm{covar}(\sigma_0)}=0,
\qquad
S\neq T.
\label{eq:centered-support-orthogonality}
\end{equation}
No corresponding orthogonality between distinct operators within the same support sector $\mathcal C_S$ is required.

For a product observable $\mathbf P=\bigotimes_{i\in X}\mathbf p_i$ with support $X$, each local factor can be written as $\mathbf p_i=\mu_iI_i+\mathbf r_i$.
Expanding the product gives
\begin{equation}
\mathbf P
=
\sum_{S\subseteq X}
\left(
\prod_{i\in X\setminus S}\mu_i
\right)
\bigotimes_{i\in S}\mathbf r_i .
\label{eq:centered-product-expansion}
\end{equation}
The cardinality $|S|$ defines the centered body order of the corresponding contribution.

Since the different centered-support sectors are covariance orthogonal, the orthogonal projection onto $\mathcal A_m$ is obtained by retaining only the terms with centered body order $|S|\leq m$,
\begin{equation}
\pi_m^{\sigma_0}(\mathbf P) = \sum_{\substack{S\subseteq X\\|S|\leq m}}
\left(
\prod_{i\in X\setminus S}\mu_i
\right)
\bigotimes_{i\in S}\mathbf r_i .
\label{eq:pim-centered-product}
\end{equation}
The map extends to arbitrary observables by linearity.
For full-rank product states, Eq.~\eqref{eq:pim-centered-product} gives the unique covariance-orthogonal projection, while for rank-deficient product states it selects a definite representative of the minimizing equivalence class.

In the original uncentered form, for an $n$-body product $\mathbf P=\bigotimes_{i=1}^{n}\mathbf p_i$ and $m<n$, Eq.~\eqref{eq:pim-centered-product} can be written as
\begin{equation}
    \pi_m^{\sigma_0}(\mathbf P) =
    \sum_{r=0}^{m}
    \beta_r^{(n,m)}
    \sum_{\substack{
        T\subseteq\{1,\ldots,n\}\\
        |T|=r
    }}
    \left(
        \prod_{j\notin T}\mu_j
    \right)
    \bigotimes_{i\in T}\mathbf p_i,
    \label{eq:pim-uncentered-expansion}
\end{equation}
where
\begin{equation}
    \beta_r^{(n,m)}
    =
    (-1)^{m-r}
    \binom{n-r-1}{m-r},
    \qquad
    0\leq r\leq m<n.
    \label{eq:pim-beta-coefficients}
\end{equation}
For $m=n$, the projection acts as the identity.
The derivation of Eqs.~\eqref{eq:pim-uncentered-expansion} and \eqref{eq:pim-beta-coefficients} is given in App.~\ref{app:projection-properties}.

It is useful to distinguish the centered body order appearing in Eq.~\eqref{eq:centered-product-expansion} from the Pauli weight of the original operator.
Pauli weight ${\rm wt(P)}$ counts the number of nonidentity factors of the uncentered Pauli string, whereas centered body order counts the number of centered local factors after resolving the operator with respect to $\sigma_0$.
Consequently, a Pauli product with original weight larger than $m$ is not discarded as a whole, since its centered decomposition can contain contributions of body order $0,1,\ldots,m$ that remain inside the retained space.
The state-adapted projection therefore removes only the higher-centered-body contribution while retaining the lower-body information resolved by the reference-state geometry.

\section{Restricted Heisenberg dynamics and dynamical compatibility}
\label{sec:projected-dynamics}
\label{sec:dynamical-compatibility}

The state-adapted projection introduced in Sec.~\ref{sec:state-adapted-projection} provides an optimal $m$-body representation of a many-body observable with respect to the covariance geometry.
Its direct application to the exact Heisenberg observable,
$\pi_m^{\sigma_0}\OH(t)$, is, however, not useful as a simulation method since it requires the full evaluation of $\OH(t)\in\mathcal A_N$.
Instead, following Ref.~\cite{Perez_short_paper_2026}, we define a restricted Heisenberg dynamics that remains in $\mathcal A_m$ throughout the evolution.
We refer to this state-adapted restricted evolution as the covariance dynamics.

The post-facto projection $\pi_m^{\sigma_0}\OH(t)$ first evolves the complete observable and only then projects it onto $\mathcal A_m$, whereas the restricted dynamics continuously projects the generated operator content during the evolution.
The exact trajectory can therefore leave $\mathcal A_m$ and later return to it, while the restricted trajectory remains in $\mathcal A_m$ at all times.

For convenience, let
\[
P_m=\pi_m^{\sigma_0},
\qquad
Q_m=I-P_m.
\]
Projecting the exact Heisenberg equation into the retained and discarded sectors gives
\begin{align}
\frac{d}{dt}P_m\OH (t)
&=
P_m\genH P_m\OH(t)
+
P_m\genH Q_m\OH(t),
\\
\frac{d}{dt}Q_m\OH(t)
&=
Q_m\genH P_m\OH(t)
+
Q_m\genH Q_m\OH(t).
\end{align}
Basic algebraic manipulation yields
\begin{align}
\frac{d}{dt}P_m\OH(t)
&=
P_m\genH P_m\OH(t)
\label{eq:nz-mbody-projected}\\
&\quad+
P_m\genH e^{tQ_m\genH Q_m}Q_m\OH(0)
\label{eq:nz-mbody-initial}\\
&\quad+
\int_0^t ds\,
P_m\genH e^{(t-s)Q_m\genH Q_m}
Q_m\genH P_m\OH(s).
\label{eq:nz-mbody-memory}
\end{align}
The second term, Eq.~\eqref{eq:nz-mbody-initial}, contains contributions that are already outside the retained space at the initial time, while the third term, Eq.~\eqref{eq:nz-mbody-memory}, accounts for operator information that leaves $\mathcal A_m$ and subsequently returns to it.

For an initially retained observable, $Q_m\OH(0)=0$.
Neglecting the remaining memory term defines the restricted $m$-body flow
\begin{equation}
\dot{\mathbf O}_m(t)
=
\pi_m^{\sigma_0}\genH\mathbf O_m(t)
\equiv
\genH_m\mathbf O_m(t),
\qquad
\mathbf O_m(0)=\pi_m^{\sigma_0}\mathbf O_S,
\label{eq:restricted-heisenberg}
\end{equation}
where $\genH_m$ denotes the restricted generator on $\mathcal A_m$.
For the unitary dynamics of Eq.~\eqref{eq:unitary}, the right-hand side is simply $i\pi_m^{\sigma_0}[\h,\mathbf O_m(t)]$.

For full-rank reference states, Eq.~\eqref{eq:restricted-heisenberg} gives the instantaneous orthogonal, in covariance geometry, projection of the Heisenberg tangent $\genH\mathbf O_m(t)$ onto $\mathcal A_m$.
The restricted $m$-body flow should nevertheless not be confused with the post-facto projection of the exact trajectory $\mathbf O_m(t) \neq \pi_m^{\sigma_0}\OH(t)$, in general, precisely because the latter retains the effect of excursions through the discarded sector.
The representation and dynamical consequences of this distinction are quantified in Sec.~\ref{sec:complexity-errors}.

For rank-deficient reference states, an additional issue arises.
As discussed in Sec.~\ref{sec:state-adapted-projection}, the covariance geometry identifies operators that differ by elements of $\mathcal N_m^{\sigma_0}$.
This static covariance equivalence need not be preserved by the restricted Heisenberg dynamics.
If two retained representatives differ by $\mathbf D\in\mathcal N_m^{\sigma_0}$, their restricted tangents remain covariance equivalent if and only if
\begin{equation}
\normsp{
P_m\genH\mathbf D
}{
\mathrm{covar}(\sigma_0)
} = 0.
\label{eq:projected-alias-amplitude}
\end{equation}

Therefore, the restricted dynamics is independent of the representative chosen within each covariance-equivalence class if and only if
\begin{equation}
P_m\genH\,\mathcal N_m^{\sigma_0}
\subseteq
\mathcal N_m^{\sigma_0}.
\label{eq:quotient-compatibility}
\end{equation}
This relation expresses a strong version of dynamical compatibility of the covariance geometry and the restricted dynamics.
When it holds, different retained representatives of the same covariance-equivalence class generate equivalent restricted dynamics and therefore the same expectation values.
When it fails, a direction that is initially unresolved by the static covariance geometry can acquire a covariance-resolved component under the restricted generator with no coordinate available for its representation.
The condition $[\h,\sigma_0]\neq0$ permits such an activation, but does not imply it for every covariance-null direction or every trajectory.

For a covariance-null direction outside $\mathcal A_m$, Eq.~\eqref{eq:projected-alias-amplitude} has a related but distinct meaning. 
It measures whether a covariance-null component of the full operator space acquires a covariance-resolved image inside the retained space, rather than the compatibility between two retained representatives.
This distinction is particularly relevant when comparing projected observable contributions at different values of $m$, since the same null direction can lie outside one retained space and inside another.

The projected-activation condition is structural and should not be interpreted as a finite-time accuracy criterion.
For a particular null direction outside $\mathcal A_m$, vanishing $P_m\genH\mathbf D$ excludes its immediate activation into a covariance-resolved retained direction but does not exclude higher-order return through the discarded sector, a question that can only be answered from knowledge of the memory kernel in Eq.\eqref{eq:nz-mbody-projected}.
Conversely, a nonzero projected image identifies an available activation channel but does not imply a failing trajectory unless the corresponding operator direction is dynamically populated.
For full-rank reference states exact covariance-null directions are absent, although near-null directions can still become strongly amplified when a rank-deficient limit is approached.
These cases, together with explicit TFIM and anisotropic-$XY$ examples, are discussed in App.~\ref{app:dynamical-compatibility}.

Finally, the Liouvillian is not in general skew-adjoint in the covariance geometry.
For Hermitian operators ${\bf A}$ and ${\bf B}$,
\begin{equation}
({\bf A},\genH{\bf B})_{\mathrm{covar}(\sigma_0)} + (\genH{\bf A},{\bf B})_{\mathrm{covar}(\sigma_0)}
= \frac{i}{2}
\Tr\!\left(
[\sigma_0,\h]\{{\bf A},{\bf B}\}
\right).
\label{eq:covariance-nonskew}
\end{equation}
Thus, unless $[\sigma_0,\h]=0$ or a trajectory-specific cancellation occurs, the covariance norm is not required to remain constant under either the exact or restricted dynamics.
This mechanism is distinct from covariance-null incompatibility and can occur also for a strictly full-rank reference state.
Consequently, covariance-norm amplification alone does not diagnose activation of a covariance-null direction.

Having introduced the $m$-body flow in Eq.~\eqref{eq:restricted-heisenberg}, in the next section we delve in detail into the adaptive restricted approximation to it that is the basis of our method.

\section{Adaptive hierarchical bases}
\label{subsec:adaptive-hierarchical-bases}

We represent the restricted dynamics using moderate-depth projected Krylov spaces as moving local coordinate systems, referred to as Krylov-like Hierarchical Bases (HBs)~\cite{perez_quantum_2024,Perez_short_paper_2026}.
Writing
\begin{equation}
\genH_m :=
\pi_m^{\sigma_0}\genH,
\end{equation}
we divide the trajectory into intervals $[T_n,T_{n+1})$ and use the current approximate observable as the seed of the $n$th chart,
\begin{equation}
\mathbf b^{(n)}_0
=
\mathbf O_{\ell,m,\varepsilon,p}(T_n).
\label{eq:hb-local-seed}
\end{equation}
The adaptive approximation $\mathbf O_{\ell,m,\varepsilon,p}(t)$ to the $m$-body flow ${\bf O}_m(t)$ depends on several parameters, $\ell$ is the HB depth, $m$ the retained body order, $\varepsilon$ the reconstruction threshold, and $p$ the number of terminal coefficients monitored for reconstruction.
The initial condition is $\mathbf O_{\ell,m,\varepsilon,p}(0)=\pi_m^{\sigma_0}\mathbf O_S$.

Using ${\bf b}_0^{(n)} = O_{\ell,m,\varepsilon,p}(T_n)$, the remaining HB vectors are generated recursively using the restricted generator $\genH_m$ defined in Eq.~\eqref{eq:restricted-heisenberg},
\begin{equation}
\mathbf b^{(n)}_{\alpha+1}
=
\genH_m\!\left(\mathbf b^{(n)}_\alpha\right),
\qquad
\alpha=0,\ldots,\ell-1.
\label{eq:hb-local-basis}
\end{equation}
The additional action $\genH_m\mathbf b^{(n)}_\ell$ required to construct the final column of $\mathcal H^{(n)}$ is evaluated as an auxiliary vector and is not included in the HB.
Thus,
\begin{equation}
\mathcal K^{(n)}_{\ell,m}
=
\operatorname{span}
\left\{
\mathbf b^{(n)}_0,
\ldots,
\mathbf b^{(n)}_\ell
\right\}.
\label{eq:hb-local-chart}
\end{equation}
Within $[T_n,T_{n+1})$, the observable is represented as
\begin{equation}
\mathbf O_{\ell,m,\varepsilon,p}(t)
=
\sum_{\alpha=0}^{\ell}
\phi^{(n)}_\alpha(t)\mathbf b^{(n)}_\alpha,
\qquad
\phi^{(n)}_\alpha(T_n)
=
\delta_{\alpha0}.
\label{eq:hb-local-expansion}
\end{equation}
The local Gram and generator matrices are the real $(\ell+1)\times(\ell+1)$ matrices
\begin{align}
\mathcal G^{(n)}_{\alpha\beta}
&=
(\mathbf b^{(n)}_\alpha,\mathbf b^{(n)}_\beta)_{\mathrm{covar}(\sigma_0)},
\nonumber\\
\mathcal H^{(n)}_{\alpha\beta}
&=
(\mathbf b^{(n)}_\alpha,\genH_m\mathbf b^{(n)}_\beta)_{\mathrm{covar}(\sigma_0)}.
\label{eq:hb-local-matrices}
\end{align}
Projecting the residual of the restricted Heisenberg equation onto the HB using the covariance bilinear form gives
\begin{equation}
\mathcal G^{(n)}\dot{\vec\phi}^{(n)}(t)
=
\mathcal H^{(n)}\vec\phi^{(n)}(t).
\label{eq:hb-local-gram-equation}
\end{equation}
For a nonsingular Gram matrix, this equation uniquely determines the reduced coordinate dynamics.
For a singular Gram matrix, however, the equation determines $\dot{\vec\phi}^{(n)}$ only up to an element of $\ker\mathcal G^{(n)}$, since $\left(\mathcal G^{(n)}\right)^+\mathcal G^{(n)}$ is the orthogonal projector onto the complement of this kernel.
Choosing the minimum-Euclidean-norm coordinate derivative gives
\begin{equation}
\dot{\vec\phi}^{(n)}(t)
=
\left(\mathcal G^{(n)}\right)^{+}
\mathcal H^{(n)}
\vec\phi^{(n)}(t).
\label{eq:hb-local-effective-generator}
\end{equation}
For exact covariance contractions, $\mathcal H^{(n)}\vec\phi^{(n)}$ belongs to the range of $\mathcal G^{(n)}$, so this prescription satisfies the original Gram equation even when the Gram matrix is singular.
Nevertheless, for a rank-deficient covariance geometry, the selection of a minimum-norm coordinate representative does not by itself guarantee that the dynamics is independent of covariance-equivalent operator representatives.
Such independence requires the dynamical compatibility condition established in Sec.~\ref{sec:dynamical-compatibility}.

Adaptivity is controlled by the last $p$ coefficients of the current chart of the HB depth $\ell$ through the normalized Euclidean tail fraction
\begin{equation}
{\cal K}_{\ell, p}^{(n)}(t)
=
\frac{
\displaystyle
\sum_{\alpha=\ell-p+1}^{\ell}
\left|\phi^{(n)}_\alpha(t)\right|^2
}{
\displaystyle
\sum_{\alpha=0}^{\ell}
\left|\phi^{(n)}_\alpha(t)\right|^2
}.
\label{eq:hb-tail-functional}
\end{equation}
This coefficient-space diagnostic measures migration of the reduced coordinate vector toward the end of the finite HB chart and should not be interpreted as a covariance-norm fraction.
A Gram-weighted tail can become artificially small when terminal directions lie in or near the covariance-null space, whereas $K_p^{(n)}(t)$ does not inherit the rank or conditioning of $\mathcal G^{(n)}$.
The criterion remains chart dependent under independent rescalings of the basis vectors, but the recursive HB convention above fixes the coordinate convention used throughout the simulations.

The next reconstruction time is determined on the simulation grid by
\begin{equation}
T_{n+1}
=
\min
\left\{
t_j>T_n:
K_p^{(n)}(t_j)\geq\varepsilon
\right\}.
\label{eq:hb-reconstruction-time}
\end{equation}
If the threshold is not reached before the final simulation time, the current chart is retained to $t_{\max}$.
At a reconstruction, the endpoint observable becomes the seed of the next chart and the new coordinates are reset to $\phi_\alpha^{(n+1)}(T_{n+1})=\delta_{\alpha0}$.
The observable is therefore continuous across chart boundaries even though the reduced coordinate generator changes when the basis is rebuilt.
The threshold $\varepsilon$ is a reconstruction parameter rather than an a priori error bound.
Smaller $\varepsilon$ produces more frequent basis updates, whereas larger values permit the trajectory to move farther toward the end of the current chart before reconstruction.

Thus, $m$ controls the structural low-body closure, whereas $\ell$, $p$, and $\varepsilon$ control the adaptive HB representation of the resulting restricted dynamics.
The corresponding error decomposition is developed in \cref{sec:complexity-errors}.
A similar idea of co-evolving Krylov spaces for real-time pure-state Schr\"odinger evolution has recently been proposed in Ref.~\cite{kessing2026paces}.

\section{Complexity, error decomposition, and body-order requirements}
\label{sec:complexity-errors}

The central question in our ansatz and covariance dynamics is related to the $m$-order that is necessary to keep, at a given time $t$, in order for the covariance dynamics to provide an accurate approximation to the exact evolution, up to some tolerance $\delta$.
Addressing this question requires separating the error induced by the $m$-body restricted evolution into components that can be controlled at the level of expectation values and at the stronger, observable level.
We will quantify these errors using the covariance seminorm, which naturally captures the physically relevant size of an observable relative to the state under consideration.
This decomposition then allows us to translate a target accuracy threshold $\delta$ into necessary and conditional sufficient conditions on the retained body order $m$ in terms of intrinsic measures of operator complexity generated by the dynamics, thereby connecting the efficiency of the truncation scheme to quantities such as operator-space entanglement entropy and measures of non-stabilizerness, including the magic.

\subsection{Retained estimate of the adaptive restricted dynamics}
\label{subsec:numerical-cost}

We first estimate the largest share of operator space that the adaptive construction may have to explore.
Let $\mathbf{O}_S$ be initially local, and let $B(t)$ be a LR region containing $\mathbf{O}_H(t)$ up to an exponentially small spatial tail.
For a local generator $\genH$ on a $d$-dimensional bounded-degree lattice $\Lambda$, the number of active sites obeys
\begin{equation}
N(t):=|B(t)|\leq C_{\mathrm{LR}}(1+t)^d.
\label{eq:cost-lightcone-volume}
\end{equation}
Since $\pi_m^{\sigma_0}$ does not enlarge spatial support, see App.\ref{app:projection-support}, the number of (at-most) $m$-body Pauli observables contained inside this region is bounded by
\begin{equation}
D_m(t)=\sum_{r=0}^{m}\binom{N(t)}{r}3^r
=\mathcal O\!\left(t^{dm}\right)
\label{eq:cost-retained-dimension}
\end{equation}
for fixed $m$.

For fixed $m$, the retained operator-space dimension therefore grows polynomially with the volume of the effective causal region.
This bounds the number of Pauli coefficients required to store each HB vector, but does not by itself bound the arithmetic cost of constructing the chart.

For local Hamiltonians with uniformly bounded interaction strengths, the LR bounds constrain the propagation of initially local observables to a finite-velocity causal region.
In regular local dynamics, it is therefore reasonable to expect the characteristic timescale of HB reconstruction to remain governed by microscopic interaction scales rather than by the growing volume of the causal region.
This motivates considering a bounded density of adaptive reconstructions, although LR bounds alone do not guarantee such behavior, particularly in the presence of ill-conditioned or rank-deficient covariance Gram matrices.

We accordingly assume that, at fixed $m$, $\ell$, $\varepsilon$, and $p$, successive reconstructions are separated by a time-independent lower bound $T_{n+1}-T_n\geq\tau_*>0$.
Under this assumption, the number of charts constructed up to time $t$ is at most $\mathcal O(t)$.
Since $D_m(T_n)\leq D_m(t)$ for all $T_n\leq t$, summing the retained-space count across the adaptive charts gives
\begin{equation}
\sum_{T_n\leq t}D_m(T_n)
\leq
N_{\mathrm{charts}}(t)D_m(t)
=
\mathcal O(t^{dm+1}),
\end{equation}
where $N_{\rm charts}(t) = O(t)$.
Consequently, the instantaneous retained-space cost and cumulative retained-space cost  \begin{equation}
\mathrm{Cost}(t)
=
\mathcal O(t^{dm}), \quad \mathrm{Cum. \, Cost}(t)
=
\mathcal O(t^{dm+1}),
\label{eq:cost-main-scaling}
\end{equation}
and the peak retained-coordinate storage satisfy
\begin{equation}
\mathrm{Mem}(t)
=
\mathcal O(t^{dm}).
\label{eq:memory-main-scaling}
\end{equation}
These estimates characterize the maximum number of available $m$-body Pauli coordinates and their cumulative count across adaptive charts, assuming fixed HB depth $\ell$.
They do not constitute bounds on the computational wall time or total implementation memory, which additionally depend on costs of contraction and calculation of pairwise covariance, projected commutators, temporary operator buffers, indexing structures, and symbolic preprocessing.
Moreover, polynomial retained-space growth at fixed $m$ does not guarantee that the same body order suffices to achieve a prescribed accuracy at arbitrarily long times.

\subsection{Representation and leakage}
\label{subsec:error-decomposition}

The total covariance error of the restricted dynamics $\mathbf{O}_{\ell,m,\varepsilon,p}(t)$ neatly separates into representation and leakage contributions,
\begin{align}
\Delta_{\ell,m,\varepsilon,p}(t)^2 &= \|\mathbf{O}_H(t)-\mathbf{O}_{\ell,m,\varepsilon,p}(t)\|_{\mathrm{covar}(\sigma_0)}^2\\
&=
\left[\Delta_{\mathrm{rep}}^{(m)}(t)\right]^2
+
\left[\Delta_{\mathrm{leak}}^{(\ell,m,\varepsilon,p)}(t)\right]^2,
\label{eq:error-decomposition-main}
\end{align}
where
\begin{align}
\Delta_{\mathrm{rep}}^{(m)}(t)
&:=\|(1- \pi_m^{\sigma_0})\mathbf{O}_H(t)\|_{\mathrm{covar}(\sigma_0)},
\label{eq:representation-error-main}\\
\Delta_{\mathrm{leak}}^{(\ell,m,\varepsilon,p)}(t)
&:=\|\pi_m^{\sigma_0}\mathbf{O}_H(t)
-\mathbf{O}_{\ell,m,\varepsilon,p}(t)\|_{\mathrm{covar}(\sigma_0)}.
\label{eq:leakage-error-main}
\end{align}
The equality follows because the representation residual $(1-\pi_m^{\sigma_0})\mathbf{O}_H(t)$ is covariance-orthogonal to $\mathcal A_m$, while the leakage term belongs to $\mathcal A_m$.
The covariance Cauchy--Schwarz inequality also makes $\Delta_{\ell,m,\varepsilon,p}(t)$ an upper bound on the corresponding expectation-value error.
The representation error is an instantaneous, post-facto question at each $t$: it asks how well the exact observable fits in the retained $m$-body space.
The leakage error, by contrast, measures the mismatch between the post-facto projected trajectory and the restricted adaptive dynamics generated by \(\bm{\mathcal L}_m\), an error source which includes feedback through discarded high-body sectors and finite-\(\ell,\varepsilon\) HB errors.

Since $I\in\mathcal A_m$, covariance orthogonality implies exact preservation of the reference-state expectation,
\begin{equation}
\Tr[\sigma_0\pi_m^{\sigma_0}\mathbf A]=\Tr[\sigma_0\mathbf A].
\label{eq:projection-expectation-preservation}
\end{equation}
The representation residual therefore does not contribute directly to the target expectation-value error.
For the propagated approximation,
\begin{equation}
\left|
\Tr\!\left[\sigma_0\left(\OH(t)-\mathbf O_{\ell,m,\varepsilon,p}(t)\right)\right]
\right|
\leq
\Delta_{\mathrm{leak}}^{(\ell,m,\varepsilon,p)}(t).
\label{eq:expectation-leakage-bound}
\end{equation}
The discarded sector remains relevant because it can feed back into the retained dynamics.

To separate physical memory from finite-chart error, let $\mathbf{O}_m(t)$ denote the closed restricted flow,
\begin{equation}
\dot{\mathbf{O}}_m(t)= \pi_m^{\sigma_0}\genH\mathbf{O}_m(t),
\qquad
\mathbf{O}_m(0)=\pi_m^{\sigma_0}\mathbf{O}_S.
\label{eq:closed-projected-exact}
\end{equation}
Then
\begin{equation}
\Delta_{\mathrm{leak}}^{(\ell,m,\varepsilon,p)}(t)
\leq E_{\mathrm{mem}}^{(m)}(t)
+E_{\mathrm{HB}}^{(\ell,m,\varepsilon,p)}(t),
\label{eq:leakage-split-main}
\end{equation}
where, in the same covariance geometry used in \cite{Perez_short_paper_2026},
\begin{align}
E_{\mathrm{mem}}^{(m)}(t)
&:=
\normsp{
\pi_m^{\sigma_0}\mathbf{O}_H(t)-\mathbf{O}_m(t)
}{\mathrm{covar}(\sigma_0)},
\\
E_{\mathrm{HB}}^{(\ell,m,\varepsilon,p)}(t)
&:=
\normsp{
\mathbf{O}_m(t)-\mathbf{O}_{\ell,m,\varepsilon,p}(t)
}{\mathrm{covar}(\sigma_0)}.
\end{align}
The first term records information that leaves the retained sector and later returns, as described by the Nakajima--Zwanzig-like decomposition in \eqref{eq:nz-mbody-projected} and App.~\ref{app:nz-projected-dynamics}, whereas the second measures the additional approximation error introduced by the finite adaptive HB representation.
The reconstruction tolerance $\varepsilon$ controls when a local basis is rebuilt and is not, by itself, an a priori bound on $E_{\mathrm{HB}}$.

Writing $P_m=\pi_m^{\sigma_0}$, $Q_m=1-P_m$, and $g=\mathrm{covar}(\sigma_0)$, define the return coupling and retained-sector growth bound by
\begin{equation}
L_{m,g}
:=
\normsp{P_m\genH Q_m}{{\rm op},g},
\qquad
\mu_{m,g}
:=
\normsp{P_m\genH P_m}{{\rm op},g}.
\label{eq:memory-coupling-definitions}
\end{equation}
For rank-deficient $\sigma_0$, these induced operators are understood on the covariance-resolved quotient whenever the corresponding maps preserve covariance-null equivalence classes.
Duhamel's formula then gives
\begin{equation}
E_{\mathrm{mem},g}^{(m)}(t) \leq
L_{m,g}\int_0^t ds\, e^{\mu_{m,g}(t-s)}
\normsp{Q_m\mathbf{O}_H(s)}{g},
\label{eq:memory-error-main-bound}
\end{equation}
which is equivalently a bound in terms of $\Delta_{\mathrm{rep}}^{(m)}(s)$.
Thus, a small discarded tail is sufficient only when its return coupling is also controlled.
The proof and the rank-deficient case are discussed in App.~\ref{app:error-decomposition-proof}.

\subsection{Necessary obstructions and conditional body-order control}
\label{subsec:body-order-requirements}

The accuracy of an \(m\)-body description depends on several distinct notions of operator complexity.
Operator-space entanglement measures the Schmidt complexity generated across spatial cuts, and is a fundamental obstruction to TN methods, while Pauli-\(\ell_1\) magic  ${\cal M}_1(t)$ measures the spread of the observable in the Pauli basis, whereas the truncation itself is controlled directly by the distribution of norm over body order.
These quantities are related but are not interchangeable.
A pedagogical explanation of the following derivations can be found in the accompanying letter \cite{Perez_short_paper_2026}.

We first obtain a state-independent necessary obstruction from operator-space entanglement.
After restricting to the relevant LR region \(B(t)\), consider a bipartition \(B(t)=L\cup R\) and let \(S_2(\mathbf{O}_H;L|R)\) denote the R\'enyi-2 operator-space entanglement entropy.
Every observable in \(\mathcal A_m\) has Schmidt rank across this cut bounded by a cut-dependent capacity \(R_m^{L|R}\), a cut-wise operator Schmidt-rank capacity, whose explicit form is derived in App.~\ref{app:osee-proof}.
If an \(m\)-body observable approximates \(\mathbf{O}_H(t)\) with relative squared HS error at most \(\Delta\), then necessarily
\begin{equation}
(1-\Delta)^2
\exp{S_2(\mathbf{O}_H(t);L|R)}
\leq
R_m^{L|R}.
\label{eq:osee-necessary-main}
\end{equation}
Consequently, violation of Eq.~\eqref{eq:osee-necessary-main} across even one cut rules out that body order at the prescribed HS accuracy.
The converse does not hold: satisfying Eq.~\eqref{eq:osee-necessary-main} does not guarantee an accurate \(m\)-body representation.
Moreover, OSEE constrains representation of the observable itself and does not independently lower-bound the leakage or memory error of the restricted dynamics.

Pauli-basis non-stabilizerness probes a different structure.
For an HS-normalized Pauli expansion
${\bf O}_H(t)=\sum_P c_P(t)P$, we use the logarithmic
coefficient-$\ell_1$ complexity
\begin{equation}
{\cal M}_1(t) = \log\!\left(\sum_P |c_P(t)|\right), 
\end{equation}
where, throughout this discussion, $\log$ denotes the natural logarithm.
For a unit-HS-norm evolution originating from a Pauli observable, this quantity is related to the $\alpha=1/2$ operator stabilizer R\'enyi entropy of Ref.~\cite{dowling2025_magic} through $2{\cal M}_1=(\log 2)\mathcal M^{(1/2)}$, where the latter is defined using base-$2$ logarithms.
The global value of ${\cal M}_1$ quantifies Pauli-basis spreading but does not determine how that weight is distributed over body order.

To obtain a sufficient estimate, additional information about the high-body tail is required.
Let \(\Theta_r^{(g)}(t)\) denote the squared norm carried by the orthogonal \(r\)-body sector in a fixed geometry \(g\).
We assume that throughout the interval of interest, the following holds
\begin{equation}
\Theta_r^{(g)}(s) \leq
C_g \,e^{2\mathcal{M}_1(s)}
e^{-2\kappa_g r},
\qquad
0\leq s\leq t,
\label{eq:covariance-tail-assumption-main}
\end{equation}
with \(C_g,\kappa_g>0\).
The exponential suppression with body order is independent dynamical information and is not implied by locality or by the magic alone.

Summing the discarded body-order sectors gives
\begin{equation}
\left\|
(1-\mathsf P_m^{(g)})
\mathbf{O}_H(t)
\right\|_g
\leq
A_{\kappa,g}
e^{\mathcal{M}_1(t)}
e^{-\kappa_g(m+1)},
\label{eq:representation-upper-main}
\end{equation}
with $A_{\kappa,g} = \sqrt{{C_g}/({1-e^{-2\kappa_g}})}$ and where \(\mathsf P_m^{(g)}\) is the orthogonal projection onto the at-most-\(m\)-body space in geometry \(g\).
For \(g=\mathrm{covar}(\sigma_0)\), whenever the covariance projection is well defined on the corresponding normed space, \(\mathsf P_m^{(g)}=\pi_m^{\sigma_0}\) and Eq.~\eqref{eq:representation-upper-main} directly bounds \(\Delta_{\mathrm{rep}}^{(m)}\).
Thus the body-order tail supplies a conditional sufficient certificate for representation accuracy, with the magic controlling the overall scale of the bound.

The same distinction applies dynamically.
Equation~\eqref{eq:memory-error-main-bound} shows that a discarded component affects the restricted trajectory only through its return coupling $L_{m,g}$.
Writing ${\cal M}_1^\star(t) = \sup_{0\leq s\leq t}{\cal M}_1(s)$, the same concentration hypothesis gives, in compatible covariance geometry,
\begin{equation}
E_{\mathrm{mem}}^{(m)}(t)
\leq 
A_{\kappa,g} \, L_{m,g} \, t \,
\exp(
{\cal M}_1^\star(t) +\mu_{m,g}t -\kappa_g(m+1)
).
\label{eq:memory-sufficient-main}
\end{equation}
For full-rank reference states, this estimate applies directly in the covariance norm.
For rank-deficient reference states, both the retained generator and the return coupling must preserve covariance-null equivalence classes, requiring the following pair of conditions, respectively
\begin{equation}
P_m\genH\mathcal P_m N^{\sigma_0}
\subseteq
\mathcal N_m^{\sigma_0},
\qquad
P_m\genH Q_m\mathcal N^{\sigma_0}
\subseteq
\mathcal N_m^{\sigma_0}.
\end{equation}
The first condition ensures that retained covariance-null directions remain null under the restricted evolution, yielding well-defined dynamics on the quotient space $\mathcal A_m/\mathcal N_m^{\sigma_0}$,
see Eq.~\eqref{eq:nz-mbody-projected}, while the second ensures that covariance-null discarded contributions cannot generate covariance-resolved return terms, 
see Eq.~\eqref{eq:nz-mbody-memory}.
If either condition fails, the covariance-quotient memory estimate is not claimed, although the underlying Duhamel comparison remains valid at the operator level and may be bounded in a suitable positive-definite geometry.

Finally, the finite-HB approximation error must be controlled independently.
Recall that the total covariance error decomposes orthogonally into representation and leakage errors Eq.~\eqref{eq:error-decomposition-main}, while the latter is bounded by the sum of the memory and finite-HB contributions, as in Eq.~\eqref{eq:leakage-split-main}.
Assigning nonnegative tolerances $\eta_{\mathrm{rep}}$, $\eta_{\mathrm{mem}}$, and $\eta_{\mathrm{HB}}$ to $\Delta_{\mathrm{rep}}^{(m)}$, $E_{\mathrm{mem}}^{(m)}$, and $E_{\mathrm{HB}}^{(\ell,m,\varepsilon,p)}$, respectively, then guarantees a total covariance error no larger than $\delta$ whenever the individual bounds hold and
\begin{equation}
\eta_{\mathrm{rep}}+\eta_{\mathrm{mem}}+\eta_{\mathrm{HB}}
\leq\delta.
\label{eq:error-budget-main}
\end{equation}

The logical roles of OSEE and magic are therefore deliberately asymmetric.
Indeed, in the HS norm, OSEE provides a necessary and natural state-independent obstruction to low-body representation by testing the Schmidt-rank capacity available across spatial cuts.
Pauli-\(\ell_1\) magic enters a sufficient body-order estimate only together with an independently controlled high-body tail, since the magic itself does not constrain the 
distribution in body order within the representation, as opposed to its prominent direct role in \cite{shao2026_topK, dowling2025_magic}.
Only when both statements are formulated in HS geometry can they be interpreted as lower and upper constraints on the same global representation problem.
For the covariance (semi-) norm, no state-independent OSEE lower bound follows without additional norm-comparison information, and the body order required to reproduce state-adapted expectation values may therefore be substantially smaller than that required to reconstruct the full observable in HS norm.
The complete proofs and the explicit cut-capacity construction are given in App.~\ref{app:osee-proof} and App.~\ref{app:tail-proof}.

\section{Numerical implementation}
\label{sec:numerical-implementation}

The numerical framework combines a symbolic preprocessing layer with \textsc{Phoenix} and the state-dependent \textsc{AdaptH} implementation.
The symbolic preprocessing determines the lattice-dependent support structure, local Pauli algebra, and sparse commutator pattern. 
\textsc{Phoenix} provides the KeyMap-based sparse instruction representation and backend generation used to lower these operations to compiled numerical kernels.
\textsc{AdaptH} implements the state-dependent covariance contractions, $m$-body projections, projected-generator matrices, Gram systems, and adaptive HB dynamics.

The implementation separates symbolic precomputation from online propagation.
Local nonvanishing commutator patterns are identified on a finite exploration domain, lifted to the physical lattice, and converted into compiled numerical instructions.
The resulting kernels act directly on sparse coefficient representations, so symbolic Pauli multiplication and support analysis are not repeated during time evolution.
State-dependent projections and covariance scalar products are then evaluated at the \textsc{AdaptH} level.

For a Hamiltonian with maximum body order $h_{\max}$, the commutator of an operator in $\mathcal A_m$ can temporarily reach body order
\begin{equation}
m_{\mathrm{int}}
=
\min\{N,m+h_{\max}-1\}.
\end{equation}
The implementation therefore evaluates the commutator in an enlarged temporary buffer before either discarding sectors above $m$ in the HS scheme or projecting them back into $\mathcal A_m$ in the covariance scheme.
For the two-body Hamiltonians considered here, $m_{\mathrm{int}}=m+1$.

The sparse representation, locality reduction, symbolic lifting procedure, compiled-kernel construction, and storage requirements are described in detail in App.~\ref{app:phoenix-precomputation} with Phoenix publicly accessible in \cite{Kost_Plenio_PHOENIX}.

\section{Results and discussion}
\label{sec:results}

We now benchmark covariance PP in interacting spin systems where Heisenberg evolution generates Pauli strings of increasing body order and complexity.
We consider the transverse-field Ising model with an additional longitudinal field,
\begin{equation}
    \h_{\mathrm{TFIM}}
    = J\sum_{\langle ij\rangle} Z_i Z_j
    + h_X\sum_i X_i
    + h_Z\sum_i Z_i,
    \label{eq:tfim-hamiltonian}
\end{equation}
on a $d$-dimensional rectangular lattice with nearest-neighbor interactions with open boundary conditions.
For a one-dimensional chain, the limit $h_Z=0$ reduces to the integrable TFIM and admits a free-fermionic solution after a Jordan--Wigner transformation.
A finite longitudinal field $h_Z\neq0$ removes this special structure and generates genuinely interacting operator dynamics.
Repeated commutation with $\h_{\mathrm{TFIM}}$ then produces many-body observables of increasing body order.

Ref.~\cite{Perez_short_paper_2026} considered the fully polarized pure reference state $\sigma_0=(\ket{\uparrow}\!\bra{\uparrow})^{\otimes N}$ in the $d=3$-dimensional TFIM with $h_Z = 0$.
That endpoint is especially instructive for the covariance construction because the covariance form is a seminorm with a nontrivial null space $\mathcal{N}_{\sigma_0}$.
In the pure-state limit, the associated null structure can provide strong compression, but it also permits statically unresolved distinctions to acquire covariance-resolved images under the restricted dynamics, as discussed in Sec.~\ref{sec:dynamical-compatibility} and in the accompanying work \cite{Perez_short_paper_2026}.
Here we extend the analysis by varying both the purity and orientation of the reference state in a TFIM with a weak longitudinal field.

Throughout the results section, we take $J$ as the unit of energy and set $h_X/J=0.75$ and $h_Z/J=0.15$.
This regime produces nontrivial operator growth while leaving a finite time window over which low-body approximations remain informative.
As already established, covariance projection cannot eliminate the representation error associated with genuinely high-body operator content, but it can change how such content is represented and compressed, after projection, onto the retained low-body space.

The initial states are chosen as homogeneous product states in the $XZ$ plane,
\begin{equation}
    \sigma_0(r,\theta)
    =
    \bigotimes_{i=1}^{N}
    \frac{1}{2}
    \left(
        I
        +r\cos\theta\,X_i
        +r\sin\theta\,Z_i
    \right),
    \label{eq:tfim-initial-state}
\end{equation}
with $0\leq r\leq1$.
The parameter $r$ controls the local purity, while $\theta$ fixes the orientation of the local Bloch vector relative to the Hamiltonian.
At the endpoint
\begin{equation*}
    \sigma_0(1,\pi/2) = (\ket{\uparrow}\!\bra{\uparrow})^{\otimes N},
\end{equation*}
the reference state is pure and the covariance geometry is rank deficient, whereas every $r<1$ gives a full-rank mixed state.
We first keep $\theta=\pi/2$ fixed and vary $r$, thereby changing the strength of the state dependence without changing its orientation.
We then fix $r=1$ and vary $\theta$, which probes how the rank-deficient covariance geometry is oriented relative to the restricted dynamics.

\paragraph{$r$ sweep.}

\begin{figure}
    \centering
    \includegraphics[width=1\linewidth]{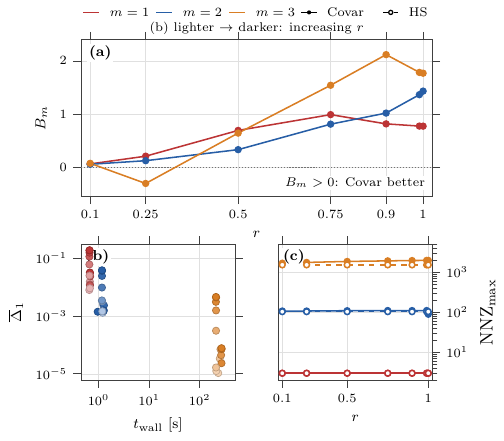}
    \caption{
\textbf{Purity dependence of accuracy and computational cost.}
Results for the $N=10$ chain at $\theta=\pi/2$, with $\ell=5$, $p=2$, $\varepsilon=10^{-10}$, and $TJ=1$.
Colors denote $m=1,2,3$.
(a) Covariance advantage $B_m$ [Eq.~\eqref{eq:covariance-advantage}], with $B_m>0$ favoring covariance over HS propagation.
(b) Time-averaged absolute error [Eq.~\eqref{eq:mean-abs-error}] versus end-to-end wall time.
Filled and open circles denote covariance and HS propagation, with lighter shades corresponding to smaller $r$.
(c) Maximum assembled-observable sparsity $\mathrm{NNZ}_{\max}$ [Eq.~\eqref{eq:hb-max-nnz}].
}
\label{fig:r-sweep}
\end{figure}

We consider an open chain of $N=10$ spins and the bulk observable $Z_5$, using zero-based site indexing.
The reference polarization is varied over
\begin{equation*}
    r\in\{0.10,\,0.25,\,0.50,\,0.75,\,0.90,\,0.99,\,1.00\},
\end{equation*}
with $\theta=\pi/2$.
The simulations are propagated beyond the main-text window, while the diagnostics in Fig.~\ref{fig:r-sweep} are evaluated over $tJ\in[0,1]$, which is the informative window.
For $r<1$, the exact benchmark is obtained by direct Schr\"odinger evolution of the mixed-state density operator, while at $r=1$ we use the equivalent pure-state evolution.

For both geometries, we fix the adaptive HB parameters to $\ell=5$, $p=2$, and $\varepsilon=10^{-10}$, and consider retained body orders $m=1,2,3$.
At fixed $m$, covariance and HS propagation use the same reconstruction rule and numerical parameters with their distinction lying in how overflowing many-body components are handled.

To benchmark each PP trajectory against the exact dynamics, we use the time-averaged absolute error
\begin{equation}
    \overline{\Delta}_1(T)
    =
    \frac{\delta t}{T}
    \sum_{0<t_n\leq T}
    \left|
        \langle Z_5(t_n)\rangle_{\mathrm{approx}}
        -
        \langle Z_5(t_n)\rangle_{\mathrm{exact}}
    \right|.
    \label{eq:mean-abs-error}
\end{equation}
For a uniform grid with $T=k\delta t$, Eq.~\eqref{eq:mean-abs-error} is the arithmetic mean of the absolute trajectory error over the first $k$ positive-time samples and approximates $T^{-1}\int_0^T|\Delta Z_5(t)|\,dt$.
We quantify the covariance advantage at fixed retained body order through
\begin{equation}
    B_m(r;T)
    =
    \log_{10}
    \left[
        \frac{
            \overline{\Delta}^{\,\mathrm{HS}}_{1,m}(r;T)
        }{
            \overline{\Delta}^{\,\mathrm{cov}}_{1,m}(r;T)
        }
    \right].
    \label{eq:covariance-advantage}
\end{equation}
Thus $B_m=0$ denotes equal time-averaged accuracy, $B_m>0$ a covariance advantage, and $B_m<0$ an HS advantage.
In particular, $B_m=1$ corresponds to a one-order-of-magnitude reduction of the time-averaged error under covariance projection.

We additionally characterize the sparsity of the propagated PP observable through its number of nonzero Pauli coefficients (NNZ).
For this diagnostic we use the counting threshold $\varepsilon_{\rm NNZ}=10^{-12}$, which is applied only when counting coefficients and does not truncate the propagated observable.
We define
\begin{equation}
    \mathrm{NNZ}_{\max}(T;\varepsilon_{\rm NNZ})
    =
    \max_{0\leq t\leq T}
    \operatorname{nnz}_{\varepsilon_{\rm NNZ}}
    \!\left[
        {\bf O}_{\ell,m,\varepsilon,p}(t)
    \right].
    \label{eq:hb-max-nnz}
\end{equation}
Here ${\bf O}_{\ell,m,\varepsilon,p}(t)=\sum_{\alpha}\phi_\alpha(t)b_\alpha(t)$ is the assembled PP observable, the trajectory defined by Eq.\eqref{eq:hb-local-expansion}, so Eq.~\eqref{eq:hb-max-nnz} measures the largest instantaneous sparse representation actually encountered by the propagated observable.

Figure~\ref{fig:r-sweep}(a) shows a pronounced dependence of the covariance advantage on the reference-state polarization.
At weak polarization, $r=0.1$, covariance and HS propagation have nearly identical time-averaged errors for all three body orders.
This is consistent with the convergence of the covariance geometry to the HS geometry on the traceless operator sector as $r\rightarrow0$.

The behavior changes as the reference state becomes more strongly polarized.
For most values of $r$, covariance PP gives smaller errors than HS at fixed $m$, although the advantage is not monotonic.
In particular, HS performs better at $m=3$ and $r=0.25$.
The largest improvement occurs near $r=0.9$, where $B_3\simeq2.1$, corresponding to a reduction of the time-averaged error by approximately two orders of magnitude.
Since this reference state is strictly full rank, the improvement cannot be attributed to exact covariance-null identifications.
Strong state adaptation therefore does not require rank deficiency.

Figure~\ref{fig:r-sweep}(b) shows the corresponding accuracy--cost comparison.
Increasing the retained body order generally reduces the time-averaged error over the displayed interval, with the largest gains obtained for covariance PP at $m=3$.
For a two-body Hamiltonian and an initially one-body observable, each nested commutator can increase the Pauli body order by at most one.
This provides a short-time rationale for the observed improvement with increasing $m$.
It does not, however, imply a monotonic ordering of independently propagated finite-$m$ trajectories.
The restricted dynamics at different body orders accumulate distinct truncation and reconstruction errors, and the longer-time trajectories in the Appendix Sec.\ref{subsec:SI_r_sweep} illustrate this limitation,

At fixed $m$, the wall time varies much less strongly with $r$ than the corresponding error.
In particular, covariance PP reaches errors in the $10^{-5}$--$10^{-4}$ range at $m=3$ across the sweep, while HS errors increase substantially toward strong polarization.
The improvement therefore does not require a comparable increase in computational time.

Figure~\ref{fig:r-sweep}(c) shows that the maximum number of nonzero Pauli coefficients is also weakly dependent on $r$ at fixed $m$.
The HS operator trajectory is independent of the reference state, whereas covariance projection changes with $r$.
Nevertheless, the large variation in covariance accuracy is not accompanied by a comparable increase in the assembled operator size.
The advantage arises from the state-adapted representation within the retained low-body space, rather than from retaining substantially more Pauli coefficients.

No expectation-bound violations are observed for either geometry over the displayed interval $tJ\in[0,1]$, including at $r=1$.
The absence of such violations does not certify spectral admissibility, but it shows that rank deficiency alone does not imply a witnessed failure.
Together, these results separate the benefits of state adaptation from the dynamical compatibility issues specific to rank-deficient covariance geometries.

The results shown in this section thus show that the covariance PP simulations can genuinely give improvements in accuracy at comparable numerical cost, compared to HS dynamics, pointing to the fact that, while their representation sizes can be comparable, they span dramatically different sections of the retained observable space ${\cal A}_m$.

\paragraph{$\theta$ sweep.}

We next fix $r=1$ and vary the orientation of the pure reference state.
The exact benchmark is obtained by Schr\"odinger-picture state-vector propagation on an open $3\times3\times2$ lattice with $N=18$.
We monitor the surface-center observable $Z_8(t)$, using zero-based C-order site indexing, for which the corresponding site has coordination number $z_8=5$.
The orientations are 
    $$\theta\in\{\pi/6,\pi/4,\pi/3,\pi/2\},$$
and the evolution extends to $T_{\max}J=1.5$.
For both geometries and all PP evolutions, we use $m=1,2,3$, $\ell=5$, $p=2$, and $\varepsilon=10^{-8}$.

All initial states in this sweep are pure and therefore have a nontrivial covariance-null space ${\cal N}_{m}^{\sigma_0}$.
Changing $\theta$ rotates this null structure relative to the Hamiltonian and the dynamically generated operator directions without changing the rank deficiency.
The sweep, thus, probes the orientation dependence of the restricted dynamics at fixed purity.

We benchmark the time-averaged absolute error $\overline{\Delta}_1(T)$, the maximum sparse representation size $\mathrm{NNZ}_{\max}$, and the expectation-bound witness
\begin{equation}
M_{m,j}(T) = \max_{0\leq t\leq T}
\left|
\langle Z_j(t)\rangle_m
\right|.
\label{eq:expval_admin}
\end{equation}
For the present benchmark, $j=8$.
Since exact unitary evolution preserves the spectrum of $Z_j$, any $M_{m,j}>1$ certifies that the approximate observable has left the exact Pauli spectral interval.
The converse does not hold: $M_{m,j}\leq1$ neither certifies spectral admissibility nor guarantees accuracy.
We therefore use $M_{m,j}$ as a geometry-independent failure witness, while the distinct covariance-null compatibility mechanism is diagnosed separately through Eq.~\eqref{eq:quotient-compatibility}.

\begin{figure}
    \centering
    \includegraphics[width=1.0\linewidth]{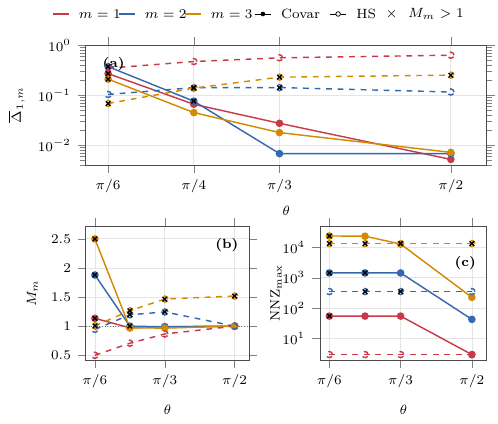}
    \caption{
\textbf{Orientation dependence of accuracy and expectation-bound violations.}
Results for the $N=18$, $3\times3\times2$ TFIM at $r=1$, with $\ell=5$, $p=2$, $\varepsilon=10^{-8}$, and $T_{\max}J=1.5$.
Colors denote $m=1,2,3$; solid/filled and dashed/open markers distinguish covariance and HS propagation.
(a) Time-averaged absolute error relative to exact dynamics.
(b) Maximum expectation-value magnitude $M_{m,8}$ [Eq.~\eqref{eq:expval_admin}], with the dotted line marking the Pauli bound $M_{m,8}=1$.
(c) Maximum assembled-observable sparsity $\mathrm{NNZ}_{\max}$.
Black crosses identify trajectories with $M_{m,8}>1$, certifying spectral inadmissibility.
}
    \label{fig:theta-sweep}
\end{figure}

Figure~\ref{fig:theta-sweep} reveals a strong orientation dependence of the restricted dynamics.
At $\theta=\pi/6$, all three covariance trajectories violate the exact Pauli expectation bound.
The largest excursions reach approximately $M_2\simeq1.9$ and $M_3\simeq2.5$, while the HS trajectories show no corresponding violations of the measured expectation bound.
The covariance errors at $m=2$ and $m=3$ are also larger than their HS counterparts over the displayed interval.
As shown by the time-resolved trajectories in the Appendix, the loss of accuracy is not limited to the intervals in which the expectation bound is violated.
This orientation is therefore an unfavorable regime for covariance PP.

The behavior changes as the reference state rotates toward the $Z$ axis.
At $\theta=\pi/4$, covariance PP gives smaller time-averaged errors than HS at all three body orders.
No covariance expectation-bound violation is observed at $m=1$ or $m=3$, while $m=2$ shows a small excursion beyond the bound.
Conversely, the HS trajectories at $m=2$ and $m=3$ violate the expectation bound.
The HS $m=1$ trajectory remains within the measured interval but has a substantially larger error.
This illustrates that expectation-bound violations and trajectory accuracy are distinct diagnostics.

At $\theta=\pi/3$ and $\theta=\pi/2$, the covariance hierarchy gives markedly smaller errors than HS without witnessed expectation-bound violations.
HS remains susceptible to violations at higher body orders, including $M_3\simeq1.5$ at $\theta=\pi/2$.
The Appendix trajectories also show that the error need not decrease monotonically with increasing $m$.

Figure~\ref{fig:theta-sweep}(c) shows a similarly pronounced dependence of the sparse representation on orientation.
The HS values of $\mathrm{NNZ}_{\max}$ are independent of $\theta$, since the restricted HS dynamics do not depend on the initial state at all.
By contrast, the covariance representation becomes substantially sparser toward the fully $Z$-polarized endpoint.
At $m=3$, $\mathrm{NNZ}_{\max}$ falls from a few times $10^4$ at $\theta=\pi/6$ to a few times $10^2$ at $\theta=\pi/2$, a reduction of approximately two orders of magnitude.
In this regime, improved accuracy is accompanied by a smaller realized sparse representation.

The coexistence of covariance and HS expectation-bound violations is important.
Such violations are not specific to state-adapted projection and can occur under either restricted dynamics.
Rank-deficient covariance propagation introduces the additional possibility of dynamical incompatibility between covariance-equivalent representatives, which must be diagnosed independently.

Together with the $r$ sweep, these results distinguish the effects of state adaptation, rank deficiency, and dynamical stability.
The largest covariance gain in the $r$ sweep occurs for a full-rank reference state, while every state in the present sweep is rank deficient.
Nevertheless, the covariance trajectories range from strong failure at $\theta=\pi/6$ to high accuracy and substantial compression at $\theta=\pi/2$.
The covariance rank alone is therefore insufficient to predict the performance of the restricted dynamics.
The observed behavior instead depends on how the state-adapted geometry couples to the operator directions generated by the restricted Liouvillian $\genH_m$.

\subsection{Scalability in larger three-dimensional models}

We finally test the method in a larger three-dimensional system for which exact propagation is unavailable.
Reliability is therefore assessed through internal convergence of the PP and tensor-network (TN) hierarchies, their mutual agreement, expectation-bound violations, and the sparse representation explored during propagation. 

The accompanying Letter~\cite{Perez_short_paper_2026} considered an $N=90$, $6\times5\times3$ nearest-neighbor $ZZ+X$ model with a corner observable.
Here we consider a complementary nonintegrable benchmark on an open $N=80$, $5\times4\times4$ lattice governed by Eq.~\eqref{eq:tfim-hamiltonian}.
We set $J=1$, $h_X=0.75$, and $h_Z=0.15$, and use the fully $Z$-polarized initial state $(r=1,\theta=\pi/2)$.
The observable is $Z_{37}(t)$, corresponding to the bulk site $(2,1,1)$ in zero-based C-order indexing, with six nearest neighbors.

The PP calculations use $\ell=5$, $p=2$, $m=1,2,3$, and $\varepsilon=10^{-9}$.
As an independent benchmark, we employ Krylov-enriched two-site TDVP--MPO evolution with $\chi=2,8,32,64$.
A complementary PP calculation at $\varepsilon=10^{-8}$ is reported in the Appendix.

Without an exact trajectory, we quantify internal convergence through
\begin{equation}
    \Delta_{\rm hier}^{(\lambda_k,\lambda_{k-1})}(t)
    =
    \left|
        \langle Z_{37}(t)\rangle_{\lambda_k}
        -
        \langle Z_{37}(t)\rangle_{\lambda_{k-1}}
    \right|,
    \label{eq:large-hierarchy-closure}
\end{equation}
Here $\lambda=m$ for PP and $\lambda=\chi$ for TN.
A small $\Delta_{\rm hier}$ indicates internal agreement within the corresponding hierarchy, but does not establish accuracy relative to the exact dynamics.
Agreement between the independently constructed PP and TN hierarchies provides an additional a posteriori consistency check.
We additionally monitor the expectation-bound witness $M_{m,j}$ of Eq.~\eqref{eq:expval_admin}, $\mathrm{NNZ}_{\max}$ of Eq.~\eqref{eq:hb-max-nnz}, and the end-to-end wall time.

\begin{figure}[t]
\centering
\includegraphics[width=\linewidth]{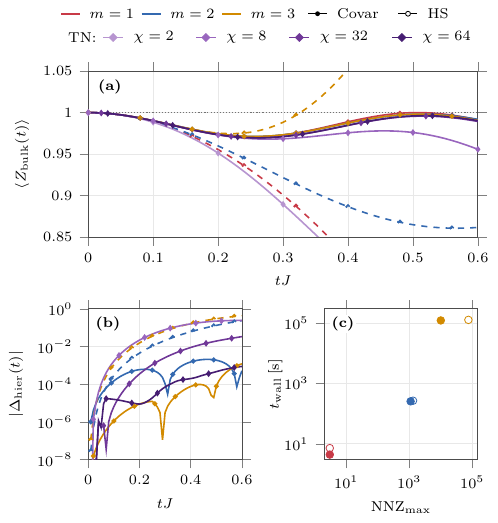}
\caption{
\textbf{Large-system comparison with an independent tensor-network hierarchy.}
Bulk polarization dynamics for the $N=80$, $5\times4\times4$ TFIM at $r=1$, $\theta=\pi/2$, $h_X=0.75$, $h_Z=0.15$, and $T_{\max}J=0.6$.
PP uses $\ell=5$, $p=2$, and $\varepsilon=10^{-9}$.
(a) Covariance PP (solid) and HS PP (dashed) for $m=1,2,3$, compared with two-site TDVP--MPO trajectories (purple, with diamonds) for $\chi=2,8,32,64$.
Darker purple denotes larger $\chi$; the dotted line marks the Pauli bound $\langle Z_{37}\rangle=1$ on a magnified vertical axis.
(b) Successive-hierarchy discrepancies [Eq.~\eqref{eq:large-hierarchy-closure}], measuring internal agreement rather than exact error.
(c) End-to-end wall time versus $\mathrm{NNZ}_{\max}$, with filled/open circles denoting covariance/HS propagation.
}
\label{fig:large-system-ex}
\end{figure}

Figure~\ref{fig:large-system-ex}(a) shows strong agreement within the covariance hierarchy.
The $m=1,2,3$ trajectories are nearly indistinguishable on the scale of the plot.
Independently, the TN hierarchy approaches the same dynamics as $\chi$ increases, with the $\chi=32$ and $\chi=64$ results lying close to the covariance trajectories.
This cross-method agreement provides strong a posteriori consistency, although no exact large-system error certificate is available.

The HS hierarchy behaves differently.
Its trajectories separate progressively with time, and the $m=3$ result violates the exact Pauli expectation bound.
No corresponding violation is observed for the displayed covariance or TN trajectories.
Thus, over the investigated interval, covariance PP exhibits a substantially cleaner body-order hierarchy than HS truncation.

Figure~\ref{fig:large-system-ex}(b) quantifies these differences.
For covariance PP, the discrepancy between $m=2$ and $m=3$ is typically smaller than the discrepancy between $m=1$ and $m=2$, with both remaining at or below the $10^{-3}$ scale over most of the displayed interval.
The TN hierarchy shows a corresponding reduction with increasing bond dimension, with the discrepancy between $\chi=32$ and $\chi=64$  reaching a comparable scale.
By contrast, the successive discrepancies for the HS grow to the $10^{-1}$-scale without showing comparable body-order closure.

Figure~\ref{fig:large-system-ex}(c) compares the realized PP representation sizes and wall times.
At $\varepsilon=10^{-9}$, both covariance and HS higher-body trajectories undergo adaptive HB reconstructions.
The corresponding reconstruction histories need not coincide, since the common reconstruction rule is evaluated on geometry-dependent HB coordinates.
At $m=1$ and $m=2$, the sparse representations remain of comparable size.
At $m=3$, covariance PP requires approximately $\mathcal{O}(10^4)$ nonzero coefficients, compared with $\mathcal{O}(10^5)$ for HS.
A substantial sparsity advantage therefore survives when both geometries reconstruct adaptively.

This reduction does not produce a comparable wall-time improvement.
Both $m=3$ calculations require wall times of order $10^5$~s, despite their different sparse representation sizes.
The computational cost also includes HB construction, commutator generation, Gram operations, and adaptive reconstruction.
The present measurements therefore establish a reduction in realized representation size, rather than a proportional computational speedup.
This speedup can possibly be realized with a more efficient numerical implementation.

Note that the NNZ diagnostic used here also differs from that of the accompanying Letter~\cite{Perez_short_paper_2026}.
The Letter characterizes the highest-index HB basis operator at reconstruction times, whereas the present $\mathrm{NNZ}_{\max}$ measures the maximum sparsity of the assembled propagated observable.
Their absolute values and trends should therefore not be compared directly.

The complementary calculation at $\varepsilon=10^{-8}$, reported in App.~\ref{subsec:SI-N80-eps1e8}, provides an additional robustness check.
At this threshold, covariance PP does not reconstruct a new chart within the simulation time while the HS does, yet the favorable covariance hierarchy and its agreement with TN remain qualitatively unchanged.
Tightening the threshold to $\varepsilon=10^{-9}$ activates covariance reconstruction without changing the central result.
The observed agreement is therefore not contingent on the absence of adaptive updates.

Together with the exact small-system benchmarks, this calculation demonstrates that state-adapted PP can yield a well-converged low-body hierarchy in a three-dimensional nonintegrable model beyond the range of exact propagation.
The result does not establish a general convergence theorem or a universal sparsity advantage.
It does show that, for the present Hamiltonian and initial state, covariance PP achieves strong internal agreement, consistency with an independent TN hierarchy, no witnessed expectation-bound violations, and a smaller high-body sparse representation than HS truncation.

\section{Conclusions and outlook}
\label{sec:conclusions}

In this work we have developed a state-adapted low-body closure for Heisenberg-picture many-body dynamics.
The method retains the conventional low-body spaces $\mathcal A_m$, but replaces state-independent Hilbert--Schmidt projection by a covariance geometry defined by a tractable reference state.
Combined with a particular choice and implementation of finite Krylov coordinates and adaptive hierarchical bases, this yields a sparse restricted dynamics tailored to the statistics of the state of interest.
For full-rank references the covariance form is an ordinary inner product, whereas rank-deficient states require the corresponding quotient construction.
For the product states considered here, the centered-support representation selects a definite representative of each covariance-equivalence class and permits a direct sparse-Pauli implementation.

Rank deficiency provides both an opportunity and an additional consistency condition.
The relevant state-adapted geometry can identify and alias statically irrelevant covariance-null directions, potentially yielding substantial compression gains, but said identification need not remain dynamically preserved.
The projected-activation criterion of Eq.~\eqref{eq:projected-alias-amplitude} tests whether a statically unresolved distinction acquires a covariance-resolved image under the retained dynamics.
It is a structural diagnostic rather than a finite-time stability criterion, covariance rank deficiency alone neither guarantees useful compression nor implies breakdown.
The relevant question is how the resolved and unresolved sectors are coupled by the Liouvillian.

The numerical benchmarks separate these effects.
In the $r$ sweep, covariance and HS propagation approach one another near the maximally mixed state, while increasing polarization produces large covariance accuracy gains at comparable body-order, sparse-representation, and computational budgets.
The strongest gains already occur for strictly full-rank states, showing that useful state adaptation does not rely on covariance singularity.
At fixed purity, the $\theta$ sweep reveals a strong orientation dependence: the same rank-deficient covariance construction ranges from an unfavorable regime with large errors and expectation-bound violations to one with high accuracy and pronounced sparse compression.
Expectation-bound violations also occur for HS propagation, confirming that loss of spectral admissibility is a generic possibility of finite-body restricted Heisenberg dynamics and is not, by itself, a signature of covariance-null incompatibility.

The large-system calculations extend this picture beyond exact propagation.
The accompanying Letter~\cite{Perez_short_paper_2026} showed favorable body-order consistency and sub-worst-case sparse growth in an $N=90$ three-dimensional TFIM.
Here, for a nonintegrable $N=80$ three-dimensional model, covariance PP exhibits clear $m=2\rightarrow3$ closure and agrees closely with an independently converging Krylov-enriched TDVP--MPO hierarchy, while the corresponding HS hierarchy does not show comparable closure and its $m=3$ trajectory violates the Pauli expectation bound.
At the common adaptive threshold used in the main benchmark, both higher-body geometries undergo HB reconstruction, yet the covariance $m=3$ trajectory explores a substantially smaller sparse representation.
This reduction does not translate into a proportional wall-time advantage in the present implementation, emphasizing that representation complexity and computational cost remain distinct.
Together, internal body-order closure, agreement with an independent TN hierarchy, and expectation-bound consistency provide strong a posteriori evidence that the state-adapted construction remains effective in a genuinely three-dimensional interacting regime where exact classical benchmarking is unavailable.

These results support a state-dependent view of operator complexity.
The difficulty of predicting a selected observable is not determined solely by global Pauli weight or operator-space entanglement, but also by how the dynamically generated operator content is resolved by the reference-state geometry.
Covariance projection exploits this structure directly.
Its performance is consequently governed by the interplay between high-body operator growth, the anisotropy of the state-dependent geometry, and the dynamical coupling between retained, discarded, and covariance-null directions.

Several questions remain open.
A central goal is to relate the body order required for fixed accuracy more quantitatively to evolution time, spatial dimension, operator-space entanglement, mana nonstabilizerness and the covariance-weighted high-body tail.
The present conditional representation and memory bounds provide a first step in this direction, but sharper dynamical estimates and practical a posteriori error indicators remain desirable.
The adaptive reconstruction rule can likewise be improved beyond the current coordinate-tail diagnostic, particularly near rank-deficient or strongly conditioned geometries.
Finally, the disparity between sparse representation size and end-to-end wall time indicates substantial room for algorithmic improvement through more efficient basis construction, commutator generation, and reuse of local operator structure.
State-adapted projection may also be combined with coefficient-based or locality-based truncation criteria, providing a possible route toward more flexible hybrid PP schemes.

The central conclusion is that low-body Heisenberg simulation need not approximate an observable uniformly in operator space.
When the prediction task concerns selected observables in a known state, the relevant complexity can itself be state dependent.
Covariance-projected Pauli propagation provides a systematic means of exploiting this structure, together with explicit diagnostics of the regimes in which the resulting reduction remains reliable.

\paragraph{Acknowledgements}
The authors would like to thank Thibaut Lacroix, Piotr Czarnik and Marek M. Rams for interesting discussions.
F.T.B.P. would like to thank Marco Cerezo for interesting discussions and comments on this line of research. 
F.T.B.P acknowledges support from CONICET of Argentina, work supported by CONICET PIP Grant No. 11220200101877CO.
G.~W. acknowledges support from the Alexander von Humboldt Foundation for support under the Humboldt Research Fellowship. 
G.~W. acknowledges the Financial Support Programmes for Early Career Researchers, Graduate and Professional Training Center, Ulm University for 2025 and 2026/2027. 
This work was supported by EU-project C-QuENS (Grant No. 101135359) and EU-Project SPINUS (Grant No. 101135699).
The authors acknowledge support by the state of Baden-Württemberg through bwHPC and the German Research Foundation (DFG) through grant no INST 40/575-1 FUGG (JUSTUS 2 cluster).
The research presented in this work was carried out primarily at Ulm University, with minor editorial revisions performed during F.T.B.P.'s subsequent appointment at Los Alamos National Laboratory.

The authors acknowledge the use of OpenAI ChatGPT, including GPT-5.6 Sol, as an assistive tool during manuscript preparation.
It was used for critical and adversarial examination of selected mathematical arguments, language revision and clarification of scientific text, consistency checks, and
assistance with debugging selected implementation code. 
The scientific ideas, conceptual framework, and principal analytical developments presented here originated with the authors. 
In particular, the state-adapted construction builds on the authors' (F.T.B.P.) earlier work on restricted maximum-entropy descriptions, developed prior to the availability of ChatGPT, while the restricted Heisenberg-picture approach underlying the present work was conceived independently of generative-AI assistance.
All AI-assisted suggestions, analytical arguments, numerical procedures, and resulting scientific claims were reviewed and independently verified by the authors, who retain full responsibility for the contents of the manuscript.

\paragraph{Data availability.}
The processed numerical data supporting the figures and quantitative comparisons are publicly available in Ref.~\cite{perez2026_pra_zenodo}.
The repository contains curated, plotting-ready datasets rather than the complete raw simulation outputs.
Part of the underlying software infrastructure is publicly available through \textsc{Phoenix}~\cite{Kost_Plenio_PHOENIX}.
The \textsc{AdaptH} implementation of the adaptive state-adapted workflow used for the present calculations remains under development and is not yet publicly available.
Further implementation details are available from the authors upon reasonable request.

\appendix

\section{Spectral admissibility and Hilbert--Schmidt norm control}

Exact unitary evolution preserves the spectrum of an observable, so for an initially Pauli observable $\mathbf O_S$,
\begin{equation}
\mathbf O_H(t)=U^\dagger(t)\mathbf O_S U(t),
\qquad
\mathbf O_H(t)^2=I,
\label{eq:si-exact-pauli-involution}
\end{equation}
and consequently,
\begin{equation}
\normsp{\mathbf O_H(t)}{\rm op}=1,
\qquad
\left|\Tr[\rho\mathbf O_H(t)]\right|\leq1
\label{eq:si-exact-spectral-bound}
\end{equation}
for every density operator $\rho$.
Approximate Heisenberg dynamics need not preserve these properties~\cite{xSPD2026,Li_2026,Angrisani_2026}, although stronger guarantees exist for certain circuit constructions~\cite{Angrisani_2025}.

Spectral admissibility should, however, be distinguished from HS norm conservation.
For the normalized HS scalar product $(\mathbf A,\mathbf B)_{\rm HS}=2^{-N}\Tr(\mathbf A\mathbf B)$ and a Pauli expansion $\mathbf O=\sum_P c_P P$,
\begin{equation}
\normsp{\mathbf O}{\rm HS}^2=\sum_P|c_P|^2.
\label{eq:si-hs-pauli-two-norm}
\end{equation}
Discrete truncation procedures, including coefficient pruning and Pauli-weight projection/truncation, remove HS-orthogonal contributions and therefore reduce this norm whenever nonzero components are discarded~\cite{shao2026_topK,Angrisani_2026}.
By contrast, the continuously restricted HS dynamics considered here and in Ref.~\cite{Perez_short_paper_2026} preserve it.
Indeed, with the HS-orthogonal projector $P_m^{\rm HS}=\pi_m^{I}$, the restricted generator
\begin{equation}
\genH_m^{\rm HS}=P_m^{\rm HS}\genH P_m^{\rm HS}
\label{eq:si-hs-restricted-generator}
\end{equation}
remains skew-adjoint, and therefore
\begin{equation}
\frac{d}{dt}\normsp{\mathbf O_m(t)}{\rm HS}^2=0.
\label{eq:si-hs-norm-conservation}
\end{equation}
Finite adaptive-HB propagation may introduce additional numerical deviations, but these are distinct from the ideal restricted dynamics.

Crucially, HS-norm conservation does not imply preservation of the operator spectrum.
For example, consider
\begin{equation}
\mathbf A=\frac{Z_1+Z_2}{\sqrt{2}},
\label{eq:si-hs-counterexample}
\end{equation}
with $\normsp{\mathbf A}{\rm HS}=1,$ but $\bra{00}\mathbf A\ket{00} = \sqrt{2}>1.$
Thus, an observable may preserve the initial HS norm of a Pauli operator while possessing an incompatible spectral range.

The distinction is also apparent from the Pauli coefficients, since for any Hermitian $\mathbf O=\sum_Pc_PP$,
\begin{equation}
\normsp{\mathbf O}{\rm op}
\leq
\sum_P|c_P|,
\label{eq:si-pauli-l2-l1-op}
\end{equation}
and an operator error $\mathbf E=\widetilde{\mathbf O}-\mathbf O=\sum_P\delta c_PP$ satisfies
\begin{equation}
\left|\Tr(\rho\mathbf E)\right|
\leq
\normsp{\mathbf E}{\rm op}
\leq
\sum_P|\delta c_P|.
\label{eq:si-l1-expectation-error}
\end{equation}
The coefficient $\ell_1$ norm therefore provides a complementary, state-independent bound on expectation-value errors, which can be tighter than the dimension-dependent HS estimate $\|\mathbf E\|_{\mathrm{op}}\leq 2^{N/2}\|\mathbf E\|_{\mathrm{HS}}$ when the error has a sufficiently sparse Pauli expansion~\cite{Angrisani_2026}.
Likewise, rescaling a truncated observable according to
\begin{equation}
\widetilde{\mathbf O}
\longrightarrow
\frac{\normsp{\mathbf O_S}{\rm HS}}
{\normsp{\widetilde{\mathbf O}}{\rm HS}}
\widetilde{\mathbf O}
\label{eq:si-hs-rescaling}
\end{equation}
restores the initial HS norm but neither the original spectrum nor the corresponding operator-norm bound~\cite{shao2026_topK}.

Consequently, an expectation value outside the exact spectral interval provides a sufficient witness of spectral inadmissibility, whereas remaining within that interval does not guarantee accuracy or preservation of the full spectrum.
Such violations can occur in both HS and covariance restricted dynamics and should not be attributed exclusively to the rank-deficient covariance geometry.
The dynamical compatibility condition of Eq.~\eqref{eq:quotient-compatibility} instead concerns the activation of covariance-null directions and is logically distinct from spectral admissibility.

\section{Covariance bilinear form, its orthogonal projection and elementary properties}
\label{app:projection-properties}

\subsection{Mean-field interpretation and low-order examples}
\label{app:mean-field-examples}

The $m=1$ projection generalizes the familiar Weiss mean-field replacement from Hamiltonians to arbitrary observables.
Consider the Ising Hamiltonian
\begin{equation}
\h=-J\sum_{\langle i,j\rangle}Z_iZ_j-h\sum_iZ_i.
\label{eq:weiss-ising-hamiltonian}
\end{equation}
For a product reference state, define $\mathfrak{m}_i=\Tr[\sigma Z_i]$.
The conventional replacement is
\begin{equation}
Z_iZ_j\longmapsto\mathfrak{m}_iZ_j+\mathfrak{m}_jZ_i-\mathfrak{m}_i\mathfrak{m}_jI.
\label{eq:weiss-replacement}
\end{equation}
This replacement removes the centered two-site fluctuation while retaining its one-site contractions and expectation value.
For a translationally invariant reference state with $\mathfrak{m}_i=\mathfrak{m}$, the projected Hamiltonian is
\begin{equation}
\h_{\mathrm{MF}}=-(Jz\mathfrak{m}+h)\sum_iZ_i+\frac{1}{2}JNz\mathfrak{m}^2I,
\label{eq:weiss-hamiltonian}
\end{equation}
where $z$ is the coordination number.
The scalar term does not affect the Heisenberg equation, but it is retained because the projection preserves expectation values.

A three-body product makes the higher-order structure explicit.
For $\mathbf{Q}=\mathbf{q}_1\mathbf{q}_2\mathbf{q}_3$ and $a_i=\Tr[\sigma_0\mathbf{q}_i]$, the $m=2$ projection is
\begin{align}
\pi_2^{\sigma_0}(\mathbf{Q})
&=a_3\mathbf{q}_1\mathbf{q}_2+a_2\mathbf{q}_1\mathbf{q}_3+a_1\mathbf{q}_2\mathbf{q}_3\nonumber\\
&\quad-a_2a_3\mathbf{q}_1-a_1a_3\mathbf{q}_2-a_1a_2\mathbf{q}_3+a_1a_2a_3I.
\label{eq:pim-threebody-m2}
\end{align}
The $m=1$ projection is
\begin{equation}
\pi_1^{\sigma_0}(\mathbf{Q})=a_2a_3\mathbf{q}_1+a_1a_3\mathbf{q}_2+a_1a_2\mathbf{q}_3-2a_1a_2a_3I.
\label{eq:pim-threebody-m1}
\end{equation}
The first expression retains all two-body contractions, whereas the second retains only one-body mean-field contributions.

\subsection{Orthogonality of the $m$-body projection}

The fact that $\pi_m^{\sigma_0}$ is the orthogonal projector with respect to the covariance inner product induced by the initial state $\sigma_0$, $(\cdot,\cdot)_{\mathrm{covar}(\sigma_0)}$, has been established and proven in \cite{perez_quantum_2024,Perez_short_paper_2026}.

\subsection{Derivation of the uncentered expansion}
\label{app:uncentered-proof}

The centered expression in Eq.~\eqref{eq:pim-centered-product} is the most compact definition of the projection for product states.
The uncentered formula in Eq.~\eqref{eq:pim-uncentered-expansion} follows by expanding every centered factor.

Start from
\begin{equation}
\pi_m^{\sigma_0}(\mathbf{P})=\sum_{\substack{S\subseteq\{1,\ldots,n\}\\|S|\leq m}}\left(\prod_{i\notin S}\mu_i\right)\bigotimes_{i\in S}(\mathbf{p}_i-\mu_iI).
\label{eq:uncentered-proof-start}
\end{equation}
Fix an uncentered support $T\subseteq\{1,\ldots,n\}$ with $|T|=r$.
This support receives contributions from every centered support $S$ satisfying $T\subseteq S$ and $|S|\leq m$.
For a given $S$, each site in $S\setminus T$ contributes $-\mu_i$, while each site outside $S$ contributes $\mu_i$.
The total coefficient multiplying $\bigotimes_{i\in T}\mathbf{p}_i$ is therefore
\begin{equation}
\left(\prod_{j\notin T}\mu_j\right)\sum_{q=0}^{m-r}(-1)^q\binom{n-r}{q}.
\label{eq:uncentered-support-coefficient}
\end{equation}
Using
\begin{equation}
\sum_{q=0}^{M}(-1)^q\binom{R}{q}=(-1)^M\binom{R-1}{M},\qquad M<R,
\label{eq:partial-binomial-identity}
\end{equation}
with $R=n-r$ and $M=m-r$ gives
\begin{equation}
\sum_{q=0}^{m-r}(-1)^q\binom{n-r}{q}=(-1)^{m-r}\binom{n-r-1}{m-r}.
\label{eq:uncentered-binomial-reduction}
\end{equation}
This is precisely the coefficient $\beta_r^{(n,m)}$ in Eq.~\eqref{eq:pim-beta-coefficients} and proves the uncentered expression.

\subsection{Support properties}
\label{app:projection-support}

The projection is local in the elementary sense that it cannot enlarge the support of a product operator.

\begin{lemma}[Support nonexpansion]
Let $\mathbf{P}=\bigotimes_{j\in X}\mathbf{p}_j$ be a product operator supported on a finite set $X$.
Then
\begin{equation}
\operatorname{supp}\!\left(\pi_m^{\sigma_0}(\mathbf{P})\right)\subseteq X.
\label{eq:appendix-support-nonexpansion}
\end{equation}
\end{lemma}

Every term in Eq.~\eqref{eq:pim-centered-product} is obtained from $\mathbf{P}$ by replacing some local factors with scalar expectations.
No term introduces an operator on a site outside the original support $X$.
The projected support is therefore contained in $X$.
\hfill$\square$

The support can shrink when the expectation values required by all terms containing a given site vanish.
The following statement gives a simple sufficient condition for retaining the complete original support.

\begin{lemma}[Support preservation]
Let \(\mathbf P\) be a product operator with support \(X\).
Write \(\mathbf P=\bigotimes_{j\in X}\mathbf p_j\) and set \(|X|=n\).
Assume \(1\leq m<n\) and \(\mu_j\neq0\) for every \(j\in X\).
Here \(\mu_j=\Tr(\sigma_j\mathbf p_j)\).
Then
\begin{equation}
\operatorname{supp}\!\left(\pi_m^{\sigma_0}(\mathbf{P})\right)=X.
\label{eq:appendix-support-preservation}
\end{equation}
\end{lemma}

In the uncentered expansion, each site $j\in X$ appears in the retained one-body term with $T=\{j\}$.
Its coefficient is proportional to $\prod_{k\in X\setminus\{j\}}\mu_k$ and is therefore nonzero under the stated assumption.
Every site in $X$ consequently appears in the projected support.
Combining this result with support nonexpansion proves Eq.~\eqref{eq:appendix-support-preservation}.
\hfill$\square$

More generally, a site $j\in X$ survives whenever there is at least one retained subset $T\subseteq X$ such that $j\in T$, $1\leq|T|\leq m$, and $\prod_{k\in X\setminus T}\mu_k\neq0$.
The preceding lemma is only a convenient sufficient condition.

\section{Dynamical compatibility of the covariance geometry}
\label{app:dynamical-compatibility}

This appendix collects the formal statements and examples underlying the dynamical-compatibility condition introduced in Sec.~\ref{sec:dynamical-compatibility}.
We distinguish the compatibility of the closed restricted dynamics with the covariance quotient from the activation of covariance-null directions in the full operator space.
We further discuss the approach to rank-deficient reference states, explicit examples of dynamically activated null directions, and the finite-time return of information through the discarded sector.

\subsection{Covariance-quotient compatibility}
\label{app:quotient-compatibility-proof}

The global compatibility condition can fail even in the favorable fully polarized TFIM regime.
For the state
\begin{equation}
    \sigma_0=(\ket{\uparrow}\!\bra{\uparrow})^{\otimes N},
\end{equation}
where the $m=1$-body operators $\mathbf D_i=I-Z_i$ are covariance null.
Nevertheless,
\begin{equation}
P_m\genH\mathbf D_i=-2h_XY_i,
\qquad
\|P_m\genH\mathbf D_i\|_{\mathrm{covar}(\sigma_0)}=2|h_X|
\end{equation}
for every $m\geq1$.
The condition therefore concerns representative independence on the full retained quotient, not the accuracy of a particular trajectory propagated with a fixed representative.

Let
\[
P_m=\pi_m^{\sigma_0},
\qquad
Q_m=I-P_m,
\qquad
\mathcal N_m^{\sigma_0}
=
\mathcal A_m\cap\mathcal N^{\sigma_0}.
\]
Consider two retained representatives $\mathbf X,\mathbf X'\in\mathcal A_m$ that belong to the same covariance-equivalence class,
\begin{equation}
\mathbf X'
=
\mathbf X+\alpha\mathbf D,
\qquad
\mathbf D\in\mathcal N_m^{\sigma_0}.
\label{eq:app-equivalent-representatives}
\end{equation}
Their projected derivatives differ by
\begin{equation}
P_m\genH\mathbf X'
-
P_m\genH\mathbf X
=
\alpha P_m\genH\mathbf D .
\label{eq:app-first-order-difference}
\end{equation}
The two derivatives therefore belong to the same covariance-equivalence class if and only if
\begin{equation}
\normsp{
P_m\genH\mathbf D
}{
\mathrm{covar}(\sigma_0)
}
=
0.
\label{eq:app-first-order-iff}
\end{equation}

Since $P_m\genH\mathbf D\in\mathcal A_m$, Eq.~\eqref{eq:app-first-order-iff} is equivalent to
\begin{equation}
P_m\genH\mathbf D
\in
\mathcal N_m^{\sigma_0}.
\end{equation}
Requiring this for every retained covariance-null direction gives
\begin{equation}
    P_m\genH\,\mathcal N_m^{\sigma_0}
\subseteq
\mathcal N_m^{\sigma_0},
\label{appeq:dyn_compat_condt}
\end{equation}
which proves Eq.~\eqref{eq:quotient-compatibility} of the main text.

The condition has a direct consequence for the complete restricted dynamics.
Writing
\[
\genH_m=P_m\genH,
\]
the invariance of $\mathcal N_m^{\sigma_0}$ implies
\[
\genH_m^r\mathcal N_m^{\sigma_0}
\subseteq
\mathcal N_m^{\sigma_0},
\qquad
r\geq1,
\]
and therefore
\[
e^{t\genH_m}\mathcal N_m^{\sigma_0}
\subseteq
\mathcal N_m^{\sigma_0}.
\]
Thus, whenever Eq.~\eqref{appeq:dyn_compat_condt} holds, the complete closed restricted flow is independent of the representative chosen from each covariance-equivalence class.

For a covariance-null defect $\mathbf D\notin\mathcal A_m$, the same quantity
\[
\normsp{
P_m\genH\mathbf D
}{
\mathrm{covar}(\sigma_0)
}
\]
has a different interpretation.
In this case $\mathbf D$ does not describe an ambiguity between two retained representatives.
Instead, it measures whether a covariance-null distinction in the full operator space is immediately activated into covariance-resolved retained directions.
This distinction will be important in the examples below.

\subsection{Full-rank states and the approach to a rank-deficient limit}
\label{app:near-null-limit}

The covariance-resolved/null distinction simplifies for a strictly full-rank reference state.
Since $\sigma_0>0$, $\mathcal N^{\sigma_0}=\{0\}$, so there are no exact covariance aliases or ambiguities associated with covariance-null directions.
The projection still removes higher-centered-body sectors, but every nonzero discarded operator has strictly positive covariance norm.

This distinction is particularly transparent for the homogeneous $z$-polarized family
\begin{equation}
\sigma_0(\mu)
=
\bigotimes_{i=1}^{N}
\frac{I+\mu Z_i}{2},
\qquad
0\leq\mu\leq1.
\label{eq:homogeneous-polarized-reference}
\end{equation}
For $\mu<1$, the state is full rank and $\mathcal N^{\sigma_0(\mu)}=\{0\}$, whereas at $\mu=1$ it becomes pure and the covariance form develops a nontrivial null space.

Approaching this endpoint can nevertheless produce a strongly anisotropic covariance geometry before exact nullity is reached.
For example, for
\[
\mathbf Q_i=I-Z_i,
\]
one obtains
\begin{equation}
\normsp{
\mathbf Q_i
}{
\mathrm{covar}(\sigma_0(\mu))
}^2
=
2(1-\mu).
\end{equation}
Thus, $\mathbf Q_i$ is not covariance null for any $\mu<1$, but its norm vanishes continuously as $\mu\rightarrow1^{-}$.
Directions of this type are near-null: they remain statistically distinguishable at every full-rank point, but only weakly so.

It is therefore useful to distinguish near-null directions, which have small but strictly positive covariance norm, from exact covariance-null directions, which occur only at rank deficiency.
Only the latter generate exact covariance equivalence between distinct operators.
Close to the rank-deficient limit, however, a near-null direction can be mapped into a finite covariance-resolved direction by the projected generator.

For a family of reference states approaching such a limit, we quantify this through
\begin{equation}
\chi_m(\mathbf D;\mu)
=
\frac{
\normsp{
P_m(\mu)\genH\mathbf D
}{
\mathrm{covar}(\sigma_0(\mu))
}
}{
\normsp{
\mathbf D
}{
\mathrm{covar}(\sigma_0(\mu))
}
},
\label{eq:app-near-null-amplification}
\end{equation}
defined away from exact nullity.
A bounded or vanishing $\chi_m$ means that the projected image weakens together with the static distinction, whereas a divergent $\chi_m$ signals singular amplification of a near-null direction.

\subsection{A bright--dark covariance alias}
\label{app:bright-dark-example}

To make the covariance aliasing mechanism explicit, consider the two product operators
\begin{equation}
\mathbf A=X_0X_2,
\qquad
\mathbf B=X_0Z_1X_2.
\label{eq:app-ab-definitions}
\end{equation}
For the reference family in Eq.~\eqref{eq:homogeneous-polarized-reference}, the covariance Gram matrix in the dictionary $\{\mathbf A,\mathbf B\}$ is
\begin{equation}
\mathcal G_{\mathbf{AB}}(\mu)
=
\begin{pmatrix}
1&\mu\\
\mu&1
\end{pmatrix}.
\label{eq:compatibility-gram-matrix}
\end{equation}
It is useful to introduce the bright and dark combinations
\begin{equation}
\mathbf S=\mathbf A+\mathbf B,
\qquad
\mathbf D=\mathbf A-\mathbf B.
\label{eq:app-bright-dark-definitions}
\end{equation}
Their covariance norms are
\begin{equation}
\begin{aligned}
\normsp{
\mathbf S
}{
\mathrm{covar}(\sigma_0(\mu))
}^2
&=
2(1+\mu),
\\
\normsp{
\mathbf D
}{
\mathrm{covar}(\sigma_0(\mu))
}^2
&=
2(1-\mu).
\end{aligned}
\label{eq:compatibility-bright-dark-norms}
\end{equation}
Thus, at $\mu=1$, the Gram matrix has rank one and
\begin{equation}
\mathbf A
\simeq_{\mathrm{covar}(\sigma_0(1))}
\mathbf B,
\qquad
\mathbf D
\simeq_{\mathrm{covar}(\sigma_0(1))}
0.
\label{eq:ab-static-alias}
\end{equation}

The rank loss removes one coefficient combination.
Indeed, writing
\begin{equation}
\begin{aligned}
\phi_A\mathbf A+\phi_B\mathbf B
&=
\frac{s}{2}\mathbf S
+
\frac{d}{2}\mathbf D,
\\
s&=\phi_A+\phi_B,
\qquad
d=\phi_A-\phi_B,
\end{aligned}
\label{eq:ab-bright-dark-coordinates}
\end{equation}
shows that the covariance geometry resolves $s$ but contains no information about $d$ at $\mu=1$.
This is also apparent from
\begin{equation}
\begin{aligned}
\mathcal G_{\mathbf{AB}}^{-1}(\mu)
&=
\frac{1}{1-\mu^2}
\begin{pmatrix}
1&-\mu\\
-\mu&1
\end{pmatrix},
\qquad
\mu<1,
\\[1mm]
\mathcal G_{\mathbf{AB}}^{+}(1)
&=
\frac14
\begin{pmatrix}
1&1\\
1&1
\end{pmatrix}.
\end{aligned}
\label{eq:ab-gram-inverse-and-pseudoinverse}
\end{equation}
The Moore--Penrose inverse therefore keeps the bright coordinate and sets the dark coordinate to zero, fixing a particular representative.

Consider now the family of covariance-equivalent operators
\begin{equation}
\mathbf O_H^{(\alpha)}(0)
=
\mathbf A+\mathbf B+\alpha(\mathbf A-\mathbf B)
=
\mathbf O_H(0)+\alpha\mathbf D.
\label{eq:ab-equivalent-representatives}
\end{equation}
At $\mu=1$, all values of $\alpha$ belong to the same covariance-equivalence class.
Their projected derivatives differ by
\begin{equation}
P_m(1)\genH\mathbf O_H^{(\alpha)}(0)
-
P_m(1)\genH\mathbf O_H(0)
=
\alpha P_m(1)\genH\mathbf D.
\label{eq:representative-derivative-difference}
\end{equation}

The defect $\mathbf D=\mathbf A-\mathbf B$ belongs to $\mathcal A_3$ but not to $\mathcal A_1$ or $\mathcal A_2$.
Consequently, for $m=1,2$, its projected image measures whether a covariance-null full-space distinction is activated into the retained equations.
Only for $m=3$ does the same defect directly test the representative independence of the retained covariance quotient.

\subsection{Transverse-field Ising and anisotropic \texorpdfstring{$XY$}{XY} examples}
\label{app:tfim-xy-compatibility}

We now evaluate the defect
\[
\mathbf D
=
X_0X_2-X_0Z_1X_2
\]
for two nearest-neighbor models, the transverse-field Ising model and the anisotropic $XY$ model.
In their spin representations, the Pauli-string dynamics do not generally close within a fixed finite-$m$ body-order sector, making them useful contrasting examples of the projected activation introduced above.

For the reference family in Eq.~\eqref{eq:homogeneous-polarized-reference},
\begin{equation}
\normsp{
\mathbf D
}{
\mathrm{covar}(\sigma_0(\mu))
}
=
\sqrt{2(1-\mu)}.
\label{eq:defect-norm-mu}
\end{equation}
The defect therefore becomes exactly covariance null at $\mu=1$.

\subsubsection{Transverse-field Ising model}

We restrict the Hamiltonian to the open three-site chain $\{0,1,2\}$, with bond sums over $i=0,1$ and onsite sums over $i=0,1,2$.
Consider the $ZZ+X$ transverse-field Ising model
\begin{equation}
\h_{\mathrm{TFIM}}
=
J\sum_i Z_iZ_{i+1}
+
h\sum_i X_i.
\label{eq:tfim-compatibility-h}
\end{equation}
The Liouvillian image of the defect is
\begin{equation}
\begin{aligned}
\genH\mathbf D
=
&-2JY_0(Z_1-I)X_2
-2JX_0(Z_1-I)Y_2
\\
&-2hX_0Y_1X_2.
\end{aligned}
\label{eq:tfim-dark-derivative}
\end{equation}

Defining
\begin{equation}
\mathbf C
=
Y_0X_2+X_0Y_2,
\label{eq:tfim-c-definition}
\end{equation}
the first three state-adapted projections are
\begin{equation}
\begin{aligned}
P_1(\mu)\genH\mathbf D
&=
0,
\\
P_2(\mu)\genH\mathbf D
&=
2J(1-\mu)\mathbf C,
\\
P_3(\mu)\genH\mathbf D
&=
\genH\mathbf D.
\end{aligned}
\label{eq:tfim-projected-images-m123}
\end{equation}
Using
\[
\normsp{
\mathbf C
}{
\mathrm{covar}(\sigma_0(\mu))
}^2
=
2(1+\mu^2),
\]
one obtains
\begin{equation}
\begin{aligned}
\normsp{
P_1(\mu)\genH\mathbf D
}{
\mathrm{covar}(\sigma_0(\mu))
}
&=
0,
\\
\normsp{
P_2(\mu)\genH\mathbf D
}{
\mathrm{covar}(\sigma_0(\mu))
}
&=
2|J|(1-\mu)\sqrt{2(1+\mu^2)},
\\
\normsp{
P_3(\mu)\genH\mathbf D
}{
\mathrm{covar}(\sigma_0(\mu))
}
&=
2\sqrt{
h^2+4J^2(1-\mu)(1+\mu^2)
}.
\end{aligned}
\label{eq:tfim-projected-norms-m123}
\end{equation}

At the pure-state endpoint,
\[
\normsp{
P_1(1)\genH\mathbf D
}{
\mathrm{covar}(\sigma_0(1))
}
=
\normsp{
P_2(1)\genH\mathbf D
}{
\mathrm{covar}(\sigma_0(1))
}
=
0,
\]
while
\[
\normsp{
P_3(1)\genH\mathbf D
}{
\mathrm{covar}(\sigma_0(1))
}
=
2|h|.
\]
Because $\mathbf D\notin\mathcal A_1,\mathcal A_2$, the first two equalities do not test the independence of two retained representatives.
Rather, they show that this particular full-space covariance-null defect is not activated into the retained $m=1$ or $m=2$ equations at first order.

For $m=2$, the corresponding near-null amplification is
\begin{equation}
\chi_2^{\mathrm{TFIM}}(\mathbf D;\mu)
=
2|J|
\sqrt{
(1-\mu)(1+\mu^2)
}
\underset{\mu\rightarrow1}{\longrightarrow}
0.
\label{eq:tfim-m2-amplification}
\end{equation}
The interpretation changes for $m=3$ because $\mathbf D$ itself belongs to the retained three-body space.
The nonzero projected image at $\mu=1$ therefore identifies a retained-null representative incompatibility.
It is not, by itself, a prediction of numerical failure, since the corresponding direction must also be dynamically reached.
Indeed, the $m=3$ covariance propagation studied in Sec.~\ref{sec:results} remains regular and substantially outperforms its HS counterpart.

\subsubsection{Anisotropic \texorpdfstring{$XY$}{XY} model}

Consider
\begin{equation}
\h_{XY}
=
\sum_i
\left(
J_xX_iX_{i+1}
+
J_yY_iY_{i+1}
\right).
\label{eq:xy-compatibility-h}
\end{equation}
Define
\begin{equation}
\mathbf C_{XY}
=
Y_1X_2+X_0Y_1,
\qquad
\mathbf C_Z
=
Z_0Y_1X_2+X_0Y_1Z_2.
\label{eq:xy-bright-directions}
\end{equation}
For the anisotropic choice $J_x=-J_y=J$, the defect evolves as
\begin{equation}
\genH\mathbf D
=
-2J
\left(
\mathbf C_Z+\mathbf C_{XY}
\right).
\label{eq:xy-dark-image}
\end{equation}

The relevant covariance overlaps are
\[
\normsp{
\mathbf C_{XY}
}{
\mathrm{covar}(\sigma_0(\mu))
}^2
=
2,
\qquad
\normsp{
\mathbf C_Z
}{
\mathrm{covar}(\sigma_0(\mu))
}^2
=
2,
\]
and
\[
(
\mathbf C_{XY},
\mathbf C_Z
)_{\mathrm{covar}(\sigma_0(\mu))}
=
2\mu.
\]
Since
\[
P_2(\mu)\mathbf C_Z
=
\mu\mathbf C_{XY},
\]
the first three projections are
\begin{equation}
\begin{aligned}
P_1(\mu)\genH\mathbf D
&=
0,
\\
P_2(\mu)\genH\mathbf D
&=
-2J(1+\mu)\mathbf C_{XY},
\\
P_3(\mu)\genH\mathbf D
&=
\genH\mathbf D.
\end{aligned}
\label{eq:xy-projected-images-m123}
\end{equation}
Their covariance norms are
\begin{equation}
\begin{aligned}
\normsp{
P_1(\mu)\genH\mathbf D
}{
\mathrm{covar}(\sigma_0(\mu))
}
&=
0,
\\
\normsp{
P_2(\mu)\genH\mathbf D
}{
\mathrm{covar}(\sigma_0(\mu))
}
&=
2\sqrt{2}|J|(1+\mu),
\\
\normsp{
P_3(\mu)\genH\mathbf D
}{
\mathrm{covar}(\sigma_0(\mu))
}
&=
4|J|\sqrt{1+\mu}.
\end{aligned}
\label{eq:xy-projected-norms-m123}
\end{equation}

At $\mu=1$, the covariance norm of the two-body projected image remains finite,
\begin{equation}
\normsp{
P_2(1)\genH\mathbf D
}{
\mathrm{covar}(\sigma_0(1))
}
=
4\sqrt{2}|J|.
\label{eq:xy-dark-to-bright}
\end{equation}
The corresponding amplification therefore diverges,
\begin{equation}
\chi_2^{XY}(\mathbf D;\mu)
=
\frac{
2|J|(1+\mu)
}{
\sqrt{1-\mu}
}
\underset{\mu\rightarrow1}{\longrightarrow}
\infty.
\label{eq:xy-alias-gain}
\end{equation}

The contrast at $m=2$ can be summarized as
\begin{equation}
\begin{aligned}
\mathrm{TFIM}:&\quad
\normsp{
\mathbf D
}{
\mathrm{covar}(\sigma_0(\mu))
}
=
O\!\left(\sqrt{1-\mu}\right),
\\
&\quad
\normsp{
P_2(\mu)\genH\mathbf D
}{
\mathrm{covar}(\sigma_0(\mu))
}
=
O(1-\mu),
\\[1mm]
XY:&\quad
\normsp{
\mathbf D
}{
\mathrm{covar}(\sigma_0(\mu))
}
=
O\!\left(\sqrt{1-\mu}\right),
\\
&\quad
\normsp{
P_2(\mu)\genH\mathbf D
}{
\mathrm{covar}(\sigma_0(\mu))
}
=
O(1).
\end{aligned}
\label{eq:tfim-xy-alias-scaling}
\end{equation}
The TFIM defect becomes dynamically less distinguishable inside the retained two-body equations as the static distinction disappears, whereas the $XY$ defect approaches a covariance-null direction while retaining a finite two-body projected Liouvillian image.
This contrast identifies the absence or presence of this particular singular activation channel.
It remains a structural diagnostic rather than a sufficient predictor of finite-time stability or accuracy, since the corresponding direction must also be dynamically populated.

\subsection{Projected equivalence beyond first order}
\label{app:all-order-projected-equivalence}

The quotient compatibility in Eq.~\eqref{eq:quotient-compatibility} concerns the closed restricted flow generated by $P_m\genH P_m$.
A distinct question is whether two covariance-equivalent full-space observables remain equivalent after exact evolution followed by post-facto projection.

For a defect $\mathbf D$, the exact projected separation is
\begin{equation}
P_m e^{t\genH}\mathbf D
=
\sum_{r=0}^{\infty}
\frac{t^r}{r!}
P_m\genH^r\mathbf D.
\label{eq:app-projected-future-expansion}
\end{equation}
If
\[
P_m\mathbf D
\simeq_{\mathrm{covar}(\sigma_0)}
0,
\]
the zeroth-order contribution is already covariance null.
The two exact post-facto projected futures remain covariance equivalent for all times if
\begin{equation}
P_m\genH^r\mathbf D
\simeq_{\mathrm{covar}(\sigma_0)}
0,
\qquad
\forall r\geq1.
\label{eq:app-all-order-condition}
\end{equation}

The $r=1$ member of Eq.~\eqref{eq:app-all-order-condition} reproduces the projected-activation quantity appearing in Eq.~\eqref{eq:projected-alias-amplitude}.
For retained covariance-null directions, however, invariance of $\mathcal N_m^{\sigma_0}$ under the restricted generator gives the stronger quotient statement proven in App.~\ref{app:quotient-compatibility-proof}.
For a covariance-null defect outside the retained space, $P_m\genH\mathbf D$ instead gives only the first-order projected activation of the exact dynamics.

A defect can therefore have vanishing first-order activation while returning to the retained sector after two or more applications of the Liouvillian.
Conversely, a nonzero first-order projected image does not imply numerical failure of a particular trajectory.
The dynamically explored operator subspace can be much smaller than the complete retained space, and a formally problematic direction is harmless when it is never reached with appreciable amplitude.

\subsection{Excursions through the discarded sector}
\label{app:nz-projected-dynamics}

The role of higher-order return is made explicit by separating the exact observable into retained and discarded components.
For compactness, in this subsection we write
\[
P_m = \pi_m^{\sigma_0},
\qquad
Q=I-P_m.
\]
Projecting the exact Heisenberg equation gives
\begin{align}
\frac{d}{dt}P_m\OH(t)
&=
P_m\genH P_m\OH(t)
+
P_m\genH Q_m\OH(t),
\label{eq:app-p-dynamics}
\\
\frac{d}{dt}Q_m\OH(t)
&=
Q_m\genH P_m\OH(t)
+
Q_m\genH Q_m\OH(t).
\label{eq:app-q-dynamics}
\end{align}
The second equation has the formal solution
\begin{equation}
\begin{split}
Q_m\OH(t)
&= e^{tQ_m\genH Q_m}Q_m\OH(0)
\\
&\quad + \int_0^t ds\, e^{(t-s)Q\genH Q_m}
Q_m\genH P_m\OH(s).
\label{eq:app-q-formal-solution}
\end{split}
\end{equation}
Substituting Eq.~\eqref{eq:app-q-formal-solution} into Eq.~\eqref{eq:app-p-dynamics} yields
\begin{align}
\frac{d}{dt}P_m\OH(t)
&=
P_m\genH P_m\OH(t)
\nonumber\\
&\quad+
P\genH Q_m
e^{tQ_m\genH Q_m}
Q_m\OH(0)
\nonumber\\
&\quad+
\int_0^t ds\,
P_m\genH Q_m
e^{(t-s)Q_m\genH Q_m}
Q_m\genH P\OH(s).
\label{eq:app-nz-projected}
\end{align}
The second term is the contribution from initially discarded components, while the integral describes information that leaves the retained sector and later returns.

For an initially retained observable,
\[
Q_m\mathbf O=0,
\qquad
P_m\mathbf O=\mathbf O,
\]
the first difference between exact evolution followed by projection and the closed restricted flow appears at second order,
\begin{equation}
P_m e^{t\genH}P_m\mathbf O - e^{tP_m\genH P_m}P_m\mathbf O = \frac{t^2}{2}
P_m\genH Q_m\genH P_m\,{\bf O} + O(t^3).
\label{eq:app-first-round-trip}
\end{equation}
Equation~\eqref{eq:app-first-round-trip} makes clear why projected activation and the finite-time truncation error are distinct.
The former asks whether a covariance-null distinction develops an immediate covariance-resolved retained image.
The latter additionally depends on excursions through the discarded sector and on their subsequent return to the dynamically explored retained space.

\subsection{Coordinate representation}
\label{app:coordinate-representation}

The operator criterion can be represented in any finite working dictionary, but the resulting matrices are coordinate objects rather than intrinsic observables.
Let $\{\mathbf B_j\}$ be such a dictionary and define
\begin{align}
\mathcal G_{ij}
&=
(\mathbf B_i,\mathbf B_j)_{\mathrm{covar}(\sigma_0(\mu))},
\nonumber\\
\mathcal H_{ij}
&=
(\mathbf B_i,\genH\mathbf B_j)_{\mathrm{covar}(\sigma_0(\mu))}.
\label{eq:app-gh-definitions}
\end{align}
For
\begin{equation}
\mathbf D_v
=
\sum_j v_j\mathbf B_j,
\label{eq:app-coordinate-defect}
\end{equation}
one has
\begin{equation}
v^T\mathcal Gv
=
\normsp{
\mathbf D_v
}{
\mathrm{covar}(\sigma_0(\mu))
}^2.
\label{eq:app-coordinate-denominator}
\end{equation}

Using the Moore--Penrose representative for the covariance projection, the projected image satisfies
\begin{equation}
\normsp{
P_m(\mu)\genH\mathbf D_v
}{
\mathrm{covar}(\sigma_0(\mu))
}^2
=
v^T
\mathcal H^T
\mathcal G^{+}
\mathcal H
v,
\label{eq:app-coordinate-numerator}
\end{equation}
provided that the matrices are evaluated in the same retained working dictionary.
Hence the near-null amplification can be written as
\begin{equation}
\chi_m(\mathbf D_v;\mu)^2
=
\frac{
v^T
\mathcal H^T
\mathcal G^{+}
\mathcal H
v
}{
v^T
\mathcal G
v
}.
\label{eq:app-coordinate-amplification}
\end{equation}

Equation~\eqref{eq:app-coordinate-amplification} is only a coordinate representation of the operator ratio in Eq.~\eqref{eq:app-near-null-amplification}.
It should not be interpreted as an independent matrix-level definition of compatibility.
In particular, statements involving the kernel of a coordinate Gram matrix are meaningful only after specifying the dictionary and the operator represented by the corresponding coefficient vector.

\subsection{Regularizing the reference geometry}
\label{app:reference-geometry-regularization}

Possible remedies for strongly singular covariance geometries lie beyond the scope of the present work.
One possibility is to distinguish the physical initial state $\rho_0$, in which expectation values are evaluated, from a full-rank state $\sigma_{\mathrm{ref}}$ used only to define the covariance geometry.
For example, one may take
\[
\sigma_{\mathrm{ref}}
=
\bigotimes_i
\frac{I+\mu_iZ_i}{2},
\qquad
|\mu_i|<1,
\]
or
\[
\sigma_{\mathrm{ref}}
=
(1-\eta)\rho_0
+
\eta\frac{I}{d},
\qquad
0<\eta<1.
\]
Both choices lift exact covariance-null directions, but they also weaken the state adaptation and do not by themselves guarantee small dynamical activation.

The physical expectation-value error remains controlled whenever
\[
\rho_0
\leq
c\,\sigma_{\mathrm{ref}},
\]
since every Hermitian error operator $\mathbf E$ then satisfies
\begin{equation}
\left|
\Tr(\rho_0\mathbf E)
\right|
\leq
\normsp{
\mathbf E
}{
\mathrm{covar}(\rho_0)
}
\leq
\sqrt{c}\,
\normsp{
\mathbf E
}{
\mathrm{covar}(\sigma_{\mathrm{ref}})
}.
\label{eq:reference-error-transfer-bound}
\end{equation}
For the depolarized choice above, one may take
\[
c=(1-\eta)^{-1}.
\]
Optimizing the reference geometry, retaining dynamically activated null directions, or approximating the memory kernel explicitly are left for future work.
In the present work, the compatibility condition and projected activation are used as structural diagnostics, while finite-time accuracy is assessed independently through the numerical benchmarks.

\section{Proofs for the complexity and error bounds}
\label{app:complexity-error-proofs}

This appendix proves the statements used in \cref{sec:complexity-errors}.
We first establish the worst-case cost count, then derive the representation--leakage decomposition, and finally prove the necessary and sufficient body-order bounds.

\subsection{Worst-case cost at fixed body order}
\label{app:cost-bound}

Let $B(t)$ be the relevant LR region containing the effective support of $\mathbf O_H(t)$ after an exponentially small spatial tail has been discarded.
For a local Hamiltonian on a $d$-dimensional bounded-degree lattice, the volume of the Heisenberg evolution satisfies
\begin{equation}
N(t):=|B(t)|\leq C_{\mathrm{LR}}(1+t)^d.
\label{eq:app-lightcone-volume}
\end{equation}
The number of qubit Pauli strings supported in $B(t)$ with body order at most $m$ is
\begin{equation}
D_m(t)=\sum_{r=0}^{m}\binom{N(t)}{r}3^r.
\label{eq:app-body-count}
\end{equation}
For fixed $m$, the elementary estimate $\binom{N}{r}\leq N^r/r!$ gives
\begin{equation}
D_m(t)
\leq
\sum_{r=0}^{m}\frac{[3N(t)]^r}{r!}
\leq C_m[1+N(t)]^m.
\label{eq:app-body-count-bound}
\end{equation}
Combining Eqs.~\eqref{eq:app-lightcone-volume} and \eqref{eq:app-body-count-bound} yields
\begin{equation}
D_m(t)=\mathcal O\!\left((1+t)^{dm}\right)
\label{eq:app-body-count-time}
\end{equation}
for fixed $m$.

We define the instantaneous retained-space cost as the maximum number of available low-body Pauli coordinates within the effective causal region,
\begin{equation}
\mathrm{Cost}(t):=D_m(t)
=\mathcal O(t^{dm}),
\end{equation}
where the scaling is understood asymptotically at large times. 

Let $T_n$ denote the beginning of the $n$th adaptive chart.
For product reference states, $\pi_m^{\sigma_0}$ does not enlarge spatial support, while each application of the local generator enlarges support by at most a finite interaction shell.
Provided the active HB seed remains within the effective causal region, the vectors generated at fixed depth $\ell$ therefore occupy a region enlarged by at most $\ell$ interaction shells.
This fixed enlargement does not change the asymptotic retained-space scaling, and the number of Pauli coordinates available to a chart satisfies
\begin{equation}
\mathrm{Cost}_n
\leq C_{\ell,m,p}(1+T_n)^{dm}.
\label{eq:app-one-chart-cost}
\end{equation}
Here $\mathrm{Cost}_n$ denotes the retained-coordinate count associated with the chart, rather than the arithmetic cost of its construction.

For local Hamiltonians with bounded interaction strengths, characteristic dynamical timescales are set by the microscopic couplings.
This motivates considering a bounded density of adaptive reconstructions, although LR locality alone does not guarantee such behavior.
We therefore assume that successive reconstruction times satisfy $T_{n+1}-T_n\geq\tau_{\rm ch}>0$.
The number of charts then obeys $N_{\rm ch}(t)\leq 1+t/\tau_{\rm ch}$.
Consequently, the cumulative retained-space count satisfies
\begin{align}
\mathrm{CumCost}(t)
&\leq
C_{\ell,m,p}
\sum_{n=0}^{N_{\rm ch}(t)-1}(1+T_n)^{dm}
\nonumber\\
&\leq
C'_{\ell,m,p}(1+t)^{dm+1}
=
\mathcal O(t^{dm+1}).
\label{eq:app-cumulative-cost}
\end{align}
Since only the current HB chart needs to be retained, the peak storage of its Pauli coefficients obeys
\begin{equation}
\mathrm{Mem}(t)
\leq C''_{\ell,m}(1+t)^{dm} = \mathcal O(t^{dm}).
\label{eq:app-peak-memory}
\end{equation}

These estimates bound the available low-body operator coordinates and their cumulative count across adaptive charts, rather than the computational wall time or total implementation memory.
The latter additionally depend on covariance contractions, projected commutators, temporary operator buffers, indexing structures, and symbolic preprocessing.
Furthermore, polynomial retained-space growth at fixed $m$ does not guarantee that a fixed body order suffices to achieve a prescribed accuracy at arbitrarily long times.

\subsection{Error decomposition and the return-coupling bound}
\label{app:error-decomposition-proof}

We work in either a HS operator Hilbert space or the quotient induced by the covariance seminorm.
In the latter case, all maps below are understood on the covariance-resolved quotient.
Writing
\begin{equation}
P_m:=\pi_m^{\sigma_0},
\qquad
Q_m:=1-P_m,
\label{eq:app-pq-definition}
\end{equation}
the exact covariance error admits the orthogonal decomposition
\begin{align}
\Delta_{\ell,m,\varepsilon,p}^2(t)
&= \left[\Delta_{\mathrm{rep}}^{(m)}(t)\right]^2
+ \left[\Delta_{\mathrm{leak}}^{(\ell,m,\varepsilon,p)}(t)\right]^2.
\label{eq:app-first-triangle}
\end{align}
The equality follows because $Q_m\mathbf{O}_H(t)$ is covariance-orthogonal to $\mathcal A_m$, whereas $P_m\mathbf{O}_H(t)-\mathbf{O}_{\ell,m,\varepsilon,p}(t)$ belongs to $\mathcal A_m$.
This is the decomposition stated in Eq.~\eqref{eq:error-decomposition-main}.

Let $\mathbf{O}_m(t)$ be the closed restricted flow,
\begin{equation}
\dot{\mathbf{O}}_m(t)
=P_m\genH P_m\mathbf{O}_m(t),
\qquad
\mathbf{O}_m(0)=P_m\mathbf{O}_S.
\label{eq:app-closed-flow}
\end{equation}
The triangle inequality inside the retained sector yields
\begin{align}
\Delta_{\mathrm{leak}}^{(\ell,m,\varepsilon,p)}(t)
&\leq
\normsp{P_m\mathbf{O}_H(t)-\mathbf{O}_m(t)}{\mathrm{covar}(\sigma_0)}
\nonumber\\
&\quad+
\normsp{\mathbf{O}_m(t)-\mathbf{O}_{\ell,m,\varepsilon,p}(t)}{\mathrm{covar}(\sigma_0)}.
\label{eq:app-second-triangle}
\end{align}
The two terms are $E_{\mathrm{mem}}^{(m)}(t)$ and $E_{\mathrm{HB}}^{(\ell,m,\varepsilon,p)}(t)$, respectively.

Define the memory difference 
\begin{equation}
\mathbf{R}_m(t):=P_m\mathbf{O}_H(t)-\mathbf{O}_m(t).
\label{eq:app-memory-difference}
\end{equation}
with ${\bf O}_m$ the restricted covariance dynamics. 
Using $\dot{\mathbf{O}}_H=\genH\mathbf{O}_H$ and $1=P_m+Q_m$, we obtain
\begin{align}
\dot{\mathbf{R}}_m(t) &= P_m\genH P_m\mathbf{R}_m(t)\\
&\quad+ P_m\genH Q_m\mathbf{O}_H(t).
\label{eq:app-memory-equation}
\end{align}
The initial condition gives $\mathbf{R}_m(0)=0$, so Duhamel's formula gives
\begin{align}
\mathbf{R}_m(t) &= \int_0^t ds\,
e^{(t-s)P_m\genH P_m} P_m\genH Q_m\mathbf{O}_H(s).
\label{eq:app-duhamel}
\end{align}
In a fixed norm geometry $g = {\rm covar} / {\rm HS}$, define
\begin{equation}
L_{m,g}:=\normsp{P_m\genH Q_m}{{\rm op},g},
\qquad
\mu_{m,g}:=\normsp{P_m\genH P_m}{{\rm op},g}.
\label{eq:app-growth-coupling}
\end{equation}
The standard semigroup estimate and submultiplicativity imply
\begin{align}
\normsp{\mathbf{R}_m(t)}{g}
&\leq
L_{m,g}\int_0^t ds\, e^{\mu_{m, g}(t-s)} \normsp{Q_m\mathbf{O}_H(s)}{g}.
\label{eq:app-memory-bound}
\end{align}
Equation~\eqref{eq:app-memory-bound} proves Eq.~\eqref{eq:memory-error-main-bound}.
If $L_{m, g}=0$, the projected exact trajectory and the closed restricted trajectory coincide even when the discarded component is nonzero.
This observation also shows why operator entanglement cannot provide an independent lower bound on leakage.

For rank-deficient $\sigma_0$, the covariance form is a seminorm on operators and a norm only on equivalence classes modulo $\mathcal N_{\sigma_0}$.
Accordingly, $P_m$, $Q_m$, and the induced operator norms above are understood on the covariance-resolved quotient, exactly as in \cite{Perez_short_paper_2026} error analysis.

\subsection{OSEE obstruction to Hilbert--Schmidt representation}
\label{app:osee-proof}

Given a finite active region \(B=L\cup R\) and a bipartition \(L|R\), every observable supported on \(B\) admits an operator Schmidt decomposition
\begin{equation}
\mathbf{O} = \sum_{i=1}^{D_{\rm cut}}
s_i\, \mathbf{A}_i\otimes\mathbf{B}_i,
\qquad
s_1\geq s_2\geq\cdots\geq0,
\label{eq:app-operator-schmidt}
\end{equation}
where the operator families on \(L\) and \(R\) are HS orthonormal.
Define the normalized weights
\begin{equation}
p_i := \frac{s_i^2}{\|\mathbf{O}\|_{\mathrm{HS}}^2},
\qquad
S_\alpha(\mathbf{O};L|R) := \frac{1}{1-\alpha} \log\!\sum_i p_i^\alpha .
\label{eq:app-renyi-osee}
\end{equation}

For \(0\leq\Delta<1\), let
\begin{equation}
\chi_\Delta(\mathbf{O};L|R) :=
\min\!\left\{ r: \sum_{i>r}p_i\leq\Delta
\right\}
\label{eq:app-approximate-rank}
\end{equation}
be the smallest Schmidt rank whose optimal truncation has relative squared HS error at most \(\Delta\).
For every \(\alpha>1\), H\"older's inequality gives
\begin{align}
1-\Delta &\leq
\sum_{i=1}^{\chi_\Delta}p_i
\nonumber \leq 
\chi_\Delta^{(\alpha-1)/\alpha}
\left( \sum_i p_i^\alpha \right)^{1/\alpha},
\end{align}
and, therefore
\begin{equation}
\chi_\Delta(\mathbf{O};L|R) \geq
(1-\Delta)^{\alpha/(\alpha-1)}
e^{S_\alpha(\mathbf{O};L|R)}.
\label{eq:app-renyi-rank-lower}
\end{equation}
At \(\alpha=2\),
\begin{equation}
\chi_\Delta(\mathbf{O};L|R)
\geq
(1-\Delta)^2
e^{S_2(\mathbf{O};L|R)}.
\label{eq:app-renyi2-rank-lower}
\end{equation}
We have employed in this manuscript $S_2$ since the analogous von-Neumann OSEE does not give a dimension-independent finite-error bound without additional constraints depending on the cut Hilbert-space dimension.

Given this bound on the bond-dimension $\chi_\Delta$, we next bound the Schmidt-rank capacity of the at-most-\(m\)-body space.
Let $N_L$ and $N_R$ denote the numbers of active qubits on either side of the cut and define
\begin{equation}
D_X(k) = \sum_{a=0}^{\min(k,N_X)} \binom{N_X}{a}3^a,
\qquad
X\in\{L,R\},
\label{eq:app-one-side-dimension}
\end{equation}
for $k \geq 0$, with $D_X(k)=0$ for \(k<0\).
Thus $D_X(k)$ counts Pauli strings of  Pauli weight at most $k$ on side \(X\).

Let $\mathcal A_m(B)$ be the span of Pauli strings of total body order at most $m$ inside \(B\).
For $\mathbf X\in\mathcal A_m(B)$, expand ${\bf X}$ in left and right Pauli bases and regard its coefficients as a matrix \(C\).
A matrix element connecting a left Pauli string of weight $a$ to a right Pauli string of weight $b$ can be nonzero only when
\begin{equation}
a+b\leq m.
\end{equation}
For any integer $k\in\{-1,0,\ldots,m\}$, split
\begin{equation}
C=C_{\leq k}+C_{>k}
\end{equation}
according to the left body order.
The first contribution occupies at most $D_L(k)$ rows and hence has rank at most $D_L(k)$.
In the second contribution, $a\geq k+1$ forces $b\leq m-k-1$, so its nonzero entries occupy at most $D_R(m-k-1)$ columns.
Subadditivity of matrix rank therefore gives
\begin{equation}
\operatorname{rank}_{L|R}(\mathbf X) 
\leq
D_L(k) + D_R(m-k-1).
\end{equation}
Minimizing over \(k\) yields the cut-capacity bound
\begin{equation}
R_m^{L|R} := \min_{-1\leq k\leq m}
\left[ D_L(k) + D_R(m-k-1)
\right].
\label{eq:app-cut-capacity}
\end{equation}
Hence, every \(\mathbf X\in\mathcal A_m(B)\) satisfies
\begin{equation}
\operatorname{rank}_{L|R}(\mathbf X) \leq R_m^{L|R}.
\label{eq:app-cut-rank-bound}
\end{equation}
The quantity $R_m^{L|R}$ should therefore be understood as a cut-dependent upper bound on the Schmidt-rank capacity available to the $m$-body space, rather than as its global dimension.

Suppose now that some $\mathbf X\in\mathcal A_m(B)$ approximates ${\bf O}$ with
\begin{equation}
\frac{ 
\|\mathbf O-\mathbf X\|_{\mathrm{HS}}^2
}{
\|\mathbf O\|_{\mathrm{HS}}^2
} \leq \Delta .
\label{eq:app-hs-approximation}
\end{equation}
By the Eckart--Young theorem, an optimal Schmidt truncation of rank $\operatorname{rank}_{L|R}(\mathbf X)$ cannot have larger HS-error than ${\bf X}$
Therefore, combining the results yields the chain inequality
\begin{equation}
{
\chi_\Delta(\mathbf O;L|R) \leq 
\operatorname{rank}_{L|R}(\mathbf X)
\leq R_m^{L|R}}.
\label{eq:app-rank-sandwich}
\end{equation}
Combining Eqs.~\eqref{eq:app-renyi-rank-lower} and \eqref{eq:app-rank-sandwich} gives the necessary condition
\begin{equation}
(1-\Delta)^{\alpha/(\alpha-1)} e^{S_\alpha(\mathbf O;L|R)} \leq R_m^{L|R}.
\label{eq:app-osee-obstruction-general}
\end{equation}
For \(\alpha=2\),
\begin{equation}
(1-\Delta)^2 e^{S_2(\mathbf O;L|R)} \leq R_m^{L|R},
\label{eq:app-osee-obstruction}
\end{equation}
which proves Eq.~\eqref{eq:osee-necessary-main}.
Violation of this inequality across any cut is therefore sufficient to rule out an \(m\)-body HS approximation at relative squared error $\Delta$

The converse is not implied.
Equation~\eqref{eq:app-osee-obstruction} tests only whether the \(m\)-body space possesses enough Schmidt-rank capacity across the selected cut.
Passing this test does not guarantee that the exact observable lies close to \(\mathcal A_m\).
Furthermore, the result concerns instantaneous representation rather than dynamical leakage.
Indeed, Eq.~\eqref{eq:app-memory-bound} shows that discarded operator weight contributes to the memory error only through the return coupling \(P_m\genH Q_m\).
If this coupling vanishes, the restricted and projected exact trajectories coincide even when the discarded component and the OSEE are nonzero.
OSEE therefore does not provide an independent lower bound on the leakage or memory error.

For a balanced cut and fixed $m$, Eq.~\eqref{eq:app-cut-capacity} grows parametrically more slowly than the global dimension of ${\cal A}_m$.
Writing \(q=\lfloor m/2\rfloor\), one has
\begin{equation}
R_m^{L|R} \leq D_L(q)+D_R(q),
\end{equation}
so that for fixed \(m\) and increasing active volume the cut capacity scales as
\begin{equation}
R_m^{L|R} = \mathcal O\!\left( N(t)^{\lfloor m/2\rfloor} \right).
\label{eq:app-cut-capacity-scaling}
\end{equation}
For the small body orders used in the numerical calculations, the exact discrete expression in Eq.~\eqref{eq:app-cut-capacity} is the appropriate form of the obstruction.

\begin{remark}
In one dimension, an MPO bond dimension bounds the operator Schmidt rank across each bond.
In higher dimensions, a PEPO with local bond dimension $D$ and $b$ virtual bonds crossing a cut has Schmidt rank at most \(D^b\).
Equation~\eqref{eq:app-osee-obstruction} therefore places tensor-network and $m$-body Pauli representations under the same cut-wise expressivity test.
It does not equate their parameter counts, computational costs, or approximation errors.
\end{remark}

\subsection{Conditional body-order bounds from Pauli-\(\ell_1\) magic and a controlled tail}
\label{app:tail-proof}

Magic and body order characterize different aspects of the observable.
For a normalized Pauli expansion
\begin{equation}
\mathbf O_H(t) = \sum_P c_P(t)P,
\end{equation}
we use the logarithmic Pauli-$\ell_1$ complexity
\begin{equation}
\mathcal M_1(t) = \log\!\left( \sum_P |c_P(t)| \right).
\label{eq:app-pauli-l1-magic}
\end{equation}
As established in the main text, this natural-logarithmic quantity differs by a constant factor from the base-$2$ operator stabilizer R\'enyi entropy of Ref.~\cite{dowling2025_magic}.
The global value of $\mathcal M_1$ does not determine how operator weight is distributed over body order.
A sufficient body-order statement, therefore, requires an independent hypothesis on the high-body tail.

Let \(g\) be a fixed positive-definite Hilbert--Schmidt geometry or a covariance geometry for which the relevant quotient construction is well defined.
Let \(\mathsf P_r^{(g)}\) denote the nested \(g\)-orthogonal projection onto the at-most-\(r\)-body space and define
\begin{equation}
\Pi_0^{(g)} = \mathsf P_0^{(g)},
\qquad
\Pi_r^{(g)} = \mathsf P_r^{(g)} - \mathsf P_{r-1}^{(g)}
\quad
(r\geq1).
\label{eq:app-sector-projectors}
\end{equation}
The squared norm carried by body-order sector \(r\) is
\begin{equation}
\Theta_r^{(g)}(t) = \left\| \Pi_r^{(g)} \mathbf O_H(t) \right\|_g^2.
\label{eq:app-sector-weight}
\end{equation}
Orthogonality of the nested sectors gives the exact identity
\begin{equation}
\left\|
(1-\mathsf P_m^{(g)})
\mathbf O_H(t)
\right\|_g^2 = \sum_{r>m}
\Theta_r^{(g)}(t).
\label{eq:app-representation-tail-identity}
\end{equation}
This construction does not require individual Pauli strings of equal body order to be mutually orthogonal in covariance geometry.

Assume that throughout \(0\leq s\leq t\),
\begin{equation}
\Theta_r^{(g)}(s)
\leq
C_g e^{2\mathcal M_1(s)} e^{-2\kappa_g r}
\label{eq:app-tail-assumption}
\end{equation}
for constants \(C_g>0\) and \(\kappa_g>0\).
The factor \(e^{-2\kappa_g r}\) is an independent concentration assumption on body order.
It is not implied by locality, by the global Pauli-\(\ell_1\) magic, or by OSEE.

Using Eq.~\eqref{eq:app-representation-tail-identity} and summing the geometric series gives
\begin{align}
\left\| (1-\mathsf P_m^{(g)}) \mathbf O_H(t)  \right\|_g^2
&\leq
C_g e^{2\mathcal M_1(t)} \sum_{r=m+1}^{\infty} e^{-2\kappa_g r}\\
&= \frac{C_g}{ 1-e^{-2\kappa_g} } e^{2\mathcal M_1(t)} e^{-2\kappa_g(m+1)}.
\label{eq:app-representation-geometric}
\end{align}
Defining
\begin{equation}
A_{\kappa,g} = \sqrt{ \frac{C_g}{ 1-e^{-2\kappa_g} } },
\label{eq:app-akg}
\end{equation}
we obtain
\begin{equation}
\left\| (1-\mathsf P_m^{(g)}) \mathbf O_H(t) \right\|_g
\leq A_{\kappa,g} e^{\mathcal M_1(t)} e^{-\kappa_g(m+1)}.
\label{eq:app-representation-final}
\end{equation}
This proves Eq.~\eqref{eq:representation-upper-main}.

When \(g=\mathrm{covar}(\sigma_0)\) and the covariance projection is well defined on the corresponding normed space,
\begin{equation}
\mathsf P_m^{(g)} = \pi_m^{\sigma_0},
\end{equation}
so Eq.~\eqref{eq:app-representation-final} directly gives
\begin{equation}
\Delta_{\mathrm{rep}}^{(m)}(t)
\leq A_{\kappa,g} e^{\mathcal M_1(t)} e^{-\kappa_g(m+1)}.
\label{eq:app-covariance-representation-final}
\end{equation}
Consequently, any body order for which the right-hand side is below a prescribed representation tolerance is certified to be sufficient under the assumed tail condition.
No claim is made that this condition is necessary.

The memory contribution requires an additional dynamical ingredient.
Recall the exact comparison
\begin{equation}
E_{\mathrm{mem}, g}^{(m)}(t) \leq
L_{m,g} \int_0^t ds\, e^{\mu_{m,g}(t-s)}
\left\| Q_m\mathbf O_H(s) \right\|_g,
\label{eq:app-memory-bound-recalled}
\end{equation}
where $P_m = \pi_m^{\sigma_0}$ and $Q_m = 1-P_m$ and
\begin{equation}
L_{m, g} = \left\| P_m\genH Q_m \right\|_{{\rm op},g},
\qquad
\mu_{m, g} = \left\| P_m\genH P_m \right\|_{{\rm op},g}.
\end{equation}
In compatible covariance geometry, $Q_m=1-\mathsf P_m^{(g)}$ and Eq.~\eqref{eq:app-representation-final} therefore controls the integrand directly.
Defining 
\begin{equation}
\mathcal M_1^\star(t) = \sup_{0\leq s\leq t} \mathcal M_1(s),
\label{eq:app-mstar}
\end{equation}
we obtain
\begin{align}
E_{\mathrm{mem},g}^{(m)}(t)
&\leq
A_{\kappa,g} L_{m,g} e^{-\kappa_g(m+1)}
\int_0^t ds\,
e^{\mu_{m,g}(t-s)+\mathcal M_1(s)}
\\
&\leq
A_{\kappa,g} L_{m,g}\, t\,
\exp\!\left[
\mathcal M_1^\star(t)
+\mu_{m,g} t
-\kappa_g(m+1)
\right].
\label{eq:app-memory-tail-final}
\end{align}
This proves Eq.~\eqref{eq:memory-sufficient-main} whenever the covariance geometry is compatible with the projected dynamics.
For this estimate to provide nontrivial scaling information, however, the constants controlling the body-order tail must themselves remain suitably bounded as the system size and retained body order increase.

For full-rank reference states, the covariance form is positive definite and the induced operator norms are well defined.
For rank-deficient states, both the retained generator and the return map must descend to the covariance quotient.
In particular, we require
\begin{equation}
P_m\genH\mathcal N_m^{\sigma_0}\subseteq\mathcal N_m^{\sigma_0},
\qquad
P_m\genH Q_m\mathcal N^{\sigma_0}\subseteq\mathcal N_m^{\sigma_0}.
\end{equation}
If either condition fails, the covariance-quotient memory bound is not claimed.
The static orthogonal error decomposition remains valid without these dynamical assumptions.
The Duhamel and induced-operator estimates in covariance geometry are, therefore, valid only when the maps entering the projected dynamics preserve covariance-null equivalence classes and descend consistently to that quotient.
If this dynamical compatibility fails, the covariance-quotient memory estimate in Eq.~\eqref{eq:app-memory-tail-final} is not claimed.
The Duhamel comparison itself may instead be formulated in a positive-definite state-independent norm, but in that case the \(g\)-orthogonal body-sector tail in Eq.~\eqref{eq:app-representation-final} is not automatically identical to $\|Q_m\mathbf O_H\|_g$.
A corresponding memory certificate therefore requires separate control of that discarded component.

Finally, we formulate a sufficient condition for a prescribed total covariance error $\delta>0$.
Let $\eta_{\mathrm{rep}}$, $\eta_{\mathrm{mem}}$, and $\eta_{\mathrm{HB}}$ denote nonnegative tolerances assigned to the representation, memory, and finite-HB errors, respectively.
Suppose that, at the time of interest, the corresponding error contributions satisfy
\begin{align}
\Delta_{\mathrm{rep}}^{(m)}(t)
&\leq \eta_{\mathrm{rep}},
\\
E_{\mathrm{mem}}^{(m)}(t)
&\leq \eta_{\mathrm{mem}},
\\
E_{\mathrm{HB}}^{(\ell,m,\varepsilon,p)}(t)
&\leq \eta_{\mathrm{HB}}.
\end{align}
The orthogonal error decomposition and the triangle inequality then imply
\begin{equation}
\Delta_{\ell,m,\varepsilon,p}(t) \leq
\sqrt{
\eta_{\mathrm{rep}}^2
+
\left(
\eta_{\mathrm{mem}}+\eta_{\mathrm{HB}}
\right)^2
}.
\end{equation}
Consequently, the sufficient condition
\begin{equation}
\eta_{\mathrm{rep}}
+
\eta_{\mathrm{mem}}
+
\eta_{\mathrm{HB}}
\leq
\delta
\label{eq:app-error-budget}
\end{equation}
guarantees that the total covariance error does not exceed $\delta$.

The OSEE and magic results therefore have complementary but different logical roles.
Equation~\eqref{eq:app-osee-obstruction} is a necessary, state-independent obstruction to low-body Hilbert--Schmidt representation.
Equation~\eqref{eq:app-representation-final} is a conditional sufficient estimate obtained only after imposing independent control of the body-order tail.
When the tail hypothesis is imposed in Hilbert--Schmidt geometry, the two results constrain the same global representation problem from opposite directions.
When it is imposed in covariance geometry, the sufficient estimate is instead state adapted, while the OSEE obstruction remains a distinct Hilbert--Schmidt statement.
Neither OSEE nor Pauli-\(\ell_1\) magic alone determines the body order required for accurate state-adapted dynamics.

\section{Symbolic precomputation and \textsc{Phoenix} backend generation}
\label{app:phoenix-precomputation}

Pauli-propagation methods repeatedly manipulate strings
$P\in\{I,X,Y,Z\}^{\otimes N}$, which are also called Pauli words in parts of the literature.
We use the term Pauli string throughout.
The multiplication of two individual strings is elementary.
The computational difficulty arises from the number of retained strings and of nonvanishing commutator couplings that can occur between them and a many-body Hamiltonian.

For an $N$-site system, define
\begin{equation}
D_m(N)=\dim\mathcal{A}_m=\sum_{k=0}^{m}3^k\binom{N}{k}.
\label{eq:phoenix-lm}
\end{equation}
For fixed $m$, $D_m(N)=\mathcal{O}(N^m)$, while the dependence on the retained body order remains combinatorial.
\Cref{tab:m-body-pauli-storage} gives representative dimensions and baseline storage requirements.
The estimates exclude indexing metadata, temporary buffers, generated instructions, and compiler overhead.
Their purpose is only to illustrate the scale of the retained space and the advantage of separating symbolic labels from numerical coefficients.

\begin{table*}[t]
\centering
\scriptsize
\caption{Dimension of the at-most-$m$-body Pauli-string space and baseline storage requirements.
Decimal units are used, with $1\,\mathrm{kB}=10^3$ bytes, $1\,\mathrm{MB}=10^6$ bytes, and $1\,\mathrm{GB}=10^9$ bytes.
The full-string estimate assumes one $N$-character ASCII string and one newline per basis element.
The real and complex columns correspond to one dense double-precision coefficient vector and exclude indexing metadata.}
\label{tab:m-body-pauli-storage}
\begin{tabular}{rrrlll}
\hline
$N$ & $m$ & $D_m(N)$ & Full-string catalogue & Real vector & Complex vector \\
\hline
10  & 1 & 31          & $341\,\mathrm{B}$    & $248\,\mathrm{B}$    & $496\,\mathrm{B}$ \\
10  & 2 & 436         & $4.80\,\mathrm{kB}$  & $3.49\,\mathrm{kB}$  & $6.98\,\mathrm{kB}$ \\
10  & 3 & 3,676       & $40.4\,\mathrm{kB}$  & $29.4\,\mathrm{kB}$  & $58.8\,\mathrm{kB}$ \\
10  & 4 & 20,686      & $228\,\mathrm{kB}$   & $165\,\mathrm{kB}$   & $331\,\mathrm{kB}$ \\
\hline
18  & 1 & 55          & $1.04\,\mathrm{kB}$  & $440\,\mathrm{B}$    & $880\,\mathrm{B}$ \\
18  & 2 & 1,432       & $27.2\,\mathrm{kB}$  & $11.5\,\mathrm{kB}$  & $22.9\,\mathrm{kB}$ \\
18  & 3 & 23,464      & $446\,\mathrm{kB}$   & $188\,\mathrm{kB}$   & $375\,\mathrm{kB}$ \\
18  & 4 & 271,324     & $5.16\,\mathrm{MB}$  & $2.17\,\mathrm{MB}$  & $4.34\,\mathrm{MB}$ \\
\hline
50  & 1 & 151         & $7.70\,\mathrm{kB}$  & $1.21\,\mathrm{kB}$  & $2.42\,\mathrm{kB}$ \\
50  & 2 & 11,176      & $570\,\mathrm{kB}$   & $89.4\,\mathrm{kB}$  & $179\,\mathrm{kB}$ \\
50  & 3 & 540,376     & $27.6\,\mathrm{MB}$  & $4.32\,\mathrm{MB}$  & $8.65\,\mathrm{MB}$ \\
50  & 4 & 19,194,676  & $979\,\mathrm{MB}$   & $154\,\mathrm{MB}$   & $307\,\mathrm{MB}$ \\
\hline
100 & 1 & 301         & $30.4\,\mathrm{kB}$  & $2.41\,\mathrm{kB}$  & $4.82\,\mathrm{kB}$ \\
100 & 2 & 44,851      & $4.53\,\mathrm{MB}$  & $359\,\mathrm{kB}$   & $718\,\mathrm{kB}$ \\
100 & 3 & 4,410,751   & $445\,\mathrm{MB}$   & $35.3\,\mathrm{MB}$  & $70.6\,\mathrm{MB}$ \\
100 & 4 & 322,029,976 & $32.5\,\mathrm{GB}$  & $2.58\,\mathrm{GB}$  & $5.15\,\mathrm{GB}$ \\
\hline
\end{tabular}
\end{table*}

\subsection{Representation and locality reduction}
\label{app:phoenix-representation}

A Pauli string is represented by its ordered support and the corresponding sequence of nonidentity local labels.
For example, $X_1Y_4Z_7$ is represented schematically by $((1,4,7),XYZ)$.
Identity factors are not stored explicitly.
A hierarchical \textsc{KeyMap} assigns each support and reduced local string a numerical offset, while fixed-layout Aligned Data Access Arrays (ADAAs) coefficient containers store the associated real or complex amplitudes.
Supports of the same cardinality can share the same reduced local-label structure, avoiding repetition of equivalent metadata.
For Hermitian Hamiltonians and observables, $i[\h,\cdot]$ preserves Hermiticity, and a real-valued coefficient backend is therefore sufficient for the calculations considered here.

A generic dense bilinear multiplication table on the retained space would contain $D_m(N)^2$ entries and would therefore scale as $\mathcal{O}(N^{2m})$ at fixed $m$.
\textsc{Phoenix} does not construct such a table.
For the local Hamiltonians considered here, only a much smaller subset of system--Hamiltonian support pairs can contribute, and overlap locality further restricts this set.

Suppose, for illustration, that the Hamiltonian contains onsite terms on $N$ sites and two-site terms on the $E$ edges of an interaction graph.
The number of support pairs between a $k$-site system string and a Hamiltonian domain that have nonempty overlap is bounded by
\begin{equation}
C_k(N,E)=N\binom{N-1}{k-1}
+E\left[\binom{N}{k}-\binom{N-2}{k}\right].
\label{eq:phoenix-overlap-count}
\end{equation}
This count still includes Pauli-label assignments whose commutator vanishes and is therefore an upper bound on candidate support embeddings rather than an arithmetic operation count.
For finite-range lattices with bounded coordination, $E=\mathcal{O}(N)$ and
\begin{equation}
\sum_{k=1}^{m}C_k(N,E)=\mathcal{O}(N^m)
\end{equation}
at fixed $m$.
Thus the geometrically admissible coupling structure remains polynomial in system size at fixed retained body order, rather than requiring a dense bilinear table on $\mathcal A_m$.

\subsection{Local exploration and lifting}
\label{app:phoenix-exploration}

The symbolic preprocessing used in this work begins on a finite exploration
domain chosen to contain the local overlap motifs required by the target
lattice geometry, retained body order, Hamiltonian body order, and interaction
range.
Its precise shape is therefore a geometrical input and is not a universal
function of $m$ alone.
The exploration domains used in the calculations reported here were chosen
and validated for their corresponding target geometries.

Our preprocessing layer enumerates retained system supports and allowed
Hamiltonian supports, applies the local Pauli multiplication rules, removes
vanishing commutators, simplifies identity factors, and identifies surviving
operations according to the relative incidence structure of the input,
Hamiltonian, and output supports.
The resulting sparse instruction structure is then represented using
\textsc{Phoenix} KeyMaps and instruction collections.

\subsection{Compiled kernels and buffered commutators}
\label{app:phoenix-compiled-kernels}

The resulting instruction families are emitted as compiled Fortran routines and exposed to Python through an \textsc{f2py} interface.
Inside a compiled commutator evaluation, the kernel executes a predetermined sequence of numerical operations on the coefficient containers.
String parsing, symbolic Pauli multiplication, support simplification, and dictionary-based symbolic dispatch are therefore absent from the compiled commutator kernel.
The symbolic construction cost is paid during precomputation and amortized over the repeated commutator evaluations required by the adaptive dynamics.

The unprojected commutator of an input in $\mathcal{A}_m$ need not itself belong to $\mathcal{A}_m$.
For a Hamiltonian with maximum body order $h_{\max}$, a nonvanishing commutator has body order at most
\begin{equation}
m_{\mathrm{int}}
=
\min\{N,m+h_{\max}-1\}.
\label{eq:phoenix-buffer-order}
\end{equation}
The subtraction by one follows from the nonempty support overlap required for a nonzero commutator.
For the two-body Hamiltonians considered here,
$m_{\mathrm{int}}=m+1$.

The compiled commutator therefore maps a retained input into a temporary enlarged representation capable of containing the complete intermediate commutator up to $m_{\mathrm{int}}$.
Hilbert--Schmidt (HS) truncation discards the components above body order $m$.
\textsc{AdaptH} instead applies the state-dependent projection
$\pi_m^{\sigma_0}$ to the intermediate result and accumulates the projected operator in the retained space.
The enlarged representation is temporary: it is not incorporated into the adaptive HB itself and is not recursively propagated as an untruncated nested-commutator hierarchy.

After symbolic precomputation, the online \textsc{AdaptH} stage therefore consists principally of compiled commutator evaluations, state-dependent projections, scalar products, and low-dimensional linear algebra within each adaptive chart.
The same compiled commutator backend can be reused for the successive HB vectors and for later chart reconstructions.
This separates the local Pauli algebra and lattice-embedding analysis performed during symbolic preprocessing from the sparse instruction representation and backend generation provided by \textsc{Phoenix}, and from the state-dependent compression and adaptive reduced dynamics implemented in \textsc{AdaptH}.

\section{Numerical protocols for the PRA benchmarks}
\label{app:benchmark-protocols}

\subsection{Campaigns, exact references, and output grids}
\label{app:exact-protocol}

All numerical benchmarks were performed using CPU computation, with no GPU or CUDA acceleration used, and all numerical results in the present article use the nonintegrable nearest-neighbor TFIM defined in Eq.~\eqref{eq:tfim-hamiltonian}, with open boundary conditions and
\begin{equation}
J=1,
\qquad h_X J=0.75, \qquad h_Z J=0.15.
\end{equation}
The reference and initial states are the homogeneous product states
\begin{equation}
\sigma_0(r,\theta) = \bigotimes_{i=1}^{N}
\frac{1}{2} \left( I+r\cos\theta\,X_i+r\sin\theta\,Z_i \right),
\end{equation}
as defined in Eq.~\eqref{eq:tfim-initial-state}.
Within each campaign, PP, and TN trajectories are exported on the same uniform time grid as well as the exact dynamics, if available for the system size in question, so every reported pointwise comparison is evaluated without temporal interpolation.
All three campaigns use the common output spacing
\begin{equation}
\Delta t\,J=0.01.
\end{equation}

For the purity sweep, we consider the open $N=10$ chain and the bulk observable $Z_5$, using zero-based site indexing.
We fix $\theta=\pi/2$ and vary
\begin{equation}
r\in
\left\{
0.10,\,
0.25,\,
0.50,\,
0.75,\,
0.90,\,
0.99,\,
1.00
\right\}.
\end{equation}
The production trajectories extend to $T_{\max}J=3$, while the main-text diagnostics are evaluated over $tJ\in[0,1]$.
At later times, the finite-body PP approximations, both covariance and HS, progressively lose predictive accuracy, with absolute errors eventually becoming $O(1)$ across the investigated truncation schemes.
Extending the time-averaged diagnostics to these longer times would therefore obscure the state-dependent differences between covariance and HS propagation, which are most clearly resolved within the shorter interval.
The longer trajectories are, nevertheless, reported in the Appendix and the datasets available in \cite{perez2026_pra_zenodo} to document the eventual breakdown of the fixed-$m$ low-body approximations.

For the orientation sweep, we use an open $3\times3\times2$ lattice with $N=18$ and monitor $Z_8$.
With the C-order lattice convention used in the simulations, site $8$ corresponds to coordinate $(1,1,0)$ and is one of the two symmetry-related central sites of the lattice.
We fix $r=1$ and use
\begin{equation}
\theta\in
\left\{
\pi/6,\,
\pi/4,\,
\pi/3,\,
\pi/2
\right\},
\end{equation}
with
\begin{equation}
T_{\max}J=1.5.
\end{equation}
Every state in this sweep is pure, so the exact reference is obtained by matrix-free Schr\"odinger-picture state-vector propagation.
For $N=18$, the state vector has dimension $2^{18}=262\,144$, whereas a generic vectorized Heisenberg operator would have dimension $4^{18}=68\,719\,476\,736$.
Neither a dense Hamiltonian matrix, a full spectral decomposition, nor a vectorized Heisenberg operator is constructed in the production exact calculation.

For the large-system benchmark, we use the open $5\times4\times4$ lattice with $N=80$ and the bulk observable $Z_{37}$.
With the C-order lattice convention, site $37$ corresponds to coordinate $(2,1,1)$ and has six nearest neighbors.
The state is fully $Z$ polarized, corresponding to $r=1$ and $\theta=\pi/2$, and the trajectories are propagated to
\begin{equation}
T_{\max}J=0.6.
\end{equation}
No exact full-Hilbert-space reference is used at $N=80$.
Reliability is instead assessed through successive convergence within the independent PP and TN hierarchies, as discussed in Sec.~\ref{sec:results}.

The exact propagators used for the $N=10$ and $N=18$ benchmarks should be distinguished from the QuTiP Krylov implementation described in Ref.~\cite{qutip5}, which is not the production backend used for the results reported here.
Here, ``exact'' denotes a numerically converged full-Hilbert-space reference rather than an analytic solution.

\subsection{Restricted PP simulation campaigns}
\label{app:PP-protocol}

Separate adaptive calculations are performed for $m=1,2,3$ using either HS or covariance projection.
At fixed $m$, the retained Pauli space is identical for both calculations.
Within each campaign, the physical Hamiltonian, initial state, adaptive reconstruction parameters, time interval, and numerical output grid are also matched.
The distinction is the geometry used to define the projection and reduced scalar products.

For covariance calculations, projection, Gram-matrix construction, and expectation values use the corresponding product state $\sigma_0(r,\theta)$ of the campaign.
For HS calculations, projection and reduced scalar products use the state-independent HS geometry, while expectation values are evaluated in the same physical initial state $\sigma_0(r,\theta)$.
Thus the HS and covariance trajectories at fixed $(N,r,\theta,m)$ differ through their projection geometry rather than through the Hamiltonian, observable, or physical state in which the final expectation value is measured.

The $N=10$ points with $r<1$ use full-rank covariance geometry.
The $N=10$ endpoint $r=1$, the complete $N=18$ orientation sweep, and the $N=80$ benchmark use pure product states, for which the covariance form is only positive semidefinite and the associated Gram matrix may contain exact null directions.
The formal reduced equations may be written on the resolved quotient space using the Moore--Penrose inverse, as discussed in Ref.~\cite{Perez_short_paper_2026}.
Within the $n$th frozen chart,
\begin{equation}
\mathcal G^{(n)}
\dot{\vec\phi}^{(n)}(t) =
\mathcal H^{(n)}
\vec\phi^{(n)}(t),
\label{eq:benchmark-reduced-generator}
\end{equation}
with
\begin{equation}
\mathcal G^{(n)}_{ij}
=
\bigl(b_i^{(n)},b_j^{(n)}\bigr),
\qquad
\mathcal H^{(n)}_{ij}
=
\bigl(b_i^{(n)},\mathcal L_m b_j^{(n)}\bigr),
\end{equation}
where the scalar product is the covariance or HS geometry appropriate to the calculation.

The production calculations do not explicitly construct the Moore--Penrose inverse of a rank-deficient $\mathcal G^{(n)}$.
Instead, each finite-dimensional Gram problem is regularized by the positive diagonal shift
\begin{equation}
\mathcal G_r^{(n)} = \mathcal G^{(n)}+r_n I,
\end{equation}
with
\begin{equation}
r_n = \varepsilon_G \max_j \left| \lambda_j\!\left(\mathcal G^{(n)}\right) \right|,
\qquad
\varepsilon_G=10^{-12}.
\label{eq:benchmark-gram-ridge}
\end{equation}
The resulting nonsingular system is solved by LU factorization, giving the numerical coordinate-space generator
\begin{equation}
\mathcal A_r^{(n)} = \left(\mathcal G_r^{(n)}\right)^{-1}
\mathcal H^{(n)}.
\end{equation}
The same Gram-solve prescription is used for the HS trajectories in order to treat ordinary numerical linear dependencies on the same footing.

At finite $r_n$, the shifted solve differs from the Moore--Penrose prescription.
For exactly evaluated covariance matrices, however, the columns of $\mathcal H^{(n)}$ belong to the range of $\mathcal G^{(n)}$.
Consequently,
\begin{equation}
\lim_{r_n\to0^+}
\left(\mathcal G^{(n)}+r_nI\right)^{-1}\mathcal H^{(n)}
=
\left(\mathcal G^{(n)}\right)^+\mathcal H^{(n)}.
\end{equation}
Finite regularization changes the response along resolved directions, while roundoff or independently thresholded contractions can introduce inconsistent components that are amplified by the shifted solve.
This numerical issue is distinct from the dynamical activation of an operator direction that is covariance null.

All PP campaigns use hierarchy depth $\ell=5$, meaning five recursive applications beyond the seed.
Consequently each frozen HB chart contains
\begin{equation}
b_0^{(n)},b_1^{(n)},\ldots,b_5^{(n)},
\end{equation}
and therefore six stored coordinate directions in total.
The additional action $\mathcal L_m b_5^{(n)}$ needed to construct the final column of $\mathcal H^{(n)}$ is evaluated as an auxiliary vector and is not stored as an additional HB basis element.

The final $p=2$ HB coordinates enter the Euclidean tail indicator
\begin{equation}
w_{\mathrm{tail}}(t) = \frac{
\displaystyle\sum_{j=d-p}^{d-1}
|\phi_j(t)|^2
}{
\displaystyle\sum_{j=0}^{d-1}
|\phi_j(t)|^2 }, \qquad d=6,
\end{equation}
and the current chart is reconstructed once
\begin{equation}
w_{\mathrm{tail}}(t)\geq
\varepsilon_{\mathrm{tail}},
\qquad
\varepsilon_{\mathrm{tail}}=10^{-8} {\rm \,\, or \, } \varepsilon_{\mathrm{tail}}=10^{-9}.
\end{equation}

The adaptive reconstruction threshold depends on the numerical campaign:
\begin{equation}
\varepsilon_{\mathrm{tail}}=
\begin{cases}
10^{-10}, & N=10,\\
10^{-8}, & N=18, \, N=80\text{ (App.)},\\
10^{-9}, & N=80\text{ (main text)},\\
\end{cases}
\end{equation}
All calculations use the same reconstruction rule, with $\ell=5$ and $p=2$.

The current approximate observable then seeds a new HB chart, for which the basis, Gram matrix, projected-generator matrix, and coordinate-space generator are recomputed.

Within a frozen chart the reduced generator is time independent.
The coordinates are therefore propagated as
\begin{equation}
\vec\phi^{(n)}(t) = \exp\!\left[
(t-T_n)\mathcal A_r^{(n)}
\right]
\vec\phi^{(n)}(T_n),
\label{eq:benchmark-coordinate-propagation}
\end{equation}
implemented through the matrix exponential in the reduced coordinate space.
No additional low-order time-integration approximation is introduced within a frozen chart.
No further modification of the projected reduced generator is applied in the production calculations.

The scalar-product and projection threshold is
\begin{equation}
\varepsilon_{\mathrm{sp}}=10^{-10}.
\end{equation}
Pauli coefficients with magnitude below
\begin{equation}
\varepsilon_{\mathrm{count}} = \epsilon_{\rm nnz} = 10^{-12}
\end{equation}
are excluded only from support-size and sparsity diagnostics and do not define the propagated body-order space.
Expectation-value contributions are retained down to
\begin{equation}
\varepsilon_{\mathrm{expect}}=10^{-15}.
\end{equation}
These numerical thresholds do not change the physical restriction to $\mathcal A_m$, although they control numerical sparsification and the treatment of poorly conditioned coordinate directions within that retained space.

As a sensitivity check for the finite Gram regularization used for rank-deficient pure reference states, tightening the Gram regularization from $10^{-12}$ to $10^{-13}$ changes the trajectories by at most $8.8\times10^{-6}$ for $m=2$ and $6.6\times10^{-6}$ for $m=3$, without changing the number of adaptive reconstructions.
The resulting dependence is small in absolute terms, although it can be comparable to the smallest reported $m=3$ trajectory errors.

\subsection{Heisenberg-picture tensor-network protocol}
\label{app:tensor-network-protocol}

The Heisenberg operator is represented as an MPO and, after fusing its two local physical legs, is evolved using an MPS representation with local physical dimension four and maximum virtual bond dimension $\chi$.
The $N=10$ problem is already a one-dimensional chain.
For the $N=18$ and $N=80$ lattices, the physical three-dimensional lattice is mapped to a one-dimensional chain using a fixed snake ordering.
All physical nearest-neighbor interactions are retained, although interactions that are local on the three-dimensional lattice can become long ranged in the one-dimensional ordering.

Left and right multiplication define the Liouvillian MPO
\begin{equation}
\mathsf{K}
=
\mathsf{R}_{\h}
-
\mathsf{L}_{\h},
\label{eq:tn-liouvillian}
\end{equation}
where
\begin{equation}
\mathsf{L}_{\h}|O\rangle\!\rangle
=
|\h O\rangle\!\rangle,
\qquad
\mathsf{R}_{\h}|O\rangle\!\rangle
=
|O\h\rangle\!\rangle.
\label{eq:tn-left-right}
\end{equation}
For the local Pauli observable $Z_j$ used in each campaign,
\begin{equation}
i\frac{d}{dt}|Z_j(t)\rangle\!\rangle
=
\mathsf{K}|Z_j(t)\rangle\!\rangle,
\label{eq:tn-vectorized-eom}
\end{equation}
which is equivalent to $\dot Z_j(t)=i[\h,Z_j(t)]$ with the convention of Eq.~\eqref{eq:heisenberg-eom}.

The vectorized operator is evolved in \textsc{TeNPy} using two-site TDVP~\cite{Haegeman_2016}.
The maximum bond dimensions used in the three PRA campaigns are
\begin{align}
N=10:&\qquad \chi_{\max}=16,32,64,128,\\
N=18:&\qquad \chi_{\max}=32,64,128,\\
N=80:&\qquad \chi_{\max}=2,8,32,64.
\end{align}
All production TN calculations use the common singular-value threshold
\begin{equation}
\varepsilon_{\mathrm{SVD}}=10^{-10}.
\end{equation}
The quoted $\chi_{\max}$ is an upper bound, and bonds are not artificially enlarged when their numerical rank is smaller.
Because $h_Z\neq0$ breaks the corresponding global spin-flip parity, that symmetry sector is not enforced in the TN calculations.

The initial local Pauli operator has bond dimension one.
Its first time derivative is generated by the onsite transverse field because the $ZZ$ interaction and the longitudinal $Z$ field commute with the initial $Z_j$ operator.
Interaction-generated spatial spreading therefore enters through subsequent nested Liouvillian actions rather than through the first derivative itself.
For the mapped three-dimensional lattices, this growth can couple sites separated by long distances in the one-dimensional ordering and thereby increase the virtual rank required across the intervening MPS cuts.

A finite-$\chi$ two-site TDVP evolution can consequently suffer from insufficiently populated virtual spaces at early times.
Increasing $\chi_{\max}$ raises the number of singular directions that may subsequently be retained, but it does not by itself guarantee that all dynamically relevant derivative-informed directions are proposed to the bond bases at the stage when they first become important.
We therefore use the global Krylov basis enrichment of Yang and White~\cite{Yang_2020}, as implemented in \textsc{TeNPy}, with
\begin{equation}
d_{\mathrm{Krylov}}^{\mathrm{TN}}=1.
\label{eq:tn-enrichment-order}
\end{equation}
One additional Liouvillian-generated direction is supplied during the enrichment step.
This parameter does not restrict the subsequent trajectory to a one-dimensional Krylov subspace.
Rather, the enrichment augments the variational bond bases with derivative-informed content, after which the two-site TDVP update and SVD truncation determine which components survive under the specified $\chi_{\max}$ and $\varepsilon_{\mathrm{SVD}}$.

\section{Appendix results}

\subsection{Complementary analysis of the $r$ sweep}
\label{subsec:SI_r_sweep}

\begin{figure}
    \centering
    \includegraphics[width=1\linewidth]{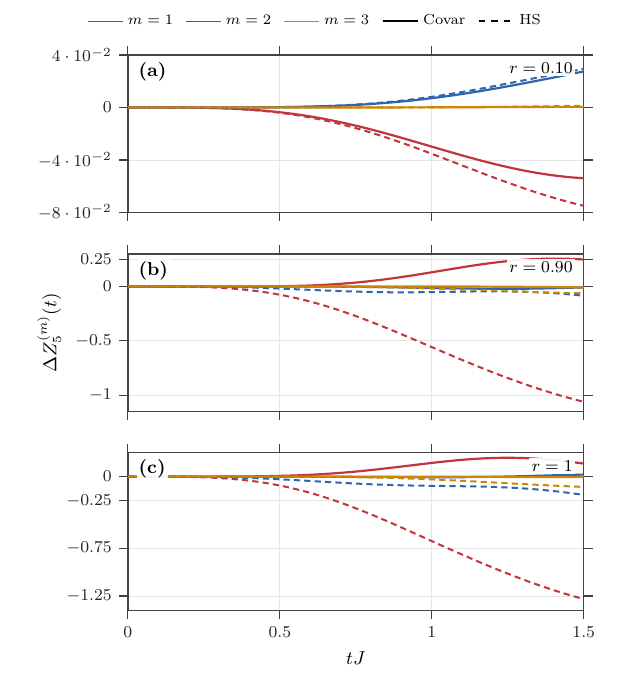}
    \caption{
\textbf{Time-resolved signed errors for the $N=10$ purity sweep.}
Signed error $\Delta Z_5^{(m)}(t)=\langle Z_5(t)\rangle_m-\langle Z_5(t)\rangle_{\mathrm{exact}}$ for (a) $r=0.1$, (b) $r=0.9$, and (c) $r=1$.
Parameters are as in Fig.~\ref{fig:r-sweep}, with trajectories shown up to $tJ=1.5$.
Colors denote $m=1,2,3$; solid and dashed curves distinguish covariance and HS propagation.
The dotted line marks zero error.
}
\label{fig:r_sweep_raw_SI}
\end{figure}

Figure~\ref{fig:r_sweep_raw_SI} shows the time-resolved signed errors for representative polarizations of the $N=10$ chain.
At $r=0.1$, covariance and HS propagation give similar deviations, which decrease rapidly with increasing body order.
At $r=0.9$, the geometries separate substantially, particularly at $m=1$, where HS develops a large negative error while covariance remains closer to the exact trajectory.
The late-time separation at $m=1$ becomes even larger at $r=1$, although the higher-body errors remain much smaller for both geometries.
These results complement the time-averaged diagnostics in Fig.~\ref{fig:r-sweep} by showing how the errors accumulate and change sign during propagation.
The larger late-time separation at $m=1$ does not imply a monotonic increase of the covariance advantage with $r$ at every body order.

\subsection{Complementary analysis of the $\theta$ sweep}
\label{subsec:SI_theta_sweep}

\begin{figure}
    \centering
    \includegraphics[width=1\linewidth]{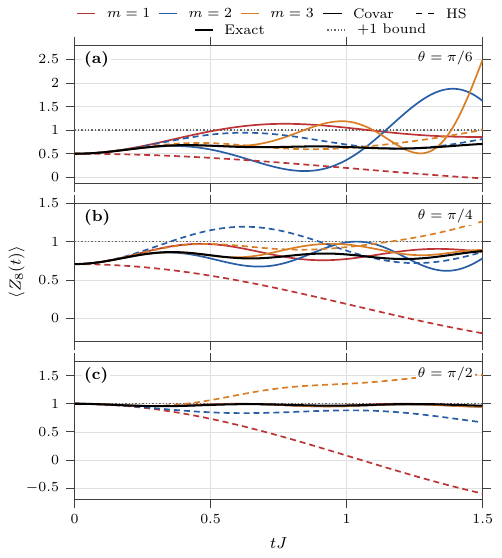}
    \caption{
\textbf{Time-resolved polarization for the $N=18$ orientation sweep.}
Evolution of $\langle Z_8(t)\rangle$ for (a) $\theta=\pi/6$, (b) $\theta=\pi/4$, and (c) $\theta=\pi/2$.
Parameters are as in Fig.~\ref{fig:theta-sweep}.
Colors denote $m=1,2,3$; solid and dashed curves distinguish covariance and HS propagation, respectively.
Black curves show exact state-vector dynamics, and the dotted line marks the upper Pauli bound $\langle Z_8\rangle=1$.
}
    \label{fig:SI_N18_theta_sweep}
\end{figure}

Figure~\ref{fig:SI_N18_theta_sweep} resolves in time the orientation dependence summarized by Fig.~\ref{fig:theta-sweep}.
At $\theta=\pi/6$, all three covariance trajectories eventually violate the exact Pauli expectation bound, while no corresponding violation is observed for HS over the displayed interval.
The deviations develop progressively, and covariance PP can remain closer to the exact trajectory over finite portions of the evolution.
An eventual expectation-bound violation therefore does not imply that the approximation was less accurate at every preceding time.

At $\theta=\pi/4$, the covariance $m=2$ trajectory exhibits a small expectation-bound violation, while no such violation is observed for covariance $m=1$ or $m=3$.
The HS $m=2$ and $m=3$ trajectories also violate the bound.
These results show that expectation-bound violations are not specific to covariance projection.

At $\theta=\pi/2$, all three covariance trajectories remain close to the exact dynamics without witnessed expectation-bound violations.
By contrast, the HS $m=3$ trajectory exceeds the upper Pauli bound.
The different behavior across body orders also illustrates that increasing $m$ does not guarantee monotonic improvement of independently propagated trajectories.

These comparisons distinguish trajectory accuracy from spectral failure.
An expectation value outside $[-1,1]$ certifies that the approximate observable has left the exact Pauli spectral interval, but remaining within this interval does not certify spectral admissibility.
Likewise, an expectation-bound violation does not by itself identify a covariance-null compatibility failure.
Establishing the latter requires the separate projected-direction analysis discussed in Sec.~\ref{sec:dynamical-compatibility}.

\subsection{Large-system threshold robustness at $N=80$}
\label{subsec:SI-N80-eps1e8}

\begin{figure}
    \centering
    \includegraphics[width=1\linewidth]{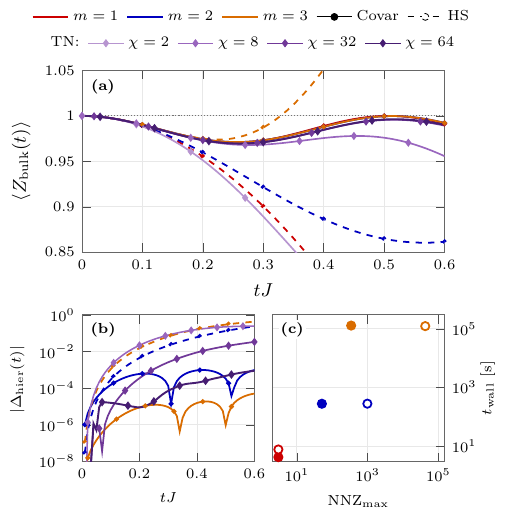}
\caption{
\textbf{Large-system benchmark at the relaxed threshold $\varepsilon=10^{-8}$.}
The calculation of Fig.~\ref{fig:large-system-ex} is repeated with a relaxed PP reconstruction threshold, leaving the physical model and TN hierarchy unchanged.
(a) Covariance PP (solid) and HS PP (dashed) for $m=1,2,3$, compared with TDVP--MPO trajectories (purple, with diamonds) for $\chi=2,8,32,64$.
Darker purple denotes larger $\chi$; the dotted line marks the Pauli bound $\langle Z_{37}\rangle=1$ on a magnified vertical axis.
(b) Successive-hierarchy discrepancies [Eq.~\eqref{eq:large-hierarchy-closure}].
(c) End-to-end wall time versus $\mathrm{NNZ}_{\max}$, with filled/open circles denoting covariance/HS propagation.
}
    \label{fig:SI-N80-eps1e8}
\end{figure}

We complement the main-text $N=80$ benchmark with a looser adaptive threshold, $\varepsilon=10^{-8}$.
The Hamiltonian, lattice, reference state, observable $Z_{37}$, retained body orders, and Krylov parameters are unchanged.
The TN trajectories are also unchanged, since the PP reconstruction threshold does not enter the TDVP--MPO evolution.
Figure~\ref{fig:SI-N80-eps1e8} compares the resulting trajectories, hierarchy discrepancies, and computational costs.

At $\varepsilon=10^{-8}$, none of the covariance trajectories undergo adaptive HB reconstruction over $tJ\in[0,0.6]$.
The HS trajectories follow different reconstruction histories.
At $m=1$, HS reconstructs at every time step, while at $m=2$ reconstructions occur at $tJ=0.23$ and $0.46$, and at $m=3$ at $tJ=0.24$ and $0.48$.
These differences arise under the same adaptive reconstruction rule.
Since the tail criterion of Eq.~\eqref{eq:hb-tail-functional} is evaluated in Euclidean HB-coordinate norm rather than the covariance Gram metric, equal thresholds need not produce identical reconstruction histories.

Despite the absence of covariance reconstructions, the body-order hierarchy remains tightly closed:
\begin{align*}
    \max_t \Delta_{\rm cov}^{(2,1)}
    &\simeq 9.8\times10^{-4},\\
    \max_t \Delta_{\rm cov}^{(3,2)}
    &\simeq 5.0\times10^{-5}.
\end{align*}
Here the superscripts identify successive retained body orders, as in Eq.~\eqref{eq:large-hierarchy-closure}.
The corresponding HS discrepancies are substantially larger:
\begin{align*}
    \max_t \Delta_{\rm HS}^{(2,1)}
    &\simeq 2.39\times10^{-1},\\
    \max_t \Delta_{\rm HS}^{(3,2)}
    &\simeq 4.26\times10^{-1}.
\end{align*}

The independent TN hierarchy shows successive improvement with increasing bond dimension:
\begin{align*}
    \max_t \Delta_{\rm TN}^{(8,2)}
    &\simeq 2.43\times10^{-1},\\
    \max_t \Delta_{\rm TN}^{(32,8)}
    &\simeq 3.36\times10^{-2},\\
    \max_t \Delta_{\rm TN}^{(64,32)}
    &\simeq 8.8\times10^{-4}.
\end{align*}
The maximum discrepancy between covariance PP at $m=3$ and TN at $\chi=64$ is approximately $3.7\times10^{-3}$.
Thus, the covariance hierarchy remains internally consistent and close to the converged TN sequence despite the absence of adaptive reconstruction.

The HS $m=3$ trajectory first violates the exact Pauli expectation bound at approximately $tJ=0.33$, reaching
\begin{equation*}
    \max_t
    \left|
        \langle Z_{37}(t)\rangle_{\rm HS,m=3}
    \right|
    \simeq 1.287.
\end{equation*}
No corresponding expectation-bound violation is observed for covariance PP or TN.
These comparisons provide a posteriori consistency evidence rather than an exact large-system error estimate.

The maximum assembled-observable sparsities for $m=1,2,3$ are
\begin{equation*}
    \mathrm{NNZ}_{\max}^{\rm cov}
    =
    (3,51,343),
    \qquad
    \mathrm{NNZ}_{\max}^{\rm HS}
    =
    (3,985,42606).
\end{equation*}
The large separation at $m=2$ and $m=3$ must be interpreted in the context of the different reconstruction histories.
Covariance PP remains in its initial HB chart, whereas HS PP reconstructs twice at each of these body orders.
The values therefore characterize the complete adaptive workflows rather than an intrinsic compression factor at fixed reconstruction history.

The reduced covariance sparsity does not produce a comparable wall-time advantage.
At $m=3$, the end-to-end wall times are approximately $1.30\times10^5$~s for covariance and $1.24\times10^5$~s for HS.
These measurements show that the total computational cost is not determined solely by the sparsity of the assembled observable.
HB construction, commutator generation, Gram operations, and reconstruction also contribute to the measured wall time.

The present calculation provides a complementary threshold-sensitivity check for the main-text result.
At $\varepsilon=10^{-8}$, covariance PP remains in its initial chart, while tightening to $\varepsilon=10^{-9}$ activates adaptive reconstruction without changing the favorable hierarchy or its agreement with TN.
The result is therefore not contingent on the absence of covariance reconstruction, although it does not constitute an exact convergence test in the adaptive threshold.

\subsection{Cold-start preprocessing costs}

To characterize the implementation-level cost of constructing the restricted Heisenberg dynamics independently of the different physical benchmarks considered above, we perform a separate controlled set of cold-start preprocessing calculations.
In these runs the Hamiltonian, reference state, projection geometry, observable type, Krylov parameters, numerical thresholds, and propagation protocol are held fixed, while only the three-dimensional lattice size and retained body order are varied.
The resulting wall times and peak process memory usages are reported in Table~\ref{tab:wall-time-scaling}. 
Because the subsequent propagation consists of a single step of duration $\delta t J=0.01$ and takes only $10^{-4}$--$10^{-3}$~s, the measured cost is overwhelmingly due to structural preprocessing.
We therefore report the preprocessing wall time directly rather than interpreting these measurements as end-to-end dynamical scaling.

\begin{table*}[t]
\centering
\caption{
\textbf{Measured cold-start preprocessing costs of the three-dimensional PP implementation.}
All calculations use open lattices and the nonintegrable Ising Hamiltonian
with $J=1$, $h_x=0.75$, and $h_z=0.15$, together with the uniform pure
$+z$ product reference state ($r=1$, $\theta=\pi/2$) and covariance
projection.
The seed observable is the local $Z_i$ listed in the third column;
$\mathbf{x}_i$ gives its lattice coordinate and $q_i$ its nearest-neighbor
coordination number.
The common numerical parameters are
$\ell=5$, $p=2$, $\zeta=1$,
$\varepsilon_{\rm tail}=10^{-8}$,
$\varepsilon_{\rm sp}=10^{-10}$,
$\varepsilon_G=10^{-12}$,
$\varepsilon_{\rm NNZ}=10^{-12}$,
and generator parameter $\eta=1$.
A $3\times3\times3$ algebra-exploration domain is used whenever supported
by the physical lattice; for $N=18$ it is restricted to the physical
$3\times3\times2$ geometry.
Each run starts from forced backend recompilation and is followed by only
one propagation step, $\delta tJ=T_{\max}J=0.01$; no adaptive rebuild occurs.
$t_{\rm pre}$ is the measured preprocessing wall time and
RSS$_{\max}$ is the process-local high-water resident-set size.
Calculations were performed on the same AMD Ryzen 9 5900X workstation
(12 physical cores/24 hardware threads, 64~GiB RAM).
A dash denotes a calculation absent from the present controlled benchmark set.
}
\label{tab:wall-time-scaling}

\begin{tabular}{rcc|rr|rr|rr}
\hline
&
&
&
\multicolumn{2}{c|}{$m=1$}
&
\multicolumn{2}{c|}{$m=2$}
&
\multicolumn{2}{c}{$m=3$}
\\
$N$
& lattice
& $Z_i:\mathbf{x}_i\;(q_i)$
& $t_{\rm pre}$ [s] & RSS$_{\max}$ [GiB]
& $t_{\rm pre}$ [s] & RSS$_{\max}$ [GiB]
& $t_{\rm pre}$ [s] & RSS$_{\max}$ [GiB]
\\
\hline

18
& $3\times3\times2$
& $Z_8:(1,1,0)\;(5)$
& 2.18 & 0.06
& 16.91 & 0.07
& 193.03 & 0.15
\\

27
& $3\times3\times3$
& $Z_{13}:(1,1,1)\;(6)$
& 5.33 & 0.06
& 107.32 & 0.09
& 2338.10 & 0.54
\\

36
& $4\times3\times3$
& $Z_{22}:(2,1,1)\;(6)$
& 5.40 & 0.06
& 112.19 & 0.13
& 2471.72 & 1.28
\\

45
& $5\times3\times3$
& $Z_{22}:(2,1,1)\;(6)$
& 5.44 & 0.06
& 118.26 & 0.19
& 4346.34 & 2.86
\\

54
& $6\times3\times3$
& $Z_{31}:(3,1,1)\;(6)$
& 5.36 & 0.07
& 129.64 & 0.29
& 9449.75 & 5.77
\\

60
& $5\times4\times3$
& $Z_{28}:(2,1,1)\;(6)$
& 5.44 & 0.07
& 152.99 & 0.37
& -- & --
\\

72
& $6\times4\times3$
& $Z_{31}:(2,2,1)\;(6)$
& 5.53 & 0.07
& 211.72 & 0.58
& -- & --
\\

80
& $5\times4\times4$
& $Z_{37}:(2,1,1)\;(6)$
& 4.23 & 0.08
& 269.75 & 0.78
& $1.30\times10^{5}$ & 26.66
\\

90
& $6\times5\times3$
& $Z_{37}:(2,2,1)\;(6)$
& 5.62 & 0.08
& 421.03 & 1.07
& -- & --
\\

120
& $6\times5\times4$
& $Z_{50}:(2,2,2)\;(6)$
& 5.92 & 0.10
& 1473.59 & 2.54
& -- & --
\\

150
& $6\times5\times5$
& $Z_{62}:(2,2,2)\;(6)$
& 6.44 & 0.12
& 4223.05 & 4.85
& -- & --
\\

\hline
\end{tabular}
\end{table*}

\clearpage
\bibliography{refs}

\end{document}